\documentclass[acmsmall, authorversion=true, nonacm=true]{acmart}

\usepackage{microtype}

\usepackage{amsfonts}
\usepackage{amsmath}
\usepackage{amsthm}
\usepackage{mathtools}

\usepackage{xspace}
\usepackage{enumitem}

\usepackage{xcolor}
\usepackage{stmaryrd} \usepackage{soul}

\usepackage{booktabs}
\usepackage{array}

\usepackage{thmtools}
\usepackage{thm-restate}

\theoremstyle{acmplain}
\newtheorem{theorem}{Theorem}[section]

\theoremstyle{remark}

\newtheorem{observation}[theorem]{Observation}

\usepackage{algpseudocode}
\usepackage{algorithm}

\usepackage{enumitem}

\usepackage{tikz}
\usetikzlibrary{automata}
\usetikzlibrary{backgrounds}
\usetikzlibrary{fit}
\usetikzlibrary{calc}
\usetikzlibrary{decorations.pathreplacing,decorations.pathmorphing}
\usetikzlibrary{positioning}

\usepackage{graphicx}

\newlength\boxwidth
\newlength\questionwidth

\newcommand{\algproblemWidth}[4]{
  \setlength\boxwidth{#1\linewidth}{
    \setlength\questionwidth{\boxwidth}\addtolength\questionwidth{-2cm}{
    \begin{center}
      \fbox{\parbox[t]{\boxwidth}{\centerline{#2}
          \vspace{1ex}
          \begin{tabular}{lp{\questionwidth}}
            Input: & #3\\[1pt]
            Question: & #4
          \end{tabular}}}
    \end{center}}}}

\newcommand{\algproblem}[3]{\algproblemWidth{.95}{#1}{#2}{#3}}

\title{Split-Correctness in Information Extraction}

\author{Johannes Doleschal}
\affiliation{\institution{University of Bayreuth \& Hasselt University}
  \country{Germany \& Belgium}}
\email{johannes.doleschal@uni-bayreuth.de}

\author{Benny Kimelfeld}
\affiliation{\institution{Technion}\country{Israel}}
\email{bennyk@cs.technion.ac.il}

\author{Wim Martens}
\affiliation{\institution{University of Bayreuth}\country{Germany}}
\email{wim.martens@uni-bayreuth.de}

\author{Frank Neven}
\affiliation{\institution{Hasselt University \& transnational University of Limburg}\country{Belgium}}
\email{frank.neven@uhasselt.be}

\author{Matthias Niewerth}
\affiliation{\institution{University of Bayreuth}\country{Germany}}
\email{matthias.niewerth@uni-bayreuth.de}

\newcommand{\eqdef}{\mathrel{{:}{=}}}
\newcommand{\eqnumber}[1]{\overset{\mathclap{(#1)}}{=}}
\newcommand{\set}[1]{\mathord{\{#1\}}}
\newcommand{\bigset}[1]{\mathord{\big\{#1\big\}}}
\newcommand{\e}[1]{\emph{#1}}

\newcommand{\join}{\mathbin{\bowtie}}

\newcommand{\doc}{\ensuremath{d}\xspace}
\newcommand{\docs}{\ensuremath{\Sigma^*}\xspace}
\newcommand{\spanner}{\ensuremath{P}\xspace}
\newcommand{\splitter}{\ensuremath{S}\xspace}
\newcommand{\splitspanner}{\ensuremath{\spanner_\splitter}\xspace}
\newcommand{\canonsplitspanner}{\ensuremath{\spanner_\splitter^{\mathrm{can}}}\xspace}
\newcommand{\canonlang}{\ensuremath{\reflang^{\canonsplitspanner}}\xspace}
\newcommand{\splitterprepost}{\ensuremath{\splitter_{\mathrm{middle}}}\xspace}

\newcommand{\toSpanner}[1]{\ensuremath{{\llbracket #1 \rrbracket}}}

\newcommand{\vars}{{\operatorname{Vars}}\xspace}
\newcommand{\svars}{\ensuremath{{\operatorname{SVars}}}\xspace}
\newcommand{\dom}{\vars}
\newcommand{\tup}{\ensuremath{\mathrm{t}}\xspace}
\newcommand{\projectTup}[2]{\ensuremath{\pi_{#2}(#1)}\xspace}
\newcommand{\emptytup}{\ensuremath{()}\xspace}

\newcommand{\spanFromTo}[2]{\ensuremath{[#1,#2\rangle}}
\newcommand{\spanij}{\spanFromTo{i}{j}\xspace}

\newcommand{\shiftSpanBy}[2]{#1\mathbin{\gg}#2}
\newcommand{\unshiftSpanBy}[2]{#1\mathbin{\ll}#2}

\newcommand{\splitcorrectness}{\textsf{Split{-}Correctness}\xspace}
\newcommand{\selfsplitcorrectness}{\selfsplittability}
\newcommand{\selfsplittability}{\textsf{Self{-}Splittability}\xspace}
\newcommand{\splittability}{\textsf{Splittability}\xspace}
\newcommand{\splitexistence}{\textsf{Split{-}Existence}\xspace}
\newcommand{\splitexistenceprepost}{\textsf{\splitterprepost{-}Split{-}Existence}\xspace}
\newcommand{\containment}{\textsf{Containment}\xspace}

\newcommand{\disjoint}{\textsf{Disjoint}\xspace}
\newcommand{\proper}{\textsf{Proper}\xspace}
\newcommand{\cover}{\textsf{Cover}\xspace}
\newcommand{\highlander}{\textsf{Highlander}\xspace}

\newcommand{\lang}{\ensuremath{\mathcal{L}}\xspace}
\newcommand{\refWord}{\ensuremath{\bold{r}}\xspace}
\newcommand{\refWordPrime}{\ensuremath{\bold{r'}}\xspace}
\newcommand{\reflang}{\ensuremath{\mathcal{R}}\xspace}
\newcommand{\pre}{\mathrm{pre}}

\newcommand{\post}{\mathrm{post}}

\newcommand{\vset}{\text{VSet}}

\newcommand{\cC}{\ensuremath{\mathcal{C}}\xspace}
\newcommand{\cCg}{\ensuremath{\mathcal{C_{\mathrm{general}}}}\xspace}
\newcommand{\cCt}{\ensuremath{\mathcal{C_{\mathrm{tractable}}}}\xspace}

\newcommand{\rgx}{\ensuremath{\mathbf{RGX}}\xspace}
\newcommand{\rgxfunc}{\ensuremath{\mathbf{fRGX}}\xspace}
\newcommand{\rgxseq}{\ensuremath{\mathbf{sRGX}}\xspace}
\newcommand{\srgx}{\rgxseq}
\newcommand{\frgx}{\rgxfunc}

\newcommand{\vsa}{\ensuremath{\mathbf{VSA}}\xspace}
\newcommand{\fvsa}{\ensuremath{\mathbf{fVSA}}\xspace}
\newcommand{\svsa}{\ensuremath{\mathbf{sVSA}}\xspace}

\newcommand{\uvsa}{\ensuremath{\mathbf{uVSA}}\xspace}
\newcommand{\ufvsa}{\ensuremath{\mathbf{ufVSA}}\xspace}
\newcommand{\usvsa}{\ensuremath{\mathbf{usVSA}}\xspace}
\newcommand{\dvsa}{\ensuremath{\mathbf{dVSA}}\xspace}
\newcommand{\dfvsa}{\ensuremath{\mathbf{dfVSA}}\xspace}
\newcommand{\dsvsa}{\ensuremath{\mathbf{dsVSA}}\xspace}

\newcommand{\varop}[1]{\Gamma_{#1}}
\newcommand{\vop}[1]{\mathop{#1{\vdash}}}
\newcommand{\vcl}[1]{\mathbin{{\dashv}#1}}

\newcommand{\clr}{\ensuremath{\operatorname{doc}}\xspace}
\newcommand{\refWordFrom}[2]{\ensuremath{\operatorname{ref}(#1,#2)}\xspace}
\newcommand{\toTuple}[1]{\ensuremath{\operatorname{tup}(#1)}\xspace}
\newcommand{\toITuple}[1]{\ensuremath{\operatorname{tup}_{#1}}\xspace}

\newcommand{\scs}{\ensuremath{\Lambda}\xspace}
\newcommand{\bbspanner}{\ensuremath{\lambda}\xspace}

\newcommand{\prim}{\ensuremath{\mathsf{Language\text{-}Primality}}\xspace}
\newcommand{\decomp}[1]{\ensuremath{#1\textrm{-}\mathsf{Decomposable}}\xspace}

\newcommand{\sconst}[3]{\ensuremath{#1 \equiv_{#3} #2}\xspace}
\newcommand{\sconstL}[2]{\sconst{#1}{#2}{\lang}}

\newcommand{\logspace}{\text{LOGSPACE}\xspace}
\newcommand{\nl}{\text{NL}\xspace}
\newcommand{\ptime}{\text{PTIME}\xspace}

\newcommand{\conp}{\text{coNP}\xspace}
\newcommand{\pspace}{\text{PSPACE}\xspace}

\newcommand{\expspace}{\text{EXPSPACE}\xspace}

\newcommand{\mop}{\odot}
\newcommand{\mzero}{e}
\newcommand{\id}{\operatorname{id}\xspace}
\newcommand{\acc}{{\mathrm{acc}}}

\begin{document}

\begin{abstract}
  Programs for extracting structured information from text, namely
  \emph{information extractors}, often operate separately on document
  segments obtained from a generic splitting operation such as
  sentences, paragraphs, $k$-grams, HTTP requests, and so on. An
  automated detection of this behavior of extractors, which we refer
  to as \emph{split-correctness}, would allow text analysis systems to
  devise query plans with parallel evaluation on segments for
  accelerating the processing of large documents.  Other applications
  include the incremental evaluation on dynamic content, where
  re-evaluation of information extractors can be restricted to revised
  segments, and debugging, where developers of information extractors
  are informed about potential boundary crossing of different semantic
  components. 
  We propose a new formal framework for split-correctness within the
  formalism of document spanners. Our analysis studies the
  complexity of split-correctness over regular spanners. We also
  discuss different variants of split-correctness, for instance, in
  the presence of black-box extractors with \emph{split
    constraints}.
\end{abstract}



\maketitle
\section{Introduction}
\e{Information extraction} (IE), the extraction of structured data
from text, is a core operation when dealing with text in data
analysis. Programming frameworks for IE, and especially
\e{declarative} ones, are designed to facilitate the development of IE
solutions. For example, IBM's
SystemT~\cite{DBLP:conf/acl/ChiticariuKLRRV10} exposes a SQL-like
declarative language, \e{AQL} (Annotation Query Language), which
provides a collection of ``primitive'' extractors (e.g., tokenizer,
dictionary lookup, Part-Of-Speech (POS) tagger, and regular-expression
matcher) alongside the relational algebra for manipulating these
relations. In Xlog~\cite{DBLP:conf/vldb/ShenDNR07}, user-defined
functions are used as primitive extractors, and Datalog is used for
relation manipulation. In
DeepDive~\cite{DBLP:journals/pvldb/ShinWWSZR15,DBLP:journals/sigmod/SaRR0WWZ16},
rules are used for generating features that are translated into the
factors of a statistical model with machine-learned parameters.  Such
rules play the role of generators of \e{noisy training data} (``labeling
functions'') in the Snorkel
system~\cite{DBLP:journals/pvldb/RatnerBEFWR17}.

When applied to a large document, an IE function may incur a high computational
cost and, consequently, an impractical execution time. However, it is frequently
the case that the program, or at least most of it, can be distributed by
separately processing smaller chunks in parallel. For instance, Named Entity
Recognition (NER) is often applied separately to different
sentences~\cite{leaman2008banner,DBLP:conf/naacl/LampleBSKD16}, and so are
instances of \e{Relation
  Extraction}~\cite{DBLP:conf/emnlp/ZengLC015,DBLP:conf/aaai/MadaanMMRS16}.
Algorithms for \e{coreference resolution} (identification of places that refer
to the same entity) are typically bounded to limited-size windows; for instance,
Stanford's well known \e{sieve}
algorithm~\cite{DBLP:conf/emnlp/RaghunathanLRCSJM10} for coreference resolution
processes separately intervals of three sentences~\cite{lee2011stanford}.
Sentiment extractors typically process individual paragraphs or even
sentences~\cite{DBLP:conf/acl/PangL04}. It is also common for extractors to
operate on windows of a bounded number $N$ of words (tokens), also known as
\e{$N$-grams} or \e{local
  contexts}~\cite{DBLP:conf/eacl/GiulianoLR06,DBLP:conf/ijcnlp/ChenJTN05a}.
Finally, machine logs often have a natural split into semantic chunks: query
logs into queries, error logs into exceptions, web-server logs into HTTP
messages, and so on.

Tokenization, $N$-gram extraction, paragraph segmentation (identifying paragraph
breaks, whether or not marked explicitly~\cite{DBLP:journals/coling/Hearst97}),
sentence boundary detection, and machine-log itemization are all examples of
what we call \e{splitters}. When IE is programmed in a development framework
such as the aforementioned ones, we aspire to deliver the premise of being
declarative---the developer specifies \e{what} end result is desired, and not
\e{how} it is accomplished efficiently. In particular, we would like the system
to automatically detect the ability to split and distribute. This ability may be
crucial for the developer (e.g., data scientist) who often lacks the expertise
in software and hardware engineering. In this paper, we embark on a principled
exploration of automated inference of \e{split-correctness} for information
extractors. That is, we explore the ability of a system to detect whether an IE
function can be applied separately to the individual segments of a given
splitter, \e{without changing the semantics}.

The basic motivation comes from the scenario where a long document is pre-split by some conventional splitters (like the aforelisted ones), and developers provide different IE functions. If the system detects that the provided IE function is correctly splittable, then it can utilize its multi-processor or distributed hardware to parallelize the computation. Moreover, the system can detect that IE programs are frequently splittable, and recommend the system administrator to materialize splitters upfront. Even more, the split guarantee facilitates \e{incremental maintenance}: when a large document undergoes a minor edit, like in the Wikipedia model, only the relevant segments (e.g., sentences or paragraphs) need to be reprocessed. Later in this section, we discuss additional motivating scenarios for split-correctness.

\subsection{Formal Framework}
Our framework adopts the formalism of \e{document spanners} (or just
\e{spanners} for short)~\cite{FaginKRV15-jacm}. In this framework, we consider
\e{documents} (strings) over a fixed finite alphabet. A spanner extracts from
every input document a set of tuples over intervals within the document. An
interval, called \e{span}, is represented simply by its starting and ending
indices in the document. An example of a spanner is a \e{regex formula}---a
regular expression with capture variables that correspond to the relational
attributes. The most studied spanner language is that of the \e{regular}
spanners, that is,  the closure of regex formulas under a subset of relational
algebra: projection, natural join, union, and
difference~\cite{FaginKRV15-jacm}.\footnote{Adding string equality selection
  would result in \emph{Core-Spanners}, which are more powerful.} Other equally expressive formalisms are non-recursive
Datalog over regex formulas~\cite{FaginKRV16} and the \e{variable-set automaton}
or (or \vset-automaton for short), which is an NFA that can open and close
variables while running~\cite{FaginKRV15-jacm}. 

The following formal concepts are the basis of our framework. A \e{splitter} is
a spanner $S$ that outputs a \e{set of intervals} (e.g., sentences, paragraphs,
$N$-grams, HTTP requests, etc.). A spanner $P$ is \e{self-splittable} by a
splitter $S$ if for all documents $d$, evaluating $P$ on $d$ gives the same
result as the union of the evaluations of $P$ on each of the chunks produced by
$S$. We also consider the more general case where we allow the spanner on the
chunks produced by $S$ to be some spanner $P_S$ different from $P$. In this
case, we say that \e{$P$ is splittable by $S$ via $P_S$}. If, for given $P$
and $S$, such a spanner $P_S$ exists, then we say that $P$ is \e{splittable} by
$S$. With these definitions, we formally define several computational problems,
each parameterized by a class $\cC$ of spanners. In the \e{\splitcorrectness}
problem, we are given $P$, $S$, and $P_S$, and the goal is to determine whether
$P$ is splittable by $S$ via $P_S$. In the \e{\splittability}
(resp.,~\e{\selfsplittability}) problem, we are given $P$ and $S$ and the goal
is to determine whether $P$ is splittable (resp.,~\e{self-splittable}) by $S$.
We also consider other settings, which we will discuss in the later sections.
In our analysis, we consider the classes of regex formulas and \vset-automata, as well as \vset-automata in known normal forms, namely \e{sequential,
  functional, unambiguous,} and \e{deterministic}. As we discuss later on, we use a slightly stronger definition of determinism than the one proposed in prior-art~\cite{MaturanaRV18}.

We show several complexity results for the studied classes of spanners. For one, the problems \splitcorrectness and \selfsplittability are PSPACE-complete for regex formulas and \vset-automata. Furthermore, we also characterize a sufficient condition for the tractability of \splitcorrectness and \selfsplittability for sequential and unambiguous \vset-automata.
This condition, which we will call the \emph{highlander condition,}\footnote{This
  is in acclimation to the tagline ``There can be only one'' of the Highlander
  movie. It will become clear why we choose this name later on.} also reduces to \pspace-completeness the complexity of \splittability, which is solvable in \expspace in general.
One key property of splitters that, most of the time, is sufficient (but not necessary) for
the highlander condition is the \e{disjointness} of the splitter. Disjointness is a natural property---it requires the splitter $\splitter$ to be such that  for all input documents, the spans produced by $\splitter$ are pairwise disjoint (non-overlapping), such as in the case of tokenization, sentence boundary detection, paragraph splitting, and paragraph segmentation.  Examples of \e{non-disjoint} splitters include $N$-grams and pairs of consecutive sentences.

Interestingly, to establish this tractability result, we needed to
revisit past notions and findings on determinism in
VSet-automata. Specifically, our notion of determinism is \e{stronger}
than that of Maturana et al.~\cite{MaturanaRV18} (without loss of
expressive power). We require that whenever the \vset-automata handle
multiple variables on the same position of the document, it does so in
a predefined order on the variables. This requirement is crucial,
since our tractability proof uses the fact that containment of
unambiguous sequential \vset-automata is solvable in polynomial time;
we prove it in Section~\ref{sec:spanner-containment}. These results
contrast some of the results of Maturana et al.~\cite{MaturanaRV18} in
the following manner. Using our stronger form of determinism, we can
show that the containment problem for deterministic VSet-automata is
solvable in polynomial time. In contrast, we prove that this problem becomes
\pspace-complete when one uses the definition of determinism of
Maturana et al.~\cite{MaturanaRV18}. The latter \pspace-completeness
result also stands in contrast with a claim that the problem is in
coNP~\cite{MaturanaRV18}.\footnote{We have contacted the authors to
  resolve the differences and verify the correctness of our proof of
  \pspace-hardness.}
Note that these results are of independent interest.

Following our analysis of \splitcorrectness and \splittability for
regular spanners, we turn to discussing additional problems that arise
in our framework.  In Section~\ref{sec:primality}, we study the
problem of \emph{\splitexistence}: given a spanner $P$, is there any
nontrivial splitter $S$ such that $P$ is splittable by $S$? Even
though we do not solve this problem, we connect it to the problem of
\e{language primality}~\cite[Problem~2.1]{Salomaa08}, a classic
problem in Formal Language Theory that is still not completely
understood. More precisely, we prove that a special case of
\splitexistence is equivalent to a variant of the language primality
problem for which the complexity is still open. In
Section~\ref{sec:reasoning}, we study the splitter framework in the
context of the relational algebra. We establish results
on the associativity of composition, the transitivity of self-splittability,
and the distributivity of composition and join.

In addition, we discuss problems that arise in natural extensions of
the basic framework. One of these problems captures the case where
some of the spanners in the query are treated as \e{black boxes} in a
formalism that we do not understand well enough to analyze (as opposed
to, e.g., regex formulas), and yet, are known to be splittable by the
splitters at hand. For example, a coreference resolver may be
implemented as a decision tree over a multitude of
features~\cite{DBLP:journals/coling/SoonNL01} but still be splittable
by sequences of three sentences, and a POS and a NER tagger may be
implemented by a bidirectional LSTM-CNN (Long short-term memory
convolutional neural network)~\cite{DBLP:journals/tacl/ChiuN16} and a
bidirectional dependency
network~\cite{DBLP:conf/naacl/ToutanovaKMS03}, respectively, but still
be splittable by sentences. Technically, our results heavily rely on
the algebraic properties of the splitter framework (associativity,
transitivity, distributivity) that we established earlier. Additional
problems we discuss are split-correctness and splittability under the
assumption that the document conforms to a regular language.

Our framework can be seen as an extension of the \e{parallel-correctness}
framework as proposed by Ameloot et al.~\cite{AmelootGKNS17-jacm,
  AmelootGKNS17-CommACM}. That work considers the parallel evaluation of
relational queries. In our terms, that work studies self-splittability where
spanners are replaced by relational queries and splitters by \e{distribution
  policies}.

\subsection{Further Motivation}
Besides the obvious, there are additional, perhaps less straightforward,
motivations. For one, even if the document is \e{not} split at evaluation time
(as opposed to pre-split), this split allows to parallelize the evaluation
\e{following} a sequential split. When the IE function is expensive, this can be
quite beneficial. For example, we have extracted $N$-grams from 1.53 GB
Wikipedia sentences and observed that this method (first split to sentences and
then distribute) gives a runtime improvement of 2.1x for $N=2$ and 3.11x for
$N=3$, all over $5$ cores. In a similar experiment on 279 MB of
PubMed\footnote{\url{https://www.ncbi.nlm.nih.gov/pubmed/}} sentences, the
speedup was 1.9x.

Another motivation comes from programming over distribution frameworks such as
Apache Hadoop~\cite{hadoop2009hadoop} and Apache
Spark~\cite{DBLP:journals/cacm/ZahariaXWDADMRV16}. In common cases, the text is
already given as a collection of small documents (e.g., tweets, reviews,
abstracts) that allow for a parallel evaluation to begin with. While we have not
seen this scenario as a motivation for our framework, it turns out that
splitting can make a considerable difference even then. For illustration, we ran
a simple event extractor of financial transactions between organizations from
sentences of around 9,000 Reuters articles over Spark. When we broke each
article into sentences, the running time reduced by 1.99x on a 5-node cluster.
We ran a similar experiment on sentences of around 570,000 reviews from the
Amazon Fine Food Reviews
dataset,\footnote{\url{https://www.kaggle.com/snap/amazon-fine-food-reviews}}
where the goal is to extract targets of a negative sentiment; we observed a
4.16x speedup. We found this remarkable, because the same amount of
parallelization was used both before and after splitting. To the best of our
understanding, this improvement can be explained by the fact that splitting
provides Spark with parallelizable tasks that are smaller in cost and larger in
number; hence, we provide Spark with considerably more (smartly exploited)
control over scheduling and resource allocation.\footnote{See
  \cite{yoavMasterThesis} for more detail on the experiments.}

Finally, another motivation comes from \e{debugging} in the
development of IE programs. For
illustration, suppose that the developer seeks HTTP requests to a
specific host on a specific date, and for that she seeks Host and Date
headers that are close to each other; the system can warn the
developer that the program is not splittable by HTTP requests like
other frequent programs over the log (i.e., it can extract the Host of
one request along with the Date of another), which is indeed a bug in
this case. In the general case, the system can provide the user with
the different splitters (sentences, paragraphs, requests, etc.) that
the program is split-correct for, in contrast with what the developer
believes should hold true.

\subsection{Organization}
The remainder of the paper is organized as follows. In Section~\ref{sec:prelim},
we give preliminary definitions and notation. We define the concepts of
splitter, split-correctness and splittability in
Section~\ref{sec:def:splitters}, and define the representations of regular
spanners in Sections~\ref{sec:prelim-regular-spanners}. We give a summary of our
main results in Section~\ref{sec:regular-results} and provide some technical
foundations in Section~\ref{sec:technical-foundations}. We discuss the upper
bounds of splitcorrectness, self-splittability and splittability in
Section~\ref{sec:upper-bounds}. The lower bounds are discussed in
Section~\ref{sec:lower-bounds}. We give the definition of split-existence and
connect it to the problem of language primality in Section~\ref{sec:primality},
and we study problems of reasoning about document spanners in
Section~\ref{sec:reasoning}. Finally, we conclude and discuss open problems in
Section~\ref{sec:conclusions}.

\subsection{Relationship to a Previous Conference Version}
This article is an extension of a previous conference
publication~\cite{DoleschalKMNN19}, and it considerably extends in
several ways. First, we establish the following significant
generalizations of the results in that version. The entire framework
is extended to the schemaless semantics of document spanners that
allows for partial answers~\cite{MaturanaRV18}.  We strengthen our
algorithmic results and generalize the previous tractable fragment of
\splitcorrectness from deterministic \vset-automata to unambiguous
\vset-automata (Section~\ref{sec:splitcorrectness}). The
\e{highlander} condition, which we introduce here, is a major
relaxation of the \e{disjointness} condition, and in some of results
we are able to replace disjointness with highlanderness.\footnote{This
  also resolves a minor mistake in the conference version~\cite[Lemma
  5.12]{DoleschalKMNN19}. That is, the Lemma does not hold if the
  spanner is not proper. This mistake was independently discovered by
  Smit~\cite{Smit20}.} Moreover, this manuscript resolves several open
problems that were stated in the conference version~\cite[Section~8]{DoleschalKMNN19}:
\begin{itemize}
\item \e{Is \splittability decidable without the assumption of
    disjointness?} We answer this question affirmatively in
  Section~\ref{sec:splittability-general-upperbound}
  (Theorem~\ref{thm:mainSplittability}).
\item \e{How are the studied problems related to the \e{language primality}
    problem?} We formalize the connection between the problems in
  Section~\ref{sec:primality}.
\item \e{In the case of \splitcorrectness and \selfsplittability, can
    we relax any of the assumptions of determinism and disjointedness
    and still retain tractability?} In
  Section~\ref{sec:regular-results}
  (Theorem~\ref{thm:splitcorrectness}), we answer this question
  affirmatively based on the new highlander condition.
\end{itemize}

We also considerably extend the study on the theory of \e{reasoning}
about the split constraints and their algebraic properties---a topic
that has been treated in only a preliminary manner in the conference
version. In particular, Section~\ref{sec:reasoning} includes several
new results
(Theorems~\ref{thm:associative},~\ref{thm:transitive},~\ref{thm:distributivity},
Section~\ref{sec:schemaConstraints}). 
Last but not least, we provide detailed and nontrivial proofs that
were absent in the abridged version.

 \section{Preliminaries on Document Spanners}\label{sec:prelim}
Our framework is within the formalism of \e{document spanners} by Fagin et
al.~\cite{FaginKRV15-jacm, FaginKRV15-sigrecord}. We first revisit some
definitions from this framework. Let $\Sigma$ be a finite set of symbols called
the \emph{alphabet}. A sequence $\doc = \sigma_1 \cdots \sigma_n$ of symbols
where every $\sigma_i \in \Sigma$ is a \emph{document (over $\Sigma$)}. If $n=0$
we denote $\doc$ by $\varepsilon$ and call $d$ \emph{empty}. By $\Sigma^*$ we
denote the set of all documents over $\Sigma$ and by $\Sigma^+$ the set of all
non-empty documents over $\Sigma$. We denote by $|\doc|$ the length $n$ of
$\doc$. A \emph{span} of $\doc$ is an expression of the form \spanij with $1
\leq i \leq j \leq n+1$. For a span \spanij of \doc, we denote by
$\doc_{\spanij}$ the string $\sigma_i \cdots \sigma_{j-1}$. A span $\spanij$ is
empty if $i=j$ which implies that $\doc_{\spanij}=\varepsilon$. For a document
\doc, we denote by \emph{Spans(\doc)} the set of all possible spans of \doc. Two
spans $\spanFromTo{i_1}{j_1}$ and $\spanFromTo{i_2}{j_2}$ are \emph{equal} if
$i_1 = i_2$ and $j_1 = j_2$. In particular, we observe that two spans do not
have to be equal if they select the same string. That is,
$\doc_{\spanFromTo{i_1}{j_1}} = \doc_{\spanFromTo{i_2}{j_2}}$ does not imply
that $\spanFromTo{i_1}{j_1} = \spanFromTo{i_2}{j_2}$. Two spans \spanij and
\spanFromTo{i'}{j'} \emph{overlap} if $i \leq i' < j$ or $i' \leq i < j'$, and
are \emph{disjoint} otherwise. Finally, \spanij \emph{covers}
\spanFromTo{i'}{j'} if $i \leq i' \leq j'\leq j$. Given a span \spanij and a
natural number $n$, we denote by $\shiftSpanBy{\spanij}{n}$ the span
$\spanFromTo{i+n}{j+n}$. Analogously, if $i > n$, we denote by
$\unshiftSpanBy{\spanij}{n}$ the span $\spanFromTo{i-n}{j-n}$.

The framework focuses on functions that extract spans from documents and assigns
them to variables. To this end, we fix a countably infinite set \svars of
\emph{span variables}, which range over spans, i.e., pairs of integers. The sets
$\Sigma$ and \svars are disjoint. A \emph{$\doc$-tuple} $\tup$ is a total
function from a finite set of variables into \emph{Spans(\doc)}. We denote the
variables of $\tup$, i.e., the set of variables on which $\tup$ is defined, by
\emph{$\dom(\tup)$}. If the document $\doc$ is clear from the context we
sometimes say \emph{tuple} instead of $\doc$-tuple. For a \doc-tuple \tup and a
set $Y \subset \svars$ we define the \doc-tuple $\projectTup{\tup}{Y}$ as the
restriction of $\tup$ to the variables in $\dom(\tup)\cap Y$. We say that a
tuple $\tup$ is \emph{empty}, denoted by $\tup = \emptytup$, if $\dom(\tup) =
\emptyset$. If $s$ is a span of $\doc$ and $\tup$ is a \doc-tuple, we say that
$s$ covers $\tup$ if $s$ covers $\tup(x)$ for every variable $x \in \dom(\tup)$.
Furthermore, let $\tup$ be a non empty $\doc$-tuple for some document $\doc \in
\docs$. We define the minimal span that covers $t$ as the span $\spanij$, where
\[
  i \;\eqdef\; \min \big\{i' \mid \spanFromTo{i'}{j'} = \tup(v), \text{ and }
    v\in\dom(\tup)\big\}\;, 
\] and 
\[
  j \;\eqdef\; \max \big\{j' \mid \spanFromTo{i'}{j'} = \tup(v), \text{ and }
    v\in\dom(\tup)\big\}\;. 
\]

If $\tup$ is a $\doc$-tuple and $n$ a natural number, we define the tuples
$\shiftSpanBy{\tup}{n}$ and $\unshiftSpanBy{\tup}{n}$ as the $\doc$-tuples that
results from shifting each span in $\tup$ by $n$. More formally, for all
variables $x \in \dom(\tup)$ we have:\footnote{Notice that when $n$ is too
  large, $\shiftSpanBy{\tup}{n}$ or $\unshiftSpanBy{\tup}{n}$ could technically
  not be a $\doc$-tuple anymore. However, we only use the operator in situations
  where this does not happen.}
\[
  (\shiftSpanBy{\tup}{n})(x)\;\eqdef\; \shiftSpanBy{\tup(x)}{n}\;,
\]
and
\[
  (\unshiftSpanBy{\tup}{n})(x)\;\eqdef\; \unshiftSpanBy{\tup(x)}{n}\;.
\]

A \emph{document spanner} (also \emph{spanner} for short) is a function \spanner
that maps every document \doc into a finite set $\spanner(\doc)$ of \doc-tuples.
By $\dom(\spanner) \eqdef \{ v \in \dom(\tup) \mid \doc \in \docs, \text{ and }
\tup \in \spanner(\doc)\}$ we denote the variables of \spanner. We note that,
following Maturana et al.~\cite{MaturanaRV18}, we do not require all tuples of a
spanner $\spanner$ assign all variables in $\dom(\spanner)$, that is, given a
document $\doc$ and a tuple $\tup$, $\dom(\tup) \subseteq \dom(\spanner)$. A
spanner $\spanner$ is called \emph{functional} if every tupel uses the same
variables, i.e., $\vars(\tup) = \vars(\spanner)$ for every document $\doc \in
\docs$ and every tupel $\tup \in \spanner(\doc)$. By $\spanner \subseteq
\spanner'$ we denote the fact that, for every document $\doc$, $\spanner(\doc)
\subseteq \spanner'(\doc)$. Furthermore, we denote by $\spanner=\spanner'$ the
fact that the spanners $\spanner$ and $\spanner'$ define the same function.

In the following, we sometimes require that a spanner only selects tuples that
use at least two different positions in $\doc$. More formally, a document
spanner $\spanner$ is \emph{proper} if for every document $\doc \in \docs$,
$\tup \in \spanner(\doc)$ implies that the minimal span that covers $\tup$ is
not empty, and $\emptytup \notin \spanner(\doc)$.

\subsection*{Algebraic Operators on Document Spanners}
We conclude this section by defining algebraic operations on spanners. We need
some basic definitions first. Two $\doc$-tuples $\tup_1$ and $\tup_2$ are
\emph{compatible} if they agree on every common variable, i.e. $\tup_1(x) =
\tup_2(x)$ for all $x\in \dom(\tup_1) \cap \dom(\tup_2)$. In this case, define
$\tup_1 \cup \tup_2$ as the tuple with $\dom(\tup_1 \cup \tup_2) = \dom(\tup_1)
\cup \dom(\tup_2)$ such that $(\tup_1 \cup \tup_2)(x) = \tup_1(x)$ for all $x
\in \dom(\tup_1)$ and $(\tup_1 \cup \tup_2)(x) = \tup_2(x)$ for all $x \in
\dom(\tup_2).$

\begin{definition}[Algebraic Operations on Spanners]\label{def:algebra}
  Let $\spanner_1, \spanner_2$ be (document) spanners and let $\doc
  \in \docs$ be a document. 
  \begin{itemize}
  \item \emph{Variable enclosing.}
    The  spanner $\spanner = x\{\spanner_1\}$ is defined by
    \[
      \spanner(\doc) \;\eqdef\; \big\{ \tup \cup \{ x \mapsto
      \spanFromTo{1}{|\doc|+1}\} \mid \tup \in 
      \spanner_1(\doc), x \in \svars \setminus \vars(t) \big\}\;.
    \]
  \item \emph{Concatenation.}
    The spanner $\spanner = \spanner_1 \cdot \spanner_2$ is defined by
    \begin{align*}
      \spanner(\doc) \;\eqdef\; \big\{ \tup_1 \cup \tup_2 \mid \
      & d=d_1 \cdot d_2,\ \tup_1 \in \spanner_1(\doc_1),
        \text{ and } \unshiftSpanBy{\tup_2}{|\doc_1|} \in \spanner_2(\doc_2)
      \big\}\;.\footnotemark
    \end{align*}
    \footnotetext{Notice that $\tup_1 \cup \tup_2$ is only defined if $\tup_1$
      and $\tup_2$ have no common variables.}
  \item \emph{Union.}
    The union $\spanner = \spanner_1 \cup \spanner_2$ is defined by
    $\spanner(\doc) \eqdef \spanner_1(\doc) \cup \spanner_2(\doc).$
  \item \emph{Projection.}
    The projection $\spanner = \pi_Y\spanner_1$ is defined by
    $\spanner(\doc) \eqdef \set{\projectTup{\tup}{Y} \mid \tup \in \spanner_1(\doc)}$.
    Recall that $\projectTup{\tup}{Y}$ denotes the restriction of $\tup$ to the variables in
    $\dom(\tup) \cap Y.$
  \item \emph{Natural Join.}
    The (natural) join $\spanner = \spanner_1 \join \spanner_2$ is defined
    such that $\spanner(\doc)$ consists of all tuples $\tup_1 \cup \tup_2$
    such that $\tup_1 \in \spanner_1(\doc)$, $\tup_2 \in \spanner_2(\doc)$, and
    $\tup_1$ and $\tup_2$ are compatible, i.e. $\tup_1(x) = \tup_2(x)$ for all $x
    \in \dom(\tup_1) \cap \dom(\tup_2)$.
  \end{itemize}
\end{definition}

 \section{General Framework and Main Problems}
\label{sec:def:splitters}
In this work, we are particularly interested in spanners that split documents
into (possibly overlapping) segments. Formally, a \emph{document splitter} (or
\emph{splitter} for short) is a functional unary document spanner $\spanner$,
that is, there is a single variable $x$ such that, for every
tuple $\tup \in \spanner(\doc)$ and $\doc \in \docs$, we have $\dom(\tup) =
\{x\}$. So, a splitter can split the document into paragraphs, sentences,
$N$-grams, HTTP messages, error messages, and so on.

In the sequel, unless mentioned otherwise, we denote a splitter by $\splitter$
and its unique variable by $x_\splitter$. Furthermore, we assume,
\mbox{w.l.o.g.}, that $x_\splitter \notin \dom(\spanner)$ for every spanner
$\spanner$. Since a splitter outputs unary span relations, its output on a
document $\doc$ can be identified with the set of spans $\{t(x_\splitter) \mid t
\in \splitter(d)\}$. We often use this simplified view on splitters and treat
their output as a set of spans. A splitter \splitter is \emph{disjoint} if the
spans extracted by \splitter are always pairwise disjoint, that is, for all
$\doc \in \docs$ and $t, t' \in \splitter(\doc)$, the spans $t(x_\splitter)$ and
$t'(x_\splitter)$ are disjoint. For instance, splitters that split documents
into spans of the form $\spanFromTo{1}{k_1}, \spanFromTo{k_1}{k_2}, \ldots$
(such as paragraphs and sentences) are disjoint, but $N$-gram extractors are not
disjoint for $N>1$.

Next, we want to define when a spanner is \emph{splittable} by a splitter, that
is, when documents can be split into components such that the operation of a
spanner can be distributed over the components. To this end, we first need some
notation. Let $\doc$ be a document, let $s\eqdef\spanij$ be a span of \doc, and
let $s_{\mathrm{local}}\eqdef\spanFromTo{i'}{j'}$ be a span of the document
$\doc_{\spanFromTo{i}{j}}$. Then, $s_{\mathrm{local}}$ also marks a span of the
original document $\doc$, namely the one obtained from $s_{\mathrm{local}}$ by
shifting it $i-1$ characters to the right. We denote this shifted span by
$s_{\mathrm{global}} \eqdef \shiftSpanBy{s_{\mathrm{local}}}{s}$, which abbreviates
$\shiftSpanBy{s_{\mathrm{local}}}{(i-1)}$ (cf., Figure~\ref{fig:shift-span}).
Hence, we have:
\[
  s_{\mathrm{global}} \;\;=\;\; \shiftSpanBy{s_{\mathrm{local}}}{s} \;\;=\;\;
  \shiftSpanBy{s_{\mathrm{local}}}{(i-1)} \;\;=\;\;
  \spanFromTo{i'+(i-1)}{j'+(i-1)}\;. 
\]
Analogously, we denote by $\unshiftSpanBy{s_{\mathrm{global}}}{s}$ the span
which is obtained from $s_{\mathrm{global}}$ by shifting it $i-1$ characters to
the left. We denote this shifted span by $s_{\mathrm{local}} =
\shiftSpanBy{s_{\mathrm{global}}}{s}$, which abbreviates
$\unshiftSpanBy{s_{\mathrm{global}}}{(i-1)}$. Hence, we have:
\[
  s_{\mathrm{local}} \;\;=\;\; \unshiftSpanBy{s_{\mathrm{global}}}{s} \;\;=\;\;
  \unshiftSpanBy{s_{\mathrm{global}}}{(i-1)} \;\;=\;\;
  \spanFromTo{i'-(i-1)}{j'-(i-1)}\;. 
\]

\begin{figure}
  \begin{tikzpicture}
    \draw [black]  (0,0)  -- (3,0);
    \draw [black]  (3,0)  -- (6,0);
    \draw [black]  (6,0)  -- (8,0);
    
    \draw [black]  (0,.5) -- (3,.5);
    \draw [black]  (3,.5) -- (6,.5);
    \draw [black]  (6,.5) -- (8,.5);

    \draw [black]  (0,0)   -- (0,.5);
    \draw [black]  (.5,0)  -- (.5,.5);
    \draw [black]  (1,0)   -- (1,.5);
    \draw [black]  (1.5,0) -- (1.5,.5);
    \draw [black]  (2,0)   -- (2,.5);
    \draw [black]  (2.5,0) -- (2.5,.5);
    \draw [black]  (3,0)   -- (3,.5);
    \draw [black]  (3.5,0) -- (3.5,.5);
    \draw [black]  (4,0)   -- (4,.5);
    \draw [black]  (4.5,0) -- (4.5,.5);
    \draw [black]  (5,0)   -- (5,.5);
    \draw [black]  (5.5,0) -- (5.5,.5);
    \draw [black]  (6,0)   -- (6,.5);
    \draw [black]  (6.5,0) -- (6.5,.5);
    \draw [black]  (7,0)   -- (7,.5);
    \draw [black]  (7.5,0) -- (7.5,.5);
    \draw [black]  (8,0)   -- (8,.5);

    \node at (.25,.25)  {$1$};
    \node at (.75,.25)  {$2$};
    \node at (1.25,.25) {$3$};
    \node at (1.75,.25)  {$4$};
    \node at (2.25,.25)  {$5$};
    \node at (2.75,.25)  {$6$};
    \node at (3.25,.25)  {$\frac{\textcolor{red}{1}}{7}$};
    \node at (3.75,.25)  {$\frac{\textcolor{red}{2}}{8}$};
    \node at (4.25,.25)  {$\frac{\textcolor{red}{3}}{9}$};
    \node at (4.75,.25)  {$\frac{\textcolor{red}{4}}{10}$};
    \node at (5.25,.25)  {$\frac{\textcolor{red}{5}}{11}$};
    \node at (5.75,.25)  {$\frac{\textcolor{red}{6}}{12}$};
    \node at (6.25,.25)  {$13$};
    \node at (6.75,.25)  {$14$};
    \node at (7.25,.25)  {$15$};
    \node at (7.75,.25)  {$16$};
    
    \node [left] at (0,.25) {$\doc$};

    \draw [dashed, red] (3,.5) -- (3,.75);
    \draw [dashed, red] (3,.75) -- (6,.75);
    \draw [dashed, red] (6,.75) -- (6,.5);
    \node [above,  red] at (4.5,.75) {$s = \spanFromTo{7}{13}$};
    
    \draw [dashed, black] (3.5,0) -- (3.5,-.25);
    \draw [dashed, black] (3.5,-.25) -- (5.5,-.25);
    \draw [dashed, black] (5.5,-.25) -- (5.5,0);
    \node [below,  black] at (4.5,-.25) {$\shiftSpanBy{s'}{s} = \spanFromTo{8}{12}$,\ \textcolor{red}{$s'=\spanFromTo{2}{6}$}};
  \end{tikzpicture}
  \vspace{-4mm}
  \caption{Visualization of the shift span operator, with $\mathbf{\spanFromTo{8}{12} =
      \shiftSpanBy{\spanFromTo{2}{6}}{\spanFromTo{7}{13}}.}$}\label{fig:shift-span}
  \vspace{-5mm}
\end{figure}

Again, we overload the notation and write $\shiftSpanBy{\tup}{s}$ (resp.,
$\unshiftSpanBy{\tup}{s}$) for the $\doc$-tuple that results from shifting each
span in $\tup$ by $s$ to the right (resp., to the left).

\begin{observation}\label{obs:sCoverst}
  Let \doc be a document, $s$ be a span of \doc, and $\tup$ be a $\doc_s$-tuple.
  Then the \doc-tuple $\tup' = \shiftSpanBy{\tup}{s}$ is covered by $s$.
  Furthermore, given a \doc-tuple $\tup$, the tuple $\unshiftSpanBy{\tup}{s}$ is
  a well defined $\doc_s$-tuple if $\tup$ is covered by $s$.
\end{observation}
  
We now define the \emph{composition} $\spanner \circ \splitter$ of a spanner
$\spanner$ and splitter $\splitter$. Intuitively, $\spanner \circ \splitter$ is
the spanner that results from evaluating \spanner on every part of the document
extracted by \splitter, with a proper shift of the indices. Recall that a
splitter $\splitter$ is functional and has exactly one variable, thus, it always
selects a set of unary tuples. In the following we abuse notation and simply
write $s$ rather than $s(x_\splitter)$ when $s\in \splitter(d)$ for some document
$\doc$. We define on every document $\doc$,
\[
  (\spanner \circ \splitter)(d) \;\eqdef \bigcup_{s\in
    \splitter(\doc)}\set{\shiftSpanBy{\tup}{s}\mid \tup\in \spanner(\doc_{s})}\;.
\] 

As an example, if $\spanner$ extracts person names and $\splitter$ is a sentence
splitter, then $\spanner \circ \splitter$ is the spanner obtained by applying
$\spanner$ to every sentence independently and taking the union of the results.
Furthermore, if $\spanner$ extracts close mentions of email addresses and phone
numbers, and $\splitter$ is the $5$-gram splitter, then $\spanner \circ
\splitter$ is obtained by applying $\spanner$ to each 5-gram individually. Since
executing \spanner on each individual output of \splitter enables
parallelization, it is interesting if there is a difference between the output
of $\spanner$ and $\spanner \circ \splitter$ on every document \doc. This
property clearly depends on the definitions of $\spanner$ and $\splitter$. We
will define it formally in Section~\ref{sec:splittability-def} under the name
\e{self-splittability}.

The following lemma gives an algebraic characterization of $\spanner \circ
\splitter$. 
\begin{lemma}\label{lem:pcircs-algebra}
  Let \spanner be a spanner and \splitter. Then $\spanner \circ \splitter =
  \pi_{\dom(\spanner)}((\Sigma^* \cdot x_\splitter\{\spanner\} \cdot \Sigma^*)
  \join \splitter)$.
\end{lemma}
\begin{proof}
  Let \spanner and \splitter be as given and let $\spanner' =
  \pi_{\dom(\spanner)}((\Sigma^* \cdot x_\splitter\{\spanner\} \cdot \Sigma^*)
  \join \splitter)$. We show both directions of the equation separately.

  ($\spanner' \subseteq \spanner \circ \splitter$): Let $\doc \in \docs$ be a
  document and $\tup' \in \spanner'(\doc)$ be a \doc-tuple. Per definition of
  $\spanner'$, there is a tuple $\tup_{x_\splitter} \in (\Sigma^* \cdot
  x_\splitter\{\spanner\} \cdot \Sigma^*) \join \splitter))(\doc)$ with $\tup'
  \eqdef \projectTup{\tup_{x_\splitter}}{\dom(\spanner)}$ and $s \eqdef
  \tup_{x_\splitter}(x_\splitter)$ covers $\tup$. Let $s \eqdef
  \tup_{x_\splitter}(x_\splitter) \in \splitter(\doc)$ and $\tup =
  \unshiftSpanBy{\tup'}{s}$ be the $\doc_s$-tuple with $\tup \in
  \spanner(\doc_s)$. Thus, due to $s \in \splitter(\doc)$ and $\tup \in
  \spanner(\doc_s)$, it must hold that $\tup' = \shiftSpanBy{\tup}{s} \in
  (\spanner \circ \splitter)(\doc)$.

  ($\spanner \circ \splitter \subseteq \spanner'$): Let $\doc \in \docs$ be a
  document, $s \in \splitter(\doc)$, and $\tup \in \spanner(\doc_s)$. Let $\tup'
  = \shiftSpanBy{\tup}{s}$, thus, by Observation~\ref{obs:sCoverst}, $s$ covers
  $\tup'$. Let $\tup_{x_\splitter}$ be the \doc-tuple defined by
  \[
    \tup_{x_\splitter}(v) \;\eqdef\;
    \begin{cases}
      \tup'(v) & \text{if } v \in \dom(\tup') \\
      s & \text{if } v = x_\splitter
      \end{cases}
  \]
  Therefore, $\tup_{x_\splitter} \in (\Sigma^* \cdot x_\splitter\{\spanner\}
  \cdot \Sigma^*) \join \splitter)(\doc)$ and $\tup' \eqdef
  \projectTup{\tup_{x_\splitter}}{\dom(\tup')} \in \spanner'(\doc)$. 
\end{proof}

\subsection{Splittability and Split-Correctness}\label{sec:splittability-def}
We say that a spanner \spanner is \e{splittable} by a splitter \splitter via a
spanner \splitspanner if evaluating \spanner on a document $\doc$ always gives
the same result as evaluating \splitspanner on every substring extracted by
\splitter (again with proper indentation of the indices). If such a
\splitspanner exists, then we say that \spanner is \e{splittable} by \splitter;
and if \splitspanner is \spanner itself, then we say that \spanner is
\e{self-splittable} by \splitter. We define these notions more formally.

\begin{definition}
  Let \spanner be a spanner and \splitter a splitter. We say that:
\begin{enumerate}
\item \spanner is \emph{splittable by \splitter via} a spanner \splitspanner, if
  $\spanner = \splitspanner \circ \splitter$;
\item \spanner is \emph{splittable by \splitter} if there exists a spanner
  \splitspanner such that $\spanner = \splitspanner \circ \splitter$;
\item \spanner is \emph{self-splittable by \splitter} if 
$\spanner = \spanner \circ \splitter$.
\end{enumerate}
We refer to $P_S$ as the \e{split-spanner}.
\end{definition}

As a simple example, suppose that we analyze a log of HTTP requests separated by
blank lines and assume for simplicity that the log only consists of \textsf{GET}
requests. Furthermore, assume that $\splitter$ splits the document into
individual requests (without the blank lines) and that $\spanner$ extracts the
request line, which is always the first line of the request. If $\spanner$
identifies the request line as the one following the blank line, then $\spanner$
is splittable by $\splitter$ via $\splitspanner$, which is the same as
$\spanner$ but replaces the requirement to follow a blank line with the
requirement of being the first line. If, on the other hand, $\spanner$
identifies the request line as being the one starting with the word
\textsf{GET}, then $\spanner$ is self-splittable by $\splitter$, since we can
apply $P$ itself to every HTTP message independently.

Other examples are as follows. Many spanners $\spanner$ that extract person
names do not look beyond the sentence level. This means that, if $\splitter$
splits to sentences, it is the case that $\spanner$ is self-splittable by
$\splitter$. Now suppose that $\spanner$ extracts mentions of email addresses
and phone numbers based on the formats of the tokens, and moreover, it allows at
most three tokens in between; if $\splitter$ is the $N$-gram splitter, then
$\spanner$ is self-splittable by $\splitter$ for $N\geq 5$ but not for $N<5$.

\subsection{Main Decision Problems}

The previous definitions and the motivating examples from the
introduction directly lead to the corresponding decision problems.  We
use \cC to denote a class of spanner representations (such as \vsa or
\rgx that we define in Section~\ref{sec:prelim-regular-spanners}).

\algproblem{$\splitcorrectness[\cC]$}{Spanners $\spanner, \splitspanner \in \cC$
  and splitter $\splitter \in \cC$.}{Is \spanner splittable by \splitter via
  \splitspanner, that is, is $\spanner = \splitspanner \circ \splitter$? }

\algproblem{$\splittability[\cC]$}{Spanner $\spanner \in \cC$ and splitter
  $\splitter \in \cC$.}{Is \spanner splittable by $\splitter$, that is, is there
  a spanner $\splitspanner \in \cC$, such that $\spanner = \splitspanner \circ
  \splitter$? }

\algproblem{$\selfsplittability[\cC]$}{Spanner $\spanner \in \cC$ and splitter
  $\splitter \in \cC$.}{Is $\spanner$ self-splittable by $\splitter$, that is,
  is $\spanner = \spanner \circ \splitter$? }

\noindent Note that $\selfsplittability[\cC]$ is a special case of 
$\splitcorrectness[\cC]$ by choosing $\splitspanner = \spanner$. It can also be
seen as a special case of $\splittability[\cC]$ in the sense that
\selfsplittability implies \splittability.

\subsection{Cover and Highlander Condition}

We now define two conditions on the interaction of spanners and splitters which
will be useful to obtain upper bounds for \splitcorrectness and \splittability.
The first condition is the cover condition, which states that, for every tuple
selected by a spanner there is at least one split covering it.

\begin{definition}[Cover Condition]
  A splitter \splitter covers a spanner $\spanner$ if for every document $\doc$
  and every non empty tuple $\tup \in \spanner(\doc)$, there exists a span $s
  \in \splitter(\doc)$ that covers $\tup$.
\end{definition}

We show now that the cover condition is indeed necessary for \splittability.
\begin{observation}\label{obs:splittabilityCover}
  Let \spanner be a spanner which is splittable by a splitter \splitter. Then
  \splitter covers \spanner.
\end{observation}
\begin{proof}
  Let $\doc$ be a document and $\tup \in \spanner(\doc)$ be a non-empty
  \doc-tuple. If \spanner is splittable by \splitter, there must be a spanner
  \splitspanner such that $\spanner = \splitspanner \circ \splitter$. By
  assumption, $\tup \in \spanner(\doc) = (\splitspanner \circ \splitter)(\doc)$,
  there is a span $s \in \splitter(\doc)$, such that $\tup' \eqdef
  \unshiftSpanBy{\tup}{s} \in \splitspanner(\doc_s)$. Thus, $\tup =
  \shiftSpanBy{\tup'}{s}$ and therefore, by Observation~\ref{obs:sCoverst}, $s$
  covers $\tup$.
\end{proof}

The second condition is the highlander condition which states that every tuple
selected by the spanner is covered by at most one split.

\begin{definition}[Highlander Condition]
  A spanner $\spanner$ and a splitter $\splitter$ satisfy the \emph{highlander
  condition} if, for every document $\doc$ and every tuple $\tup \in
  \spanner(\doc)$, there exists at most one span $s \in \splitter(\doc)$ that
  covers the tuple $\tup$.
\end{definition}

Recall from the introduction that disjointness is a natural property that
splitters often satisfy in real life (e.g., tokenization, sentence boundary
detection, paragraph splitting and segmentation). Given a disjoint splitter, it
is easy to see that the highlander condition is almost guaranteed to be
satisfied. The only case in which the highlander condition is not satisfied on
a disjoint splitter is if the spanner selects a tuple which does not cover a
nonempty part of the document, that is, \spanner is not proper.

\begin{observation}\label{obs:disjoint-highlander}
  Let $\spanner$ be a proper spanner and let \splitter be a disjoint splitter.
  Then $\spanner$ and $\splitter$ satisfy the highlander condition.
\end{observation}
\begin{proof}
  For the sake of contradiction, assume that $\spanner$ is proper and \splitter
  is disjoint but the highlander condition is not satisfied. Therefore there is
  a document $\doc \in \docs$ and a tuple $\tup \in \spanner(\doc)$, such that
  $\tup$ is covered by $\spanFromTo{i_1}{j_1}, \spanFromTo{i_2}{j_2} \in
  \splitter(\doc)$. We assume, \mbox{w.l.o.g.}, that $i_1 \leq i_2$. Let
  $\spanij$ be the minimal span covering $\tup$, which is well defined as $\tup$
  can not be empty due to $\spanner$ being proper. Therefore, $\spanij$ is
  covered by $\spanFromTo{i_1}{j_1}$ and $\spanFromTo{i_2}{j_2}$, that is $i_1
  \leq i \leq j\leq j_1$ and $i_2 \leq i \leq j\leq j_2$. Due to disjointness of
  \splitter, $\spanFromTo{i_1}{j_1} $ and $\spanFromTo{i_2}{j_2}$ must be
  disjoint, that is, $i_1 \leq j_1 \leq i_2 \leq j_2$. Thus, $\spanij$ can only
  be covered by both $\spanFromTo{i_1}{j_1}$ and $\spanFromTo{i_2}{j_2}$ if $i_1
  = i = j = j_2$. Therefore, the tuple $\spanij$ is empty, leading to the
  desired contradiction as $\spanij$ is the minimal span covering $\tup \in
  \spanner(\doc)$, which can not be empty if $\spanner$ is proper.
\end{proof}

We conclude this section by defining the corresponding decision problems. As
before, we use $\cC$ to denote a class of spanner representations (such as
$\vsa$ or $\rgx$ that we define in Section~\ref{sec:prelim-regular-spanners}). 

\noindent
\begin{center}
\noindent
\begin{minipage}{.3\linewidth}
  \algproblemWidth{1}{$\disjoint[\cC]$}{Splitter $\splitter \in \cC$.}{Is \splitter
    disjoint?}
\end{minipage}
\hspace{4.5pt}
\begin{minipage}{.6\linewidth}
  \algproblemWidth{1}{$\cover[\cC]$}{Spanner $\spanner \in \cC$ and
    splitter $\splitter \in \cC$.}{Do \spanner and \splitter satisfy the cover
    condition?}
\end{minipage}

\noindent
\begin{minipage}{.3\linewidth}
  \algproblemWidth{1}{$\proper[\cC]$}{Spanner $\spanner \in \cC$.}{Is \spanner proper?}
\end{minipage}
\hspace{4.5pt}
\begin{minipage}{.6\linewidth}
  \algproblemWidth{1}{$\highlander[\cC]$}{Spanner $\spanner \in \cC$ and splitter
    $\splitter \in \cC$.}{Do \spanner and \splitter satisfy the highlander
    condition?}
\end{minipage}
\end{center}

 \section{Representations of Regular Document
  Spanners}\label{sec:prelim-regular-spanners}

In this section, we recall the terminology and definition of regular languages
and \emph{regular} spanners~\cite{FaginKRV15-jacm}. We assume that the reader is
familiar with (non)deterministic finite state automata (abrev. NFA and DFA). By
$\lang(A)$ we denote the language accepted by a (non)deterministic finite state
automaton $A$.

We use two main models for representing spanners: \e{regex-formulas} and
\emph{\vset-automata}. Furthermore, following
Freydenberger~\cite{Freydenberger19}, we introduce so-called \e{ref-words},
which connect spanner representations with regular languages. We also introduce
various classes of \vset-automata, namely deterministic and unambiguous
\vset-automata, that have properties essential to the tractability of some
problems we study in the paper. Figure~\ref{fig:modelcontainment} provides an
overview over all representations of regular document spanners we use throughout
the paper.

\subsection{Regex Formulas}
A \emph{regex-formula (over $\Sigma$)} is a regular expression that may include
variables (called capture variables). Formally, we define the syntax with the
recursive rule
\[
  \alpha \;\eqdef\; \emptyset \mid \varepsilon \mid \sigma \mid (\alpha \lor
  \alpha)\mid (\alpha \cdot \alpha) \mid \alpha^* \mid x\{\alpha\}\,,
\]
where $\sigma \in \Sigma$ and $x \in \svars$. We use $\alpha^+$ as a shorthand
for $\alpha \cdot \alpha^*$ and $\Sigma$ as a shorthand for $\bigvee_{\sigma \in
  \Sigma} \sigma$. The set of variables that occur in $\alpha$ is denoted by
$\dom(\alpha)$ and the size $| \alpha |$ is defined as the number of symbols in
$\alpha$.
The spanner $\toSpanner{\alpha}$ represented by a regex formula $\alpha$ is
given by the following inductive definition that uses the algebraic operations
from Definition~\ref{def:algebra}:
\begin{align*}
  \toSpanner{\emptyset} &\eqdef \emptyset & \toSpanner{\varepsilon} &\eqdef \big\{ \varepsilon \mapsto \{\emptytup\}\big\} & \toSpanner{(\alpha_1 \lor \alpha_2)} &\eqdef \toSpanner{\alpha_1} \cup \toSpanner{\alpha_2} & \toSpanner{\alpha^*} &\eqdef \cup_{i \geq 0} \toSpanner{\alpha^i}\\
  \toSpanner{x\{\alpha\}} &\eqdef x\big\{\toSpanner{\alpha}\big\} & \toSpanner{\sigma} &\eqdef \big\{ \sigma \mapsto \{\emptytup\}\big\} & \toSpanner{(\alpha_1 \cdot \alpha_2)} &\eqdef \toSpanner{\alpha_1} \cdot \toSpanner{\alpha_2} & \end{align*}
We say that a regex formula $\alpha$ is \emph{sequential} if
\begin{itemize}
\item no variable occurs under the Kleene star,
\item for every subformula of the form $x\{\alpha_1\}$ it holds that $x$ does
  not occur in $\alpha_1$, and
\item for every subformula of the form $\alpha_1 \cdot \alpha_2$ it holds that
  the sets of variables used in $\alpha_1$ and $\alpha_2$ are disjoint.
\end{itemize}
A regex formula $\alpha$ is \emph{functional} if
$\alpha$ is sequential and the spanner $\toSpanner{\alpha}$ is functional.

The set of all regex formulas is denoted by \rgx. Similarly, the sequential (or
functional) regex formulas are denoted by \rgxseq (\rgxfunc, respectively). It
follows immediately from the definitions that every functional regex-formula is
also sequential, but not vice versa. For instance, the regex-formula $\alpha =
x_1\{a\} \lor x_2\{b\}$ is sequential, but not functional (and therefore,
$\rgxfunc \subsetneq \rgxseq$).

Maturana et al.~\cite{MaturanaRV18} showed that the class of spanners defined by
regex-formulas is the same as the class of spanners defined by sequential
regex-formulas. However, using the same technique as
Freydenberger~\cite[Proposition 3.9]{Freydenberger19}, it can be shown that the
smallest sequential regex-formula equivalent to a given regex formula $\alpha$
can be exponentially larger then $\alpha$.

\subsection{Ref-words}
For a finite set $V \subseteq \svars$ of variables, ref-words are defined over
the extended alphabet $\Sigma \cup \varop{V}$, where $\varop{V} \eqdef \{\vop{x},
\vcl{x} \mid x \in V \}$. We assume that $\varop{V}$ is disjoint with $\Sigma$
and $\svars$. Ref-words extend strings over $\Sigma$ by encoding opening
($\vop{x}$) and closing ($\vcl{x}$) of variables.

A ref-word $\refWord \in (\Sigma \cup \varop{V})^*$ is \emph{valid} if every
occurring variable is opened and closed exactly once. More formally, for each $x
\in V$, the string \refWord has precisely one occurrence of $\vop{x}$ and
precisely one occurrence of $\vcl{x}$,which is after the occurrence of
$\vop{x}$. For every valid ref-word $\refWord$ over $(\Sigma \cup \varop{V})$ we
define $\dom(\refWord)$ as the set of variables $x \in V$ which occur in the
ref-word. More formally,
\[
  \dom(\refWord) \;\eqdef\; \{ x \in V \mid \exists \refWord_x^{\pre},
  \refWord_x, \refWord_x^{\post} \in (\Sigma \cup \varop{V})^* \text{ such that }
  \refWord = \refWord_x^{\pre} \cdot \vop{x} \cdot \refWord_x \cdot \vcl{x}
  \cdot \refWord_x^{\post}\}.
\]

Intuitively, each valid ref-word $\refWord$ encodes a $\doc$-tuple for some
document $\doc$, where the document is given by symbols from $\sigma$ in
$\refWord$ and the variable markers encode where the spans begin and end.
Formally, we define functions $\clr$ and $\operatorname{tup}$ that, given a
valid ref-word, output the corresponding document and
tuple.\footnote{The function \clr is sometimes also called
  $\operatorname{clr}$ in literature (cf. Freydenberger et
  al.~\cite{FreydenbergerKP18}).} The morphism $\clr \colon (\Sigma \cup
\varop{V})^* \to \Sigma^*$ is defined as:
\[
  \clr(\sigma) \;\eqdef\; \begin{cases}
    \sigma & \text{if }\sigma \in \Sigma \\
    \varepsilon & \text{if } \sigma \in \varop{V}
  \end{cases}
\]

By definition, every valid ref-word $\refWord$ over
$(\Sigma \cup \varop{V})$ has a unique factorization
\[
  \refWord \;=\; \refWord_x^{\pre} \cdot \vop{x} \cdot \refWord_x \cdot \vcl{x}
  \cdot \refWord_x^{\post}
\]
for each $x \in \dom(\refWord)$. We are now ready to define the function
$\operatorname{tup}$ as
\[
  \operatorname{tup}(\refWord) \;\eqdef\; \{ x \mapsto
  \spanFromTo{i_x}{j_x} \mid x \in \dom(\refWord), i_x =
  |\clr(\refWord_x^{\pre})|, j_x=i_x + |\clr(\refWord_x)| \}\;.
\]
The usage of the \clr morphism ensures that the indices $i_x$ and $j_x$ refer to
positions in the document and do not consider other variable operations.

A \emph{ref-word language} $\reflang$ is a language of ref-words. The spanner
$\toSpanner{\reflang}$ represented by a ref-word language $\reflang$ is given by
\[
  \toSpanner{\reflang}(\doc) \;\eqdef\; \big\{\toTuple{\refWord} \mid \refWord
  \in \reflang, \refWord \text{ is valid, and } \clr(\refWord) = \doc \big\}\;.
\]

A ref-word language $\reflang$ is sequential if every ref-word $\refWord \in
\reflang$ is valid. It is functional if it is sequential and
$\toSpanner{\reflang}$ is functional.

\subsection{Variable order condition}
Observe that multiple ref-words can encode the same tuple. For instance, the
ref-words $\refWord = \vop{x_1}\vop{x_2}a \vcl{x_1}\vcl{x_2}$ and $\refWordPrime
= \vop{x_1}\vop{x_2}a \vcl{x_2}\vcl{x_1}$ both encode the tuple which selects
the span $\spanFromTo{1}{2}$ in both variables $x_1,x_2$ on document $a$. Thus,
there can be multiple ref-word languages, representing the same spanner.
Sometimes it is convenient to have a one-to-one mapping between ref-words and
tuples. To this end, we fix a total, linear order $\prec$ on the set
$\varop{\svars}$ of variable operations, such that $\vop{v} \prec \vcl{v}$ for
every variable $v \in \svars$. We say that a ref-word satisfies the
\emph{variable order condition} if all adjacent variable operations in
$\refWord$ are ordered according to the fixed linear order $\prec$. Observe
that, for every document $\doc$ and every tuple $\tup$, there is exactly one
ref-word $\refWord$, with $\doc = \clr(\refWord)$ and $\tup =
\toTuple{\refWord}$, that satisfies the variable order condition. We define
$\operatorname{ref}$ as the function that, given a document $\doc$ and a
\doc-tuple $\tup$, returns this unique ref-word that satisfies the variable
order condition.

The following observation shows the connections between the functions $\clr$,
$\operatorname{ref},$ and $\operatorname{tup}$.
\begin{observation}\label{obs:canonicalRefWord}
  Let $\refWord$ be a valid ref-word and let $\refWordPrime \eqdef
  \refWordFrom{\clr(\refWord)}{\toTuple{\refWord}}$. Then $\toTuple{\refWord} =
  \toTuple{\refWordPrime}$. Furthermore, $\refWord = \refWordPrime$ if and only
  if $\refWord$ satisfies the variable order condition.\qed
\end{observation}

Analogous to sequentiallity, we say that a ref-word language $\reflang$
satisfies the variable order condition if every ref-word $\refWord \in \reflang$
satisfies the variable order condition. The following lemma connects
spanners and sequential ref-word languages which satisfy the variable order
condition.

\begin{lemma}\label{lem:seqvopReflangContainment}
  Let $\reflang_1, \reflang_2$ be sequential ref-word language which satisfies
  the variable order condition. Then $\reflang_1 \subseteq \reflang_2$ if and
  only if $\toSpanner{\reflang_1} \subseteq \toSpanner{\reflang_2}$.
\end{lemma}
\begin{proof}
  (If): Let $\refWord \in \reflang_1$. Thus, $\toTuple{\refWord} \in
  \toSpanner{\reflang_1}(\clr(\refWord)) \subseteq
  \toSpanner{\reflang_2}(\clr(\refWord))$ and therefore $\refWord \in
  \reflang_2$.
  
  (Only if): Let $\doc \in \docs$ be a document and $\tup \in
  \toSpanner{\reflang_1}(\doc)$. Thus, there must be a valid ref-word $\refWord
  \in \reflang_1$ with $\clr(\refWord) = \doc$ and $\toTuple{\refWord} = \tup$.
  Due to $\reflang_1$ satisfying the variable order condition, $\refWord$ must
  satisfy the variable order condition and therefore, $\refWordFrom{\doc}{\tup}
  = \refWord \in \reflang_1 \subseteq \reflang_2$ and thus $\tup \in
  \toSpanner{\reflang_2}(\doc)$, concluding the proof.
\end{proof}

\subsubsection*{Connection between ref-words and regex formulas}

Every regex-formula can be interpreted as a generator of a (regular) ref-word
language $\reflang(\alpha)$ over the extended alphabet $\Sigma \cup
\varop{\dom(\alpha)}$ using the usual semantics for regular expressions and
interpreting every subformula of the form $x\{\beta\}$ as $\vop{x} \cdot \beta
\cdot \vcl{x}$.

A straightforward induction shows that $\toSpanner{\alpha} =
\toSpanner{\reflang(\alpha)}$ for every regex-formula $\alpha$. Furthermore
$\alpha$ is sequential (functional) if and only if $\reflang(\alpha)$ is
sequential (functional).

\subsection{Variable Set-Automata}
A \emph{variable-set automaton (\vset-automaton)} with variables from a finite
set $V \subseteq \svars$ can be understood as an $\varepsilon$-NFA that is
extended with edges that are labeled with variable operations $\varop{V}$.
Formally, a \vset-automaton is a sextuple $A \eqdef (\Sigma, V, Q, q_0, Q_F,
\delta)$, where $\Sigma$ is a finite set of alphabet symbols, $V$ is a finite
set of variables, $Q$ is a finite set of states, $q_0 \in Q$ is an initial
state, $Q_F \subseteq Q$ is a set of final states, and $\delta \colon Q \times
(\Sigma \cup \{\varepsilon\} \cup \varop{V}) \to 2^Q$ is the transition
function. The \emph{size} of a \vset-automaton $A$ is defined by $|A| = |Q| +
|Q_F| + |\delta| + 1$. By $\vars(A) \eqdef V$ we denote the variables of $A$. To
define the semantics of $A$, we first interpret $A$ as an $\varepsilon$-NFA over
the terminal alphabet $\Sigma \cup \varop{V}$, and define its \emph{ref-word
  language} $\reflang(A)$ as the set of all ref-words $\refWord \in \lang(A)
\subseteq (\Sigma \cup \varop{V})^*$ that are accepted by the
$\varepsilon$-NFA $A$.

Analogous to runs of $\varepsilon$-NFAs, we define a \emph{run} $\rho$ of $A$
on a ref-word $\refWord = \sigma_1 \cdots \sigma_n$ as the sequence
\[
  \rho \;\;\eqdef\;\; q_0 \xrightarrow{\sigma_1} q_1 \;\cdots\; q_{n-1}
  \xrightarrow{\sigma_n} q_n\;,
\]
where $q_{i+1} \in \delta(q_i,\sigma_{i+1})$ for all $0 \leq i < n$, and $q_n
\in Q_F$. Observe that $\refWord \in \reflang(A)$ if and only if there is a
run $\rho$ of $A$ on $\refWord$. Furthermore, a run $\rho$ of $A$ on ref-word
$\refWord$ accepts a \doc-tuple $\tup$ if $\clr(\refWord) = \doc$ and $\tup =
\toTuple{\refWord}$.

We define $\toSpanner{A}$ as $\toSpanner{\reflang(A)}$ and say that $A$ is
\emph{sequential} if $\reflang(A)$ is sequential. Furthermore, we say that $A$
is \emph{functional} if $\reflang(A)$ is functional and $\vars(\toSpanner{A}) =
V$. Two \vset-automata $A_1,A_2$ are \emph{equivalent} if they define the same
spanner, i.e., if $\toSpanner{A_1} = \toSpanner{A_2}$. Furthermore, a
\vset-automaton $A$ satisfies the \emph{variable order condition} if
$\reflang(A)$ satisfies the variable order condition.

We refer to the set of all \vset-automata as \vsa and to the set of all
sequential (or functional) \vset-automata as \svsa (or \fvsa, respectively).

We observe, that given a \vset-automaton $A$, $\varepsilon$-transitions can be
removed in \ptime, using the classical $\varepsilon$-removal algorithm for
$\varepsilon$-NFAs.

\begin{observation}\label{obs:varepsilonRemoval}
  Given a \vset-automaton $A$ an equivalent \vset-automaton $A'$ which
  does not use $\varepsilon$-transitions can be constructed in \ptime. 
\end{observation}

\subsubsection*{Deterministic and Unambiguous \vset-Automata}

We use the notion of determinism as introduced by Maturana et
al.~\cite{MaturanaRV18}, but refer to it as \emph{weakly} deterministic because,
as we will show in Theorem~\ref{theorem:maturana-error}, weakly deterministic
\vset-automata still have sufficient non-determinism to make the containment
problem \pspace-hard, which is equally hard than for general
\vset-automata.\footnote{Assuming $\conp \neq \pspace$, this result contradicts
  Theorem 6.6 in Maturana et al.~\cite{MaturanaRV18}, where it is argued that
  containment for weakly deterministic sequential \vset-automata is in \conp.
  There is an error in the upper bound of Maturana et al.~\cite{MaturanaRV18},
  as can be seen in the version that includes the proofs~\cite{MaturanaRV17}.
  The specific error is in the pumping argument for proving a polynomial size
  witness property for non-containment. The polynomial size witness property is
  not necessarily true, due to the non-determinism entailed in the ability of the
  automaton to open variables in different orders. At every specific position in
  the string, the execution can be in $\Theta(n)$ possible states, where $n$ is
  the number of states, implying that a minimal witness may require a length of
  $2^{\Theta(n)}$.} We therefore define a stronger notion of determinism, which
will lead to an \nl-complete containment problem
(Theorem~\ref{thm:spannerContainment-det}). Furthermore, we define unambiguous
\vset-automata, which utilize a relaxed notion of determinism that preserves
tractability of containment (Theorem~\ref{thm:spannerContainment-det}).

Formally, a \vset-automaton $A = (\Sigma, V, Q, q_0, \allowbreak
Q_F,\allowbreak \delta)$ is \emph{weakly deterministic}, if
\begin{enumerate}
\item $\delta(q,\varepsilon) = \emptyset$ for every $q \in Q$, i.e. it does not
  use $\varepsilon$-transitions, and
\item $|\delta(q,v)| \leq 1$ for every $q \in Q$ and every $v \in \Sigma \cup
  \varop{V}$.
\end{enumerate}

At last we define deterministic and unambiguous \vset-automata. To this end, we
define the following three conditions:
\begin{enumerate}[label=(C\arabic*)]
\item $A$ is weakly deterministic; \label{cond:weaklydet}
\item $A$ satisfies the variable order condition; \label{cond:voc}
\item there is exactly one accepting run of $A$ on every $\refWord \in
  \reflang(A)$.\label{cond:unambig}
\end{enumerate}
We then say that a \vset-automaton $A$ is \emph{deterministic} if it satisfies
conditions~\ref{cond:weaklydet} and~\ref{cond:voc} and it is \emph{unambiguous}
if it satisfies conditions~\ref{cond:voc} and~\ref{cond:unambig}. The following
observation is obvious, as~\ref{cond:weaklydet} clearly
implies~\ref{cond:unambig}.

\begin{observation}\label{obs:det-vsa-unambig}
  Every deterministic \vset-automaton is also unambiguous.
\end{observation}

We note that for Boolean spanners the definitions coincide with the classical
unambiguity/deter\-minism definitions of finite state automata. That is, a
\vset-automaton with $\vars(A) = \emptyset$ is deterministic (unambiguous) if it
is a deterministic (unambiguous) finite state automaton.

We show in Lemma~\ref{lem:vset-determinization} that none of the conditions
(C1),~(C2), and~(C3) restrict the expressiveness of regular spanners. We discuss
complexity of deterministic \vset-automata in
Section~\ref{sec:spanner-containment}. In the following, we denote by \dvsa
(resp., \dfvsa and \dsvsa) the class of deterministic (resp., deterministic and
functional, deterministic and sequential) \vset-automata and by \uvsa (resp.,
\ufvsa and \usvsa) the class of unambiguous (resp., unambiguous and functional,
unambiguous and sequential) \vset-automata.

Deterministic \vset-automata are similar to the \emph{extended deterministic
  \vset-automata} by Florenzano et al.~\cite{FlorenzanoRUVV18}, which allow
multiple variable operations on a single transition and force each variable
transition to be followed by a transition processing an alphabet symbol.
However, deterministic \vset-automata can be exponentially more succinct than
extended deterministic \vset-automata. An example class of automata where this
blowup occurs is depicted in Figure~\ref{fig:eVSA-Blowup}.

\begin{figure}
  \begin{tikzpicture}[>=latex]
    \node[state,initial] (q0) at (0,0){$q_0$};
    \node[] (dots2) at (5,0){$\vdots$};
    \node[state, double] (qn) at (10,0){$q_n$};
    \draw (q0) edge[->, bend left=24] node[above] {$\{\emptyset\}$}(qn);
    \draw (q0) edge[->, bend left=12] node[above] {$\{\vop{x_1},\vcl{x_1}\}$}(qn);
    \draw (q0) edge[->, bend right=24] node[below] {$\{\vop{x_1},\vcl{x_1}, \ldots, \vop{x_n}, \vcl{x_n} \}$}(qn);
  \end{tikzpicture}
  \begin{tikzpicture}[>=latex]
    \node[state,initial above] (q0) at (0,0){$q_0$};
    \node[state] (qx1) at (1.5,.5){$q_{x_1}$};
    \node[state] (q1) at (3,0){$q_1$};
    \node[state] (qx2) at (4.5,.5){$q_{x_2}$};
    \node[state] (q2) at (6,0){$q_2$};
    \node[] (dots1) at (7.3,.5){};
    \node[] (dots2) at (8,0){$\cdots$};
    \node[] (dots3) at (8.7,.5){};
    \node[state] (qn1) at (10,0){$q_{n-1}$};
    \node[state] (qxn) at (11.5,.5){$q_{x_n}$};
    \node[state, double] (qn) at (13,0){$q_n$};
    \draw (q0) edge[->] node[above,sloped] {$\vop{x_1}$}(qx1);
    \draw (qx1) edge[->] node[above,sloped] {$\vcl{x_1}$}(q1);
    \draw (q0) edge[->, bend right] node[below] {$\varepsilon$}(q1);
    \draw (q1) edge[->] node[above,sloped] {$\vop{x_2}$}(qx2);
    \draw (qx2) edge[->] node[above,sloped] {$\vcl{x_2}$}(q2);
    \draw (q1) edge[->, bend right] node[below] {$\varepsilon$}(q2);
    \draw (q2) edge[->,dashed] node[above,sloped] {$\vop{x_3}$} (dots1);
    \draw (q2) edge[->, bend right, dashed] node[below] {$\varepsilon$}(dots2);
    \draw (dots3) edge[->,dashed] node[above,sloped] {$\vcl{x_{n-1}}$}(qn1);
    \draw (dots2) edge[->, bend right,dashed] node[below] {$\varepsilon$}(qn1);
    \draw (qn1) edge[->] node[above,sloped] {$\vop{x_n}$}(qxn);
    \draw (qxn) edge[->] node[above,sloped] {$\vcl{x_n}$}(qn);
    \draw (qn1) edge[->, bend right] node[below] {$\varepsilon$}(qn);
  \end{tikzpicture}
  \caption{Class of example spanners where the smallest deterministic extended
    VSAs (top) are exponentially larger than the smallest deterministic
    \vset-automata (bottom). The automaton on the top has a transition
    $\delta(q_0,\varop{V}) = \{q_n\}$ for every $V \subseteq
    \{x_1,\ldots,x_n\}$, thus, it has $2^n$ transitions. The automaton on the
    bottom has $3n$ transitions in total.}\label{fig:eVSA-Blowup}
\end{figure}

The following proposition shows that deterministic \vset-automata are equally
expressive as \vset-automata in general. 
\begin{lemma}\label{lem:vset-determinization}
  For every \vset-automaton $A$ there is an equivalent sequential deterministic
  \vset-automaton $A'$, i.e. $\toSpanner{A} = \toSpanner{A'}$.
\end{lemma}
\begin{proof}
  We have to show that we can find a \vset-automaton $A'$, such that $A'$ is
  equivalent to $A$ and $A'$ satisfies~\ref{cond:weaklydet} and~\ref{cond:voc}.

  Maturana et al~\cite[Proposition 5.6]{MaturanaRV18} show that for every \vset-automaton
  there is an equivalent sequential \vset-automaton. Therefore, we can assume,
  \mbox{w.l.o.g.}, that $A$ is sequential. Florenzano et al.~\cite[Theorem 3.1,
  Proposition 3.2]{FlorenzanoRUVV18} show that every \vset-automaton can be
  transformed into an equivalent extended \vset-automaton and vice versa. The
  model of extended \vset-automata allows to annotate a set of variable
  operations to a single edge. For the construction of a \vset-automaton from a
  given extended \vset-automaton they fix an order on the variables and replace
  each transition, containing multiple variable operations by a sequence of
  edges. Therefore, the variable order condition~\ref{cond:voc} can be achieved
  by using $\prec$ as variable order for the transformation from extended to
  normal \vset-automata.\footnote{Fagin et al.~\cite{FaginKRV15-jacm} already
    gave a similar construction on so called \emph{lexicographic
      \vset-automata}, i.e., \vset-automata in which consecutive variable
    operations always follow a given linear order.} We note that all involved
  constructions preserve sequentiality.

  We can achieve~\ref{cond:weaklydet} by interpreting the \vset-automaton
  as an $\varepsilon$-NFA that accepts ref-words and using the classical
  $\varepsilon$-NFA determinization construction. This construction also
  preserves sequentiality as it does not change the involved ref-word language.
\end{proof}

Finally, we recall that it is well known that the class of regex-formulas (\rgx)
is less expressive than the class of \vset-automata (\vsa)
\cite{FaginKRV15-jacm, MaturanaRV18}. In order to reach the expressiveness of
\vset-automata, \rgx needs to be extended with projection, natural join, and
union. Figure~\ref{fig:modelcontainment} gives an overview of the expressiveness
and inclusions between the introduced classes of document
spanners.

We denote the set of all representations depicted in
Figure~\ref{fig:modelcontainment} by $\cCg$ and the unambiguous and sequential
subset by $\cCt$, that is $\cCt = \{\usvsa, \allowbreak \ufvsa, \allowbreak
\dsvsa, \allowbreak \dfvsa\}$.

\begin{figure}
  \begin{tikzpicture}[shorten >=1pt,every node/.style={},align=center,node distance=1em]
      \tikzstyle{state}=[minimum width=5em]
      \tikzstyle{supseteq}=[draw=none, edge node={node [scale=1,sloped]{$\supseteq$}}]
      \tikzstyle{supsetneq}=[draw=none, edge node={node [scale=1,sloped]{$\supsetneq$}}]

      \node[state] (rgx) at (0,0){\rgx};
      \node[state, below=of rgx] (srgx){\srgx};
      \node[state, below=of srgx] (frgx){\frgx};
      
      \node[state, right=of rgx] (vsa){\vsa};
      \node[state, right=of vsa] (uvsa){\uvsa};
      \node[state, right=of uvsa] (dvsa){\dvsa};
      \node[state, below=of vsa] (svsa){\svsa};
      \node[state, right=of svsa] (usvsa){\usvsa};
      \node[state, right=of usvsa] (dsvsa){\dsvsa};
      \node[state, below=of svsa] (fvsa){\fvsa};
      \node[state, right=of fvsa] (ufvsa){\ufvsa};
      \node[state, right=of ufvsa] (dfvsa){\dfvsa};

      \draw (rgx) edge[supseteq] (srgx);
      \draw (srgx) edge[supseteq] (frgx);
      
      \draw (vsa) edge[supseteq] (uvsa);
      \draw (uvsa) edge[supseteq] (dvsa);
      \draw (vsa) edge[supseteq] (svsa);
      \draw (svsa) edge[supseteq] (usvsa);
      \draw (uvsa) edge[supseteq] (usvsa);
      \draw (dvsa) edge[supseteq] (dsvsa);
      \draw (usvsa) edge[supseteq] (dsvsa);
      \draw (svsa) edge[supseteq] (fvsa);
      \draw (fvsa) edge[supseteq] (ufvsa);
      \draw (usvsa) edge[supseteq] (ufvsa);
      \draw (dsvsa) edge[supseteq] (dfvsa);
      \draw (ufvsa) edge[supseteq] (dfvsa);

      \begin{pgfonlayer}{background}
        \draw [line width=.5mm,join=round,blue!10]
          (dvsa.north -| dvsa.east) rectangle (svsa.south  -|
          svsa.west);
        \draw [line width=.5mm,join=round,blue!10]
          (rgx.north -| rgx.east) rectangle (srgx.south  -|
           srgx.west);
        \draw [line width=.5mm,join=round,red!10]
          (dfvsa.north -| dfvsa.east) rectangle (fvsa.south  -|
          fvsa.west);
        \draw [line width=.5mm,join=round,red!10]
          (frgx.north -| frgx.east) rectangle (frgx.south  -|
           frgx.west);
      \end{pgfonlayer}
    \end{tikzpicture}
    \caption{Expressiveness and inclusion relations of classes of regular document
      spanners. All formalisms within the same box are equally
      expressive.}\label{fig:modelcontainment}
\end{figure}

 \section{Complexity Results for Regular Document
  Spanners}\label{sec:regular-results}  
We now give the main results for the decision problems we introduced in
Section~\ref{sec:def:splitters} in the case of regular spanners. The following
two theorems summarize the main complexity results. The upper bounds are given
in Section~\ref{sec:upper-bounds} and are heavily based on containment of
regular document spanners, which is discussed in
Section~\ref{sec:spanner-containment}. The lower bounds are given in
Section~\ref{sec:lower-bounds}.

\begin{theorem}\label{thm:splitcorrectness}
  Let $\cC \in \cCg$ be a class of document spanners. Then
  $\splitcorrectness[\cC]$ and $\selfsplitcorrectness[\cC]$ are
  \pspace-complete. Furthermore, $\splitcorrectness[\cC]$ and
  $\selfsplitcorrectness[\cC]$ are in \ptime if
  \begin{itemize}
  \item $\cC \in \cCt$, and
  \item the spanner is proper and the splitter is disjoint, or the highlander
    condition is satisfied by the spanner and splitter.
  \end{itemize} 
\end{theorem}

\begin{theorem}\label{thm:mainSplittability}
  Let $\cC \in \cCg$ be a class of document spanners. Then deciding
  $\splittability[\cC]$ is in \expspace and \pspace-hard. Furthermore, it is
  \pspace-complete if one of the following two conditions is satisfied:
  \begin{itemize}
  \item the highlander condition is satisfied by spanner and splitter, or
  \item the spanner is proper and the splitter is disjoint.
  \end{itemize} 
\end{theorem}

 \section{Technical Foundations}\label{sec:technical-foundations}

In this section, we provide the technical foundation for our main results. We
show that, given a spanner $\spanner$ and a splitter $\splitter$, represented by
\vset-automata $A_\spanner, A_\splitter \in \vsa$, the spanner $\spanner \circ
\splitter$ can be constructed as a \vset-automaton $A_{\spanner \circ
  \splitter}$. If $A_\spanner$ and $A_\splitter$ are unambiguous and sequential,
and $\spanner \circ \splitter$ and \splitter satisfy the highlander condition,
then the constructed \vset-automaton for $A_{\spanner \circ \splitter}$ is also
unambiguous and sequential. In Section~\ref{sec:spanner-containment}, we study
the complexity of containment of regular document spanners. In particular, we
show that containment of regex-formulas and \vset-automata is \pspace-complete
(Corollary~\ref{cor:spannerContainment}), even under some determinism
assumptions introduced in past work~\cite{MaturanaRV18}
(Theorem~\ref{theorem:maturana-error}), but is it solvable in \ptime for
unambiguous and even in \nl for deterministic \vset-automata
(Theorem~\ref{thm:spannerContainment-det}). Furthermore, we provide upper bounds
for the complexity of $\disjoint, \proper, \highlander$
(Proposition~\ref{prop:test-highlander}) and $\cover$
(Proposition~\ref{prop:cover-pspace}).

\subsection{Spanner/Splitter Composition}

We begin by showing that, given $A_\spanner, A_\splitter \in \vsa$, a
\vset-automaton that represents the spanner $\toSpanner{A_\spanner} \circ
\toSpanner{A_\splitter}$ can be constructed. Furthermore, if $A_\spanner$ and
$A_\splitter$ are unambiguous and sequential, and $\toSpanner{A_\spanner} \circ
\toSpanner{A_\splitter}$ and \splitter satisfy the highlander condition, then
the constructed \vset-automaton is also unambiguous and sequential.

\begin{proposition}\label{prop:pcircs-algebra-construction}
  Given VSet-automata $A_\spanner$ and $A_\splitter$ representing a spanner and
  a splitter, respectively, a VSet automaton $A_{\spanner \circ \splitter}$ can
  be constructed in polynomial time, such that
  \begin{itemize}
  \item $\toSpanner{A_{\spanner \circ \splitter}} = \toSpanner{A_\spanner}
    \circ \toSpanner{A_\splitter}$;
  \item $A_{\spanner \circ \splitter} \in \svsa$ if $A_\spanner \in
    \svsa$; and
  \item $A_{\spanner \circ \splitter} \in \usvsa$ if $A_\spanner, A_\splitter
    \in \usvsa$, and $A_{\spanner \circ \splitter}$ and $A_\splitter$ satisfy
    the highlander condition.
  \end{itemize}
\end{proposition}

Peterfreund et al.~\cite{PeterfreundFKK19} showed that the join of sequential
\vset-automata can be computed in polynomial time, if the number of shared
variables is bounded by a constant. Furthermore, for sequential \vset-automata,
projection can be computed in polynomial time. The proof extends to arbitrary
\vset-automata, if the number of removed variables is bounded by a constant.
This shows the first two bullet points. We show the last bullet point, using an
explicit construction, that also proves the first two bullet points.

\begin{proof}
  We assume,\mbox{w.l.o.g.}, that $x_\splitter \notin\vars(A_\spanner)$, where
  $x_\splitter$ is the variable of $A_\splitter$.\footnote{This is possible as
    the composition $\spanner \circ \splitter$ does not depend on the variable
    $x_\splitter$. If $x_\splitter \in \vars(A_\spanner)$, we can therefore
    modify $A_\splitter$ to use a variable $x \notin \vars(A_\spanner)$ instead.
    We observe that this obviously can be done in polynomial time.} We use the
  algebraic characterization from Lemma~\ref{lem:pcircs-algebra} that states
  that
  \[
    \spanner \circ \splitter \;=\; \pi_{\dom(\spanner)}\big((\Sigma^* \cdot
    x_\splitter\{\spanner\} \cdot \Sigma^*) \join \splitter\big)
  \]
  for a spanner $\spanner$ and a splitter $\splitter$. Let $A_\spanner =
  (\Sigma, V, Q_\spanner, q_{0,\spanner}, Q_{F,\spanner}, \delta_\spanner)$ and
  $A_\splitter = (\Sigma, \set{x_\splitter}, Q_\splitter, q_{0,\splitter},
  Q_{F,\splitter}, \delta_\splitter)$ be \vset-automata representing a spanner
  $\spanner$ and a splitter $\splitter$. By
  Observation~\ref{obs:varepsilonRemoval}, we assume, \mbox{w.l.o.g.}, that
  $A_\spanner$ and $A_\splitter$ do not use $\varepsilon$-transitions. We
  construct the \vset-automaton
  \[
    A_{\spanner \circ \splitter} \;\;\eqdef\;\; \Big(\Sigma,\; V,\; Q_\splitter \times
    (Q_\spanner \cup \set{\bot}) \times \set{1,2,3},\; (q_{0,\splitter},\bot,1),\;
    Q_{F,\splitter} \times \set{(\bot,3)},\; \delta\Big)\;.
  \]
  The construction is similar to a product construction for the
  automata $A_\spanner$, $A_\splitter$, and a three state automaton that accepts
  the language $\Sigma^* \cdot x_\splitter\{ (\Sigma \cup \Gamma_V)^* \} \cdot
  \Sigma^*$. The main idea of the construction is simulation in three phases. In
  phase one, $A_\splitter$ runs. Whenever $A_\splitter$ can open its variable it
  is decided non-deterministically whether the simulation stays in phase one or
  continues with phase two. At the beginning of phase two, $A_\spanner$ is
  initialized with its start state and runs in parallel to the simulation of
  $A_\splitter$. Whenever $A_\splitter$ allows to close its variable and
  $A_\spanner$ is in an accepting state, the simulation non-deterministically
  decides to stay in phase two or continue with phase three. In phase three, the
  simulation of $A_\splitter$ is finished. The simulation can end at every point
  in which $A_\splitter$ is in an accepting state.

  Thus, the transition function is defined by
  \[
    \begin{array}{r@{\;}l@{\;}l@{\;}r}
      \delta\; \eqdef
      & \big\{\big((q,\bot,1),\sigma,(q',\bot,1)\big)
      &\mid (q,\sigma,q') \in \delta_{\splitter}, \sigma \in \Sigma\big\} \;\; \cup
      & \text{$A_\splitter$ runs}\\[.2ex]
      & \big\{\big((q,\bot,1),\varepsilon, (q',q_{0,\spanner},2)\big)
      &\mid (q,\vop{x_\splitter},q') \in \delta_{\splitter}\big\} \;\; \cup
      & \text{$A_\spanner$ starts}\\[.2ex]
      & \big\{\big((q,p,2),\sigma,(q',p',2)\big)
      &\mid (q,\sigma,q') \in \delta_{\splitter}, (p,\sigma,p') \in
        \delta_{\spanner}, \sigma \in \Sigma\big\} \;\; \cup
      & \text{$A_\splitter$ and $A_\spanner$ run}\\[.2ex]
      & \big\{\big((q,p,2),v,(q,p',2)\big)
      &\mid (p,v,p') \in \delta_{\spanner}, v \in \Gamma_V \big\}\;\; \cup
      & \text{variable operation of $A_\spanner$}\\[.2ex]
      & \big\{\big((q,p,2),\varepsilon, (q',\bot,3)\big)
      &\mid (q,\vcl{x_\splitter},q') \in \delta_{\splitter}, p \in
        Q_{F,\spanner}\big\} \;\; \cup
      & \text{$A_\spanner$ stops}\\[.2ex]
      & \big\{\big((q,\bot,3),\sigma,(q',\bot,3)\big)
      &\mid (q,\sigma,q') \in \delta_{\splitter}, \sigma \in \Sigma \big\}.
      & \text{$A_S$ runs}
    \end{array}
  \]
  By construction, every run of $A_{\spanner \circ \splitter}$ on a valid
  ref-word $\refWord = \sigma_1 \cdots \sigma_n$ uses exactly two
  $\varepsilon$-transitions and is of the form
  \begin{multline*}
    (q_0,\bot,0) \xrightarrow{\sigma_1} (q_1,\bot,0) \xrightarrow{\sigma_2}
    \cdots \xrightarrow{\sigma_{i-1}} (q_{i-1},\bot,0) \xrightarrow{\varepsilon}
    (q_i,p_0,1) \xrightarrow{\sigma_i} (q_{i+1},p_1,1)
    \xrightarrow{\sigma_{i+1}} \cdots \\ \cdots \xrightarrow{\sigma_{j-1}}
    (q_j,p_{j-i},1) \xrightarrow{\varepsilon} (q_{j+1},\bot,2)
    \xrightarrow{\sigma_j} \cdots \xrightarrow{\sigma_n} (q_{n+2},\bot,2)
  \end{multline*}
  where
  \[
    p_0 \xrightarrow{\sigma_{i}} p_1 \xrightarrow{\sigma_{i+1}} \cdots
    \xrightarrow{\sigma_{j-1}} p_{j-i}
  \]
  is a run of $A_\spanner$ on $\sigma_i \cdots \sigma_{j-1}$ and
  \[
    q_0 \xrightarrow{\sigma_1} q_1 \cdots q_{i-1} \xrightarrow{\sigma_{i-1}}
    q_{i-1} \xrightarrow{\vop{x_\splitter}} q_i \xrightarrow{\clr(\sigma_i)} \cdots
    \xrightarrow{\clr(\sigma_{j-1})} q_j \xrightarrow{\vcl{x_\splitter}} q_{j+1}
    \xrightarrow{\sigma_j} q_{j+2} \cdots q_{n+1} \xrightarrow{\sigma_n} q_{n+2}
  \]
  is a run of $A_\splitter$ on $\doc \eqdef \clr(\refWord) = \sigma_1 \cdots
  \sigma_{i-1} \cdot \clr(\sigma_i \cdots \sigma_{j-1}) \cdot \sigma_j \cdots
  \sigma_n$.\footnote{We note that by construction of $A_{\spanner \circ
      \splitter}$ the first component of the state does not change, when
    $\clr(\sigma_i)=\varepsilon$.} Furthermore, the span $\spanFromTo{i}{j'}$
  with $j'=i+|\clr(\sigma_i \cdots \sigma_{j-1})|$, which is defined by the
  positions of the $\varepsilon$-transitions in the run, is in $\splitter(\doc)$
  and covers $\toTuple{\refWord}$.

  We can therefore conclude that $\toSpanner{A_{\spanner \circ \splitter}}
  = \pi_{\dom(\spanner)}((\Sigma^* \cdot x_\splitter\{\spanner\} \cdot \Sigma^*) \join
  \splitter)$ and $A_{\spanner \circ \splitter}$ is sequential if $A_\spanner$
  is sequential. 
 
  It remains to show that $A_{\spanner \circ \splitter}$ is unambiguous if
  \begin{itemize}
  \item $A_\spanner$ and $A_\splitter$ are unambiguous, and
  \item $A_{\spanner \circ \splitter}$ and $A_\splitter$ satisfy the highlander
    condition.
  \end{itemize}
  
  To this end, observe that:
  \begin{enumerate}
  \item a run of $A_{\spanner \circ \splitter}$ that witnesses the violation of
    the variable order condition~\ref{cond:voc} of $A_{\spanner \circ \splitter}$ implies that
    there is a run of $A_\spanner$ that witnesses the violation of the condition
    for $A_\spanner$;
  \item two distinct runs of $A_{\spanner \circ \splitter}$ that violate
    unambiguity condition~\ref{cond:unambig} of $A_{\spanner \circ \splitter}$ must either
    \begin{itemize}
    \item have $\varepsilon$-transitions at different positions and therefore
      witness the existence of two distinct spans in $S$ that both cover
      $\toTuple{\refWord}$, which violates the highlander condition, or
    \item have $\varepsilon$-transitions at the same positions and therefore
      witness that either $A_\spanner$ has two distinct runs on $\refWord$ or
      $A_\splitter$ has two distinct runs on the unique ref-word corresponding
      to the span indicated by the positions of the $\varepsilon$-transitions.
    \end{itemize}
  \end{enumerate}
  Altogether, this shows that $A_{\spanner \circ \splitter}$ being not
  unambiguous leads to a contradiction to the assumption that $A_\spanner$ and
  $A_\splitter$ are unambiguous and that $\spanner \circ \splitter$ and
  $\splitter$ satisfy the highlander condition.
\end{proof}

\subsection{Containment of Regular Document
  Spanners}\label{sec:spanner-containment}
Recall that the containment is the problem that asks, given two spanners
$A_\spanner, A_{\spanner'} \in \cC$, whether
$\toSpanner{A_\spanner}(\doc)\subseteq \toSpanner{A_\spanner'}(\doc)$ for every
document $\doc$. As we will see later, deciding containment is essential for
deciding split-correctness and splittability.

\algproblemWidth{.4}{$\containment[\cC]$}{Spanner $\spanner, \spanner' \in \cC$.}{Is
  $\spanner \subseteq \spanner'$?}

The next theorem establishes the complexity of containment in the general case.

\begin{theorem}[{Maturana et
    al.~\cite{MaturanaRV18}}]\label{thm:spannerContainment} 
  Containment is \pspace-hard for \frgx and \fvsa and in \pspace for \rgx and
  \vsa.
\end{theorem}

Since we know from Figure~\ref{fig:modelcontainment} that $\frgx \subseteq \srgx
\subseteq \rgx$ and $\fvsa \subseteq \svsa \subseteq \vsa$, we have the
following corollary.

\begin{corollary}\label{cor:spannerContainment}
  Containment of regex-formulas (\rgx, \rgxseq, \rgxfunc) and \vset-automata
  (\vsa, \fvsa, \svsa) is \pspace-complete.
\end{corollary}

We now consider containment of deterministic and weakly deterministic
\vset-automata. We first show that containment of weakly deterministic
\vset-automata is \pspace-complete, which contradicts Maturana et
al.~\cite[Theorem 6.6]{MaturanaRV18} if $\conp \neq \pspace$. As we will see in
the proof, the hardness of containment is due to the fact that multiple
variables operations can occur without reading any alphabet symbols and
therefore, multiple different orderings of variable operations can be used to
introduce non-deterministic choice.

\begin{theorem}\label{theorem:maturana-error}
  Containment of weakly deterministic functional \vset-automata is
  \pspace-complete. 
\end{theorem}
\begin{proof}
  The upper bound follows directly from Theorem~\ref{thm:spannerContainment}.

  For the lower bound we reduce from the \pspace~complete problem of DFA union
  universality~\cite{kozen77}. Given deterministic finite automata
  $A_1,\ldots,A_n$ over the alphabet $\Sigma$, the union universality problem
  asks whether
  \begin{equation}  
    \lang(\Sigma^*) \;\subseteq\; \bigcup_{1 \leq i \leq n} \lang(A_i)\;. \tag{\dag}
  \end{equation}
  We construct \vset-automata $A, A'$ using the variable set $V =
  \{x_1,\ldots,\allowbreak x_n\}$, such that $A(\doc) \subseteq A^\prime(\doc)$ for
  all documents $\doc\in\docs$ if and only if $(\dag)$ holds. To this end, let
  $A$ accept the language defined by the regex-formula
  \[
    \alpha_A \;\eqdef\; x_1\Big\{x_2\big\{\cdots x_n\set{\Sigma^*}\cdots\big\}\Big\}\;,
  \]
  selecting the whole document with every variable. Clearly, the regex-formula
  $\alpha_A$ can be represented by a weakly deterministic functional
  \vset-automaton $A$. We now abuse notation and describe the language accepted
  by $A^\prime$ by a hybrid regex-formula
  \[
    \alpha_{A^\prime} \;\eqdef\; x_1\{\alpha_1\} + \cdots + x_n\{\alpha_n\},
  \]
  where the DFAs $A_i$ are plugged in. In particular,
  \[
    \alpha_i \;\eqdef\; x_1\Big\{\cdots x_{i-1}\big\{x_{i+1}\{\cdots
    \{x_n\{A_i\}\big\}\cdots\Big\}\;,
  \]
  for $1 \leq i \leq n$. Term $i$ in $\alpha_{A'}$ starts by first opening
  variable $x_i$, continues to open all other variables in increasing order, and
  finally selects the whole document \doc for every variable if $\doc \in
  \lang(A_i)$. Clearly, as every term starts with a different variable symbol,
  this hybrid formula can be transformed into an equivalent weakly deterministic
  functional \vset-automaton $A^\prime$ in linear time.

  It remains to argue that $\toSpanner{A}(\doc) \subseteq
  \toSpanner{A^\prime}(\doc)$ for every document $\doc\in\docs$ if and only if
  $(\dag)$ holds. 
 
  (if): Assume that $\lang(\Sigma^*) \subseteq \bigcup_{1 \leq i \leq n}
  \lang(A_i)$ holds. Let $\doc \in \docs$ be a document and $\tup \in
  \toSpanner{A}(\doc)$ be a $\doc$-tuple. Per definition of $A$, we have
  $\tup(v) = \spanFromTo{1}{|\doc|+1}$ for all variables $v \in V$. By
  assumption, there is an automaton $A_i$ such that $\doc \in \lang(A_i)$.
  Therefore, the tuple $\tup$ is accepted by term $i$ of $A^\prime$, thus
  $\tup \in \toSpanner{A'}(\doc)$.

  (only if): Assume that $\forall \doc \in \docs: A(\doc) \subseteq A'(\doc)$.
  Let $\doc \in \docs$ be an arbitrary document and $\tup \in A(\doc)$. Per
  assumption, $\tup$ is also in $\toSpanner{A^\prime}(\doc)$. Thus there is a
  run of $A^\prime$ on $\doc$ selecting $\tup$. Let $x_i$ be the first variable
  which is opened in this run. Per construction of $A^\prime$ it follows, that
  $\doc \in \lang(A_i)$.
\end{proof}

The question is now whether there exists a satisfactory notion of determinism
for \vset-automata that allows for efficient containment testing without loss of
expressiveness. Our definitions of unambiguity and determinism resolves this
complexity issue, without loss of expressiveness (cf.
Lemma~\ref{lem:vset-determinization}). Towards tractability results for
containment, we first show that the emptiness problem is decidable in \nl.
\begin{proposition}\label{prop:svsaEmptiness}
  Given a sequential \vset-automaton $A$, it can be checked in \nl whether
  $\toSpanner{A}(\doc) \neq \emptyset$ for some document $\doc$. 
\end{proposition}
\begin{proof}
  Let $A \in \svsa$. Due to $A$ being sequential, all ref-words $\refWord \in
  \reflang(A)$ must be valid. Thus, $\toSpanner{A}(\doc) \neq \emptyset$ if and
  only if $\reflang(A) \neq \emptyset$. The result follows from the fact that
  emptiness of $\varepsilon$-NFAs can be checked in \nl.\footnote{Emptiness of
    $\varepsilon$-NFAs is the same problem as Reachability in graphs, which is
    well known to be \nl-complete (cf. Papadimitriou~\cite[Theorem
    16.2]{Papadimitriou94}).}
\end{proof}

Now, we can show that containment is tractable for deterministic and unambiguous
\vset-automata.

\begin{theorem}\label{thm:spannerContainment-det}
  Containment for \usvsa is in \ptime and containment for \dsvsa is in \nl.
\end{theorem}
\begin{proof}
  As we will see next, the \nl upper bound for \dsvsa follows from containment
  of deterministic finite state automata. The \ptime upper bound for \usvsa
  follows from containment of unambiguous finite state automata. In the
  following, we only give the proof for \dsvsa. The proof for \usvsa is
  analogous (using the fact that containment for unambiguous finite automata is
  in \ptime~\cite[Corollary 4.7]{StearnsH-sicomp85}).
  
  To this end, let $A_1,A_2$ be deterministic sequential \vset-automata (i.e.
  $A_1, A_2 \in \dsvsa$). By Lemma~\ref{lem:seqvopReflangContainment},
  $\toSpanner{A_1} \subseteq \toSpanner{A_2}$ if and only if $\reflang(A_1)
  \subseteq \reflang(A_2)$. Let $i \in \{1,2\}$. Observe that due to $A_i$ being
  weakly deterministic, it must hold that $A_i$, interpreted as
  $\varepsilon$-NFA, is deterministic. The result follows since containment for
  deterministic finite automata is well known to be in \nl.
\end{proof}

\subsection{Complexity of Checking Cover and Highlander
  Condition}\label{sec:prelim-highlander-cover}

The next proposition shows that deciding the highlander condition is tractable
if spanner and splitter are sequential.
\begin{proposition}\label{prop:test-highlander}
  $\proper[\svsa], \disjoint[\svsa],$ and $\highlander[\svsa]$ are in $\nl$.
  Furthermore, $\proper[\srgx], \disjoint[\srgx],$ and $\highlander[\srgx]$ are
  in $\ptime$.
\end{proposition}
\begin{proof}
  For every regex-formula an equivalent $\vset$-automaton can be constructed in
  polynomial time using the usual constructions that convert a regular
  expression into an NFA. Thus it suffices to show that the problems are in
  $\nl$ for $\cC = \svsa$.

  Let $A_\spanner,A_\splitter \in \svsa$ be automata representing a spanner and
  a splitter, respectively. Let $\spanner = \toSpanner{A_\spanner}$ and
  $\splitter = \toSpanner{A_\splitter}$. We denote the variables of $A_\spanner$
  by $V$, the single variable of $A_S$ with $x$, and a fresh variable not
  used by $A_\spanner$ or $A_\splitter$ by $y$. We provide logspace
  constructions for \svsa{s} $A_\text{prop}$, $A_\text{disjoint}$, and
  $A_\text{highlander}$, such that the ref-word languages of the 
  automata are empty if and only if $\spanner$ is proper, $\splitter$ is
  disjoint, and $\spanner$ and $\splitter$ satisfy the highlander condition,
  respectively. The result follows, as emptiness of \svsa can be checked in \nl
  (cf. Proposition~\ref{prop:svsaEmptiness}). 

  To ease readability, we abbreviate $\toTuple{\refWord}(x)$ by
  $\toITuple{\refWord}(x)$ in the remainder of this proof.

  \medskip
  
  $A_\text{proper}$: The automaton $A_\text{proper}$ is the intersection of
  $A_\spanner$ and an automaton $A'$ such that $\reflang(A')=\Sigma^* \cdot
  (\Gamma_V)^* \cdot \Sigma^*$.

  The automaton $A_\text{proper}$ is sequential, since
  $\reflang(A_\text{proper}) \subseteq \reflang(A_\spanner)$ and $A_\spanner$ is
  sequential. Thus, all ref-words $\refWord \in \reflang(A_\text{proper})$ are
  valid. Assume that $\refWord \in \reflang(A_\text{proper})$. Let $\doc \eqdef
  \clr(\refWord)$. Observe that $\toTuple{\refWord} \in \spanner(\doc)$, due to
  $\reflang(A_\text{proper}) \subseteq \reflang(A_\spanner)$.
  Furthermore, due to $\reflang(A_\text{proper}) \subseteq \reflang(A')
  = \Sigma^* \cdot (\Gamma_V)^* \cdot \Sigma^*$ it must hold that
  $\toTuple{\refWord}$ is either empty or the minimal span covering it is empty. In
  both cases $\toTuple{\refWord}$ is a witness that $A_\spanner$ is not proper. For
  the other direction, assume there is a document $\doc$ and a tuple $\tup \in
  \spanner(\doc)$ such that $\tup$ is empty or the minimal span covering $\tup$
  is empty, than the ref-word $\refWord \in \reflang(A_\spanner)$ with
  $\toTuple{\refWord}=\tup$ is in $\reflang(A')$ and therefore in
  $\reflang(A_\text{proper})$.

  \medskip
  
  $A_\text{disjoint}$: We define the automaton $A_\text{disjoint}$ as the
  intersection of the following four automata such that $\tup \in
  \toSpanner{A_\text{disjoint}}(\doc)$ is a tuple over variables $x,y$. The
  automaton $A_\splitter^x$ (resp., $A_\splitter^y)$ selects all $(x,y)$ tuples
  such that $\tup(x) \in \splitter(\doc)$ (resp., $\tup(y) \in \splitter(\doc)$)
  for a document $\doc \in \docs$. The automaton $A_\text{distinct}$ verifies
  whether $\tup(x) \neq \tup(y)$ and $A_\text{overlap}$ verifies whether
  $\tup(x)$ and $\tup(y)$ overlap. More formally, we define the automata as
  follows:
  \begin{itemize}
  \item $A_\splitter^x$ is derived from $A_\splitter$ by adding self-loops for
    every label from $\Gamma_{\{y\}} = \{\vop{y}, \vcl{y}\}$ to every state.
  \item $A_\splitter^y$ is derived from $A_\splitter$ by changing every label
    $\vop{x}$ to $\vop{y}$, every label $\vcl{x}$ to $\vcl{y}$, and afterwards
    adding self loops for every label from $\Gamma_{\{x\}} = \{\vop{x},
    \vcl{x}\}$ to every state.
  \item $A_\text{distinct}$ ensures that $\toITuple{\refWord}(x) \neq
    \toITuple{\refWord}(y)$ for every ref-word $\refWord \in
    \reflang(A_\text{distinct})$.\footnote{Note that $A_\text{distinct}$ does
      not select all ref-words with $\toITuple{\refWord}(x) \neq
      \toITuple{\refWord}(y)$, as it does not consider cases where one variables
      is opened and closed before the other variable is opened.} This automaton
    is depicted in Figure~\ref{fig:adistinct}.
  \item $A_\text{overlap}$ is a three state automaton with
    $\reflang(A_\text{overlap})\eqdef(\Sigma \cup \{\vop{x},\vop{y}\})^* \cdot
    \Sigma \cdot (\Sigma \cup \{\vcl{x}, \vcl{y}\})^*$ that ensures that at
    least one symbol is read while both variables are open. We note that
    $A_\splitter^x$ and $A_\splitter^y$ already ensure that both variables are used.
  \end{itemize}
  
  \begin{figure}[t]
    \begin{tikzpicture}[>=latex]
    \node[state,initial] (q0) at (0,0){$q_0$};
    \node[state] (q1) at (2,0){$q_1$};
    \node[state] (q2) at (4,1){$q_2$};
    \node[state] (q3) at (4,-1){$q_3$};
    \node[state] (q4) at (6,1){$q_4$};
    \node[state] (q5) at (6,-1){$q_5$};
    \node[state] (q6) at (8,0){$q_6$};
    \node[state, double] (q7) at (10,0){$q_7$};
    \draw (q0) edge node[above] {$\vop{x}, \vop{y}$}(q1);
    \draw (q1) edge node[above,sloped] {$\vop{x},\vop{y}$}(q2);
    \draw (q1) edge node[above] {$\Sigma$}(q3);
    \draw (q2) edge node[above] {$\vcl{x}, \vcl{y}$}(q4);
    \draw (q3) edge node[above] {$\vop{x}, \vop{y}$}(q5);
    \draw (q4) edge node[above] {$\Sigma$}(q6);
    \draw (q5) edge node[above,sloped] {$\vcl{x}, \vcl{y}$}(q6);
    \draw (q6) edge node[above] {$\vcl{x}, \vcl{y}$}(q7);
    \draw (q0) edge[loop above] node[above] {$\Sigma$} (q0);
    \draw (q1) edge[loop above] node[above] {$\Sigma$} (q1);
    \draw (q2) edge[loop above] node[above] {$\Sigma$} (q2);
    \draw (q3) edge[loop above] node[above] {$\Sigma$} (q3);
    \draw (q4) edge[loop above] node[above] {$\Sigma$} (q4);
    \draw (q5) edge[loop above] node[above] {$\Sigma$} (q5);
    \draw (q6) edge[loop above] node[above] {$\Sigma$} (q6);
    \draw (q7) edge[loop above] node[above] {$\Sigma$} (q7);
  \end{tikzpicture}
  \caption{The automaton $A_\text{distinct}$.}\label{fig:adistinct}
  \end{figure}
 
  The constructions of all automata are in \logspace. Note that
  $A_\text{disjoint}$ is sequential, since $A_\splitter^x$, (resp.,
  $A_\splitter^y$) ensures that $x$ (resp., $y$) is not opened or closed several
  times and that it is closed if and only if it is opened. A ref-word $\refWord
  \in \reflang(A_\text{disjoint})$ witnesses that $\splitter$ is not disjoint,
  since $A_\splitter^x$ verifies that $\toITuple{\refWord}(x) \in
  \splitter(\clr(\refWord))$, $A_\splitter^y$ verifies that $\toITuple{\refWord}(y)
  \in \splitter(\clr(\refWord))$, $A_\text{distinct}$ verifies that
  $\toITuple{\refWord}(x) \neq \toITuple{\refWord}(y)$, and $A_\text{overlap}$ verifies that
  $\toITuple{\refWord}(x)$ and $\toITuple{\refWord}(y)$ overlap. For the other direction, a
  document $\doc$ where $\splitter(\doc)$ has two overlapping spans $s_1$ and
  $s_2$ ensures that $\refWord \in \reflang(A_\text{disjoint})$, where
  $\refWord$ is derived from $\doc$ by inserting opening and closing operations
  for $x$ and $y$ at the positions indicated by $s_1$ and $s_2$.
  
  \medskip
  
  $A_\text{highlander}$: We define the automaton $A_\text{highlander}$ as the
  intersection of the following automata, such that $\vars(A_\text{highlander})
  = V \cup \{x,y\}$. $A_\splitter^x$ (and $A_\splitter^y$) again ensures that
  $\tup \in \toSpanner{A_\text{highlander}}(\doc)$ implies that $\tup(x) \in
  \splitter(\doc)$ (resp., $\tup(y) \in \splitter(\doc)$). Following the same
  idea, $A_\spanner'$ ensures that $\projectTup{\tup}{V} \in \spanner(\doc)$.
  The automaton $A_\text{distinct}$ ensures that $\tup(x) \neq \tup(y)$ if $\tup
  \in \toSpanner{A_\text{highlander}}(\doc)$. The last automaton,
  $A_\text{enclosed}$ ensures that both, $\tup(x)$ and $\tup(y)$, contain
  $\projectTup{\tup}{V}$. More formally:
  \begin{itemize}
  \item $A_\splitter^x$, $A_\splitter^y$, and $A_\text{distinct}$ are as above
    but with additional self-loops for every symbol from $\Gamma_V$ at each
    state.
  \item $A_\spanner'$ is derived from $A_\spanner$ by adding self-loops for
    every label from $\Gamma_{\{x,y\}}$ to every state.
  \item $A_{\text{enclosed}}$ is an automaton with
    $\reflang(A_\text{enclosed})=(\Sigma \cup \{\vop{x},\vop{y}\})^* \cdot
    (\Sigma \cup \Gamma_V)^*\cdot (\Sigma \cup \{\vcl{x},\vcl{y}\})^*$ that
    ensures that no variable operation for variables from $V$ is used
    outside of the spans defined by $x$ and $y$.
  \end{itemize}
  We compute $A_\text{highlander}$ as the intersection of $A_\spanner'$,
  $A_\splitter^x$, $A_\splitter^y$, $A_\text{distinct}$, and $A_{\text{enclosed}}$.
  We note that even if the five automata are not sequential, the automaton
  $A_\text{highlander}$ is sequential. For every variable, one of the automata
  $A_\spanner'$, $A_\splitter^x$, and $A_\splitter^y$ ensures that it is not
  opened or closed several times and that it is closed if and only if it is
  opened.

  We explain why a ref-word $\refWord \in \reflang(A_\text{highlander})$
  witnesses a violation of the highlander condition. By the construction of
  $A_{\text{enclosed}}$, the (different) spans $\toITuple{\refWord}(x)$ and
  $\toITuple{\refWord}(y)$ are both in $\splitter(\clr(r))$ and both cover
  $\projectTup{\toITuple{\refWord}}{V} \in \spanner(\clr(r))$. For the other
  direction, a document $\doc$, spans $s_1,s_2 \in \splitter(\doc)$, and a tuple
  $\tup \in \spanner(\doc)$ witnessing the violation of the highlander condition
  ensure that \[\refWordFrom{\doc}{\tup \cup \{x \mapsto s_1, y \mapsto s_2\}}
    \;\in\; \reflang(A_\text{highlander})\;.\] Therefore, the language is not
  empty.
\end{proof}

We proceed by studying the complexity of testing the cover condition. Here, we
only give an upper bound, a matching lower bound is established in
Lemma~\ref{lem:lowerbounds-det}.

\begin{proposition}\label{prop:cover-pspace}
  $\cover[\vsa]$ is in \pspace.
\end{proposition}
\begin{proof}
  Let $\spanner$ be a spanner and $\splitter$ be a splitter, given as
  $A_\spanner, A_\splitter \in \vsa$. We assume, \mbox{w.l.o.g.}, that $x_S
  \notin V$. We define a spanner $A_V\in \vsa$ that selects every possible
  tuple. More formally, $A_V \eqdef (\Sigma,V, \{q_0\}, q_0, \{q_0\}, \delta)$
  is the \vset-automaton with a single state $q_0$, where $\delta \eqdef
  \{(q_0,c,q_0) \mid c \in \Sigma \cup \Gamma_V\}.$ We argue next that
  $\splitter$ covers $\spanner$ if and only if $\spanner \subseteq
  \toSpanner{A_V} \circ \splitter$.

  \medskip
  
  (if): Assume that the cover condition does not hold. Then there is a document
  $\doc \in \docs$ and a tuple $\tup \in \spanner(\doc)$, such that there is no
  split $s \in \splitter(\doc)$ which covers $\tup$. Even though $A_V$
  selects every possible tuple, we have $\tup \notin (\toSpanner{A_V} \circ
  \splitter)(\doc)$.

  \medskip
  
  (only if): Assume that the cover condition holds. Let $\doc \in \docs$ be a
  document and $\tup \in \spanner(\doc)$ be a \doc-tuple. Since \splitter covers
  \spanner, there is a split $s \in \splitter(\doc)$ which covers $\tup$. Thus,
  per definition of $A_V$, it must hold that $\unshiftSpanBy{\tup}{s} \in
  \toSpanner{A_V}(\doc),$. Therefore $\spanner \subseteq \toSpanner{A_V} \circ
  \splitter$ also holds.

  \medskip
  
  The \pspace upper bound follows from
  Proposition~\ref{prop:pcircs-algebra-construction} (bullet point 1), which
  shows that a \vset-automaton $A \in \vsa$ with $\toSpanner{A} =
  \toSpanner{A_V} \circ \splitter$ can be constructed in \ptime, and
  Theorem~\ref{thm:spannerContainment} which states that containment of
  \vset-automata is in \pspace.
\end{proof}

 \section{Complexity Upper Bounds}\label{sec:upper-bounds}
In this section, we show upper bounds for $\splitcorrectness, \splittability,$
and $\selfsplittability$. In Section~\ref{sec:splitcorrectness}, we show that
\splitcorrectness and \selfsplittability are in \pspace for regex-formulas and
\vset-automata, while both problems are in \ptime if $\spanner, \splitspanner,$
and $\splitter$ are given as $\ufvsa$ and $\spanner$ and \splitter satisfy the
highlander condition.

Sections~\ref{sec:splittabilityCharacterization}-\ref{sec:splittabilityHighlander}
are devoted to the \splittability problem.

\subsection{Split-Correctness and
  Self-Splittability}\label{sec:splitcorrectness}  

It follows directly from Proposition~\ref{prop:pcircs-algebra-construction} that
split-correctness is decidable in \ptime when the highlander condition is
satisfied and the \vset-automata are unambiguous and sequential.

\begin{lemma}\label{lem:splitcorrectness-upper-bound}
  Deciding $\splitcorrectness[\vsa]$ is in \pspace. Furthermore,
  $\splitcorrectness[\usvsa]$ is in \ptime if \spanner and \splitter
  satisfy the highlander condition.
\end{lemma}
\begin{proof}
  Let $A_\spanner, A_{\splitspanner}, A_\splitter \in \vsa$ with $\spanner =
  \toSpanner{A_\spanner}, \splitter = \toSpanner{A_\splitter}$, and $\splitspanner
  = \toSpanner{A_{\splitspanner}}.$ Furthermore, let $A_{\splitspanner \circ
    \splitter}$ be as constructed in
  Proposition~\ref{prop:pcircs-algebra-construction}, that is
  $\toSpanner{A_{\splitspanner \circ \splitter}} = \splitspanner \circ \splitter$.
  Thus, $\spanner$ is splittable by $\splitter$ via $\splitspanner$ if and only
  if $\toSpanner{A_\spanner} = \toSpanner{A_{\splitspanner
      \circ \splitter}}$. It follows from Theorem~\ref{thm:spannerContainment}
  that this equivalence can be checked in \pspace.

  Assume that $A_\spanner, A_{\splitspanner}, A_\splitter \in \ufvsa$ and that
  $\spanner$ and $\splitter$ satisfy the highlander condition. We begin by
  checking whether $\splitspanner \circ \splitter$ and $\splitter$ satisfy the
  highlander condition, which can be done in \ptime due to
  Proposition~\ref{prop:test-highlander}. If this is the case, it can be checked
  in \ptime whether $\spanner$ is splittable by $\splitter$ via $\splitspanner$
  as shown in Theorem~\ref{thm:spannerContainment-det}. Otherwise, if
  $\splitspanner \circ \splitter$ and $\splitter$ do not satisfy the highlander
  condition, there must be a document $\doc \in \docs$ and a tuple $\tup \in
  ({\splitspanner \circ \splitter})(\doc)$ such that at least two splits $s,s'
  \in \splitter(\doc)$ cover $\tup$. Therefore, due to $\spanner$ and
  $\splitter$ satisfying the highlander condition, it must hold that $\tup
  \notin \spanner(\doc)$, which implies that $\spanner$ is not splittable by
  $\splitter$ via $\splitspanner$.
\end{proof}

\begin{corollary}\label{cor:splitcorrectness-upper-bound-disjoint}
  $\splitcorrectness[\usvsa]$ is in \ptime if \spanner is proper and \splitter
  is disjoint.
\end{corollary}
\begin{proof}
  Follows directly from Observation~\ref{obs:disjoint-highlander} and
  Lemma~\ref{lem:splitcorrectness-upper-bound}.
\end{proof}

\begin{corollary}
  Deciding $\selfsplittability[\vsa]$ is in \pspace. Furthermore,
  $\selfsplittability[\usvsa]$ is in \ptime if either \spanner and \splitter
  satisfy the highlander condition or \spanner is proper and \splitter is
  disjoint.
\end{corollary}

\subsection{Characterization of
  Splittability}\label{sec:splittabilityCharacterization} 

We begin by giving a characterization of \splittability. To this end, we show
that a spanner is splittable by a splitter if and only if it is splittable via a
specific canonical split-spanner. Note that this characterization also holds for
spanners and splitter which are not regular.

The following example illustrates that there can be different split-spanners
witnessing splittability.

\begin{example}\label{ex:nondisjoint:splitter}
  Consider $\spanner \eqdef a y\set{b} b$ and $\splitter \eqdef x\set{ab}\, b
  \lor a x\set{b b}$. Then, both $\spanner = \splitspanner \circ \splitter$ and
  $\spanner = \splitspanner' \circ \splitter$ for $\splitspanner \eqdef a
  y\set{b}$ and $\splitspanner' \eqdef y\set{b} b$ but
  $\splitspanner\neq\splitspanner'$. The reason why this happens is that
  $\splitter$ selects two different spans $s=\spanFromTo{1}{3}$ and
  $s'=\spanFromTo{2}{4}$ that both cover the span $\spanFromTo{2}{3}$ selected
  by $P$ on $abb$. Since the selected spans are different, the split-spanners
  $\splitspanner$ and $\splitspanner'$ need to be different as well to be able
  to simulate $P$. Notice that $S$ is not a disjoint splitter, as
  $\spanFromTo{1}{3}$ and $\spanFromTo{2}{4}$ are not disjoint. \qed
\end{example}

We show, that there is a canonical split-spanner $\canonsplitspanner$ for every
spanner \spanner and splitter \splitter such that \spanner is splittable
by \splitter if and only if it is splittable via \canonsplitspanner:
\[
  \canonsplitspanner(\doc) \;\eqdef\; \big\{\tup \mid \,
  \forall \doc' \in \docs, \forall s \in \splitter(\doc') \text{ such that }
  \doc'_s = \doc, \text{ it holds that }
  (\shiftSpanBy{\tup}{s})\in \spanner(\doc')\big\}\;.
\]
Intuitively, a tuple is selected by \canonsplitspanner if and only if it is
``safe'' to be selected. A $\doc$-tuple $\tup$ is not safe if there is a
document $\doc'$ and a split $s \in \splitter(\doc')$ with $\doc'_s = \doc$ and
$\shiftSpanBy{\tup}{s} \notin \spanner(\doc)$. As we will show in
Lemma~\ref{lem:splittabilityCanonSubset}, $\canonsplitspanner \circ \splitter
\subseteq \spanner$. 

Note that the definition of \canonsplitspanner is not the same as in
\cite{DoleschalKMNN19}, where \canonsplitspanner is defined with an existential
quantifier instead of the second universal quantifier in the present definition.
The present canonical split-spanner can be used more generally.

\begin{lemma}\label{lem:splittabilityCanonSubset}
  Let $\spanner$ be a document spanner and $\splitter$ be a document splitter.
  Then $\canonsplitspanner \circ \splitter \subseteq \spanner$.
\end{lemma}
\begin{proof}
  Let \spanner and \splitter be as stated. Recalling the definition of the
  $\circ$ operator, we have that
  \[
    (\canonsplitspanner \circ \splitter)(\doc) \;\;\eqdef\; \bigcup_{s\in
      \splitter(\doc)}\set{\shiftSpanBy{\tup}{s}\mid \tup\in
      \canonsplitspanner(\doc_{s})}\,.
  \]
  Let $\doc$ be a document and $\tup \in (\canonsplitspanner \circ
  \splitter)(\doc)$ be a \doc-tuple. Then, there is a split $s \in
  \splitter(\doc)$, such that $\tup'  \eqdef \unshiftSpanBy{\tup}{s} \in
  \canonsplitspanner(\doc_s)$. Per definition of $\canonsplitspanner$ it must
  hold that $\tup = \shiftSpanBy{\tup'}{s} \in \spanner(\doc)$, concluding the
  proof.
\end{proof}

\begin{theorem}\label{thm:splittability}
  Let $\spanner$ be a document spanner and $\splitter$ be a document splitter.
  Then \spanner is splittable by \splitter if and only if \spanner is splittable
  by \splitter via $\canonsplitspanner$.
\end{theorem}
\begin{proof}
  We only have to show the ``only if'' direction, since the other direction is
  trivial. Due to Lemma~\ref{lem:splittabilityCanonSubset}, it suffices to show
  that $\spanner \subseteq \canonsplitspanner \circ \splitter$.

  To this end, assume that $\spanner$ is splittable by $\splitter$ via some
  spanner $\splitspanner$. We begin by showing that $\splitspanner \subseteq
  \canonsplitspanner$. Let \doc be a document and $\tup \in \spanner(\doc)$ be a
  \doc-tuple. As $\spanner = \splitspanner \circ \splitter$ there is a split
  $s \in \splitter(\doc)$, such that $\tup' \eqdef \unshiftSpanBy{\tup}{s} \in
  \splitspanner(\doc_s)$. For the sake of contradiction, assume that $\tup'
  \notin \canonsplitspanner(\doc_s)$. By definition of $\canonsplitspanner$,
  there is a document $\doc' \in \docs$ and a split $s' \in \splitter(\doc')$
  with $\doc_s = \doc'_{s'}$ such that $\shiftSpanBy{\tup'}{s'} \notin
  \spanner(\doc').$ Therefore, $s' \in \splitter(\doc')$ and $\tup' \in
  \splitspanner(\doc'_{s'})$ but $\shiftSpanBy{\tup'}{s'} \notin
  \spanner(\doc')$, leading to the desired contradiction as $\spanner =
  \splitspanner \circ \splitter$. Therefore, $\splitspanner \subseteq
  \canonsplitspanner$.

  It remains to show that $\spanner \subseteq \canonsplitspanner \circ
  \splitter$. Recalling the definition of $\spanner \circ \splitter$,
  \[
    (\splitspanner \circ \splitter)(\doc) \;\;=\; \bigcup_{s\in
      \splitter(\doc)}\set{\shiftSpanBy{\tup}{s}\mid \tup\in
      \splitspanner(\doc_{s})} \;\;\subseteq\; \bigcup_{s\in
      \splitter(\doc)}\set{\shiftSpanBy{\tup}{s}\mid \tup\in
      \canonsplitspanner(\doc_{s})} \;\;=\;\; (\canonsplitspanner \circ
    \splitter)(\doc). 
  \]
  Therefore, $\spanner = \splitspanner \circ \splitter \subseteq
  \canonsplitspanner \circ \splitter$, concluding the proof. 
\end{proof}

\subsection{Constructing the Canonical Split-Spanner}\label{sec:canonregular}
In this section, we will show that $\canonsplitspanner$ is regular if \spanner
and \splitter are regular (Corollary~\ref{cor:canonregular}). To this end, we show
that the language of valid ref-words in $\canonsplitspanner$ is regular by
defining a finite monoid $M$ such that $\canonsplitspanner$ is exactly the
spanner represented by the language recognized by $M$.

A \emph{monoid} is a triple $(M, \mop, \mzero)$ consisting of a set $M$, an
associative binary operation $\mop \colon M \times M \to M$, and a neutral
element $\mzero$. We say that a monoid $M$ \emph{recognizes} a language $\lang$
over the alphabet $\Xi$ if there is a homomorphism $h \colon \Xi^* \to M$ and a
set $M^{\acc} \subseteq M$ such that $w \in \lang$ if and only if $h(w) \in
M^{\acc}$. A function $h$ is a (string) homomorphism if and only if
$h(\varepsilon)=\mzero$ and $h(w_1 \cdot w_2) = h(w_1) \mop h(w_2)$ for all
strings $w_1,w_2 \in \Xi^*$. It is well known that a language $\lang$ is regular
if and only if it is recognized by a finite monoid $M$. All monoids that we
define will be finite.

Given a \vsa $A = (\Sigma, V, Q, q_0, Q_F, \delta)$, the
\emph{transition monoid} $M_A$ of $A$ is $(2^{Q\times Q}, \mop,
\id_Q)$, where $2^{Q\times Q}$ is the set of all possible binary relations over
$Q$, the operation $\mop$ is the composition of relations, i.e.
\[
  m_1 \mop m_2 \;\eqdef\; \big\{ (x,z) \mid \exists y \in Q \text{, such that } (x,y)
  \in m_1 \text{ and } (y,z) \in m_2\big\}\; ,
\]
and $\id_q \eqdef \{(q,q) \mid q \in Q\}$ is the identity relation over $Q$. The
canonical homomorphism $h_A$ for the transition monoid is defined by
\[
  h_A(\refWord) \;\eqdef\; \big\{ (p,q) \mid q \in \delta^*(p,\refWord)\big\}\;.
\]
For reasons
that become apparent later, we define $h_A(\alpha)=\id_Q$ for every
variable operation $\alpha \in \Gamma_{\svars \setminus \dom(A)}$ that
does not belong to a variable used by $A$. This has the effect that
$h_A$ ignores all ``foreign'' variables, which is helpful when combining
the transition monoids of different spanners.

\begin{observation}\label{obs:spannerMonoid}
  Let $A \in \vsa$. The language $\reflang(A)$ is recognized by $M_A$.
\end{observation}
\begin{proof}
  Let $\refWord$ be a ref-word over the alphabet $\Sigma \cup \varop{\dom(A)}$,
  that is $\refWord \in (\Sigma \cup \varop{\dom(A)})^*$. Furthermore, let
  $M_A^\acc \eqdef \{m \mid m \cap (\{q_0\} \times Q_F) \neq \emptyset\}$ be the
  set of accepting monoid elements. As $\refWord \in \reflang(A)$ if and only if
  there is an accepting run of $A$ on $\refWord$ and by the definition of $h_A$,
  we get that $\refWord \in \reflang(A)$ if and only if $h_A(\refWord) \in
  M_A^\acc$, concluding the proof.
\end{proof}
As we will show, given a spanner \spanner, one can also construct a monoid that
recognizes the language of all valid ref-words, satisfying the variable order
condition, which correspond to a tuple selected by \spanner. More formally, we
define the language $\reflang^\spanner$, where $\spanner$ is a document spanner:
\[
  \reflang^\spanner \;\eqdef\; \big\{ \refWordFrom{\doc}{\tup} \mid \exists \doc
  \in \docs, \text{ such that } \tup \in \spanner(\doc)\big\}\;. 
\]
Observe that $\reflang^\spanner = \reflang(A)$ if $\spanner$ is given as a
sequential \vset-automaton $A$ which satisfies the variable order condition. We
generalize this and show that, for every document spanner $\spanner$ given by a
\vset-automaton $A$, there is a monoid $M$ of size exponential in $A$ which
recognizes $\reflang^\spanner$.

\begin{restatable}{lemma}{monoidConstructionLemma}\label{lem:monoidConstruction}
  Let $A\in \vsa$. There is a monoid $M_A^\prec$ of exponential size that recognizes
  $\reflang^{\toSpanner{A}}$. Furthermore, $M_A^\prec$ can be constructed in \pspace.
\end{restatable}

We note that if $A$ is sequential and satisfies the variable order condition,
then $\reflang^{\toSpanner{A}} = \reflang(A)$ and the transition monoid $M_A$ of
$A$ can be used for $M_A^\prec$. In the general case, the construction of
$M_A^\prec$ is quite involved. To meet the exponential size restriction it is
not possible to compute an equivalent sequential \vset-automaton that complies
with the variable order condition. Instead, sequentiality and the variable order
condition have to be dealt with in the monoid construction itself. We give a
proof for Lemma~\ref{lem:monoidConstruction} in
Appendix~\ref{sec:generalMonoid}.

In the following, given a set $X$, we denote by $2^X$ the power set of $X$.
Given a set $V$ of variables, we define the monoid $M_V$ that can test whether a
ref-word (using variables from $V$) satisfies the variable order condition:
\begin{align*}
  M_V \;\eqdef{}&\; \Big(2^{\varop{V}} \cup \{0\},\mop_V,\emptyset\Big) \\
  X \mop_V Y \;\eqdef{}&
  \begin{cases}
    X \cup Y & \text{if } X \cap Y = \emptyset \text{ and } x \prec y \text{ for
      all } x \in X, y \in Y \\
    0 & \text{otherwise}
  \end{cases}
\end{align*}

\begin{lemma}
  For every finite set $V \subseteq \svars$ of variables, $M_V$ recognizes the
  set $\reflang^V$ of all valid ref-words over $V$ which satisfy the variable
  order condition.
\end{lemma}
\begin{proof}
  Let $M_V^{\acc}= \{X \neq 0 \mid \forall v \in V, \text{it
      holds that} \vop{v} \in X \Leftrightarrow \vcl{v} \in X\}$ and $h_V
  \colon (\Sigma \cup V)^* \to M_V$ be the homomorphism induced by
  \[
    h_V(a) \;\eqdef\;
    \begin{cases}
      {a} & \text{if } a \in \varop{V}\\
      \emptyset & \text{otherwise}
    \end{cases}
  \]

  It remains to show that $\refWord \in (\Sigma \cup \varop{V})^*$ is valid and
  satisfies the variable order condition if and only if $h(\refWord) \in
  M_V^{\acc}$. To this end, let $h(\refWord) \in M_V^{\acc}$.
  Observe that, per definition of $\mop_V$, $\refWord$ must satisfy the variable
  order condition. Furthermore, per definition of $\prec$, it must hold that
  $\vop{v} \prec \vcl{v}$ for all variables $v \in \svars.$ Thus, $\refWord$
  must be valid, as all variables $v \in \dom(\refWord)$ must be opened and
  closed exactly once and opened before they are closed. For the other
  direction, assume that $\refWord$ is valid and satisfies the variable order
  condition. It is straightforward to verify that $h(\refWord) \neq 0$ and
  furthermore, $h(\refWord) \in M_V^{\acc}$.
\end{proof}

Let $\spanner$ be a regular document spanner, \splitter be a regular document
splitter, and $V = \vars(\spanner)$. We now use
Lemma~\ref{lem:monoidConstruction} to show that the Cartesian product of the
monoids $M_V$, $M_\spanner^\prec$, and $M_\splitter^\prec$ contains enough
structure to recognize $\canonlang$. Therefore, $\canonsplitspanner$ is indeed a
regular document spanner.

\begin{proposition}\label{prop:MrecL}
  For every regular document spanner $\spanner$ and every regular document
  splitter $\splitter$, the monoid $M\eqdef M_V \times M_\spanner^\prec \times
  M_\splitter^\prec$ recognizes $\canonlang$.
\end{proposition}
\newcommand{\macc}{
  \begin{align*}
    M^{\acc} \eqdef \Big\{ (m_{V},m_\spanner,m_\splitter) \mid {}
    & m_{V} \in M_{V}^{\acc} \text{ and for all } \doc_1,\doc_2 \in \docs
      \text{ it holds that } m'_\splitter \in M_\splitter^{\acc} \Rightarrow m'_\spanner \in M_\spanner^{\acc}, \\
    & \text{where } (m'_{V},m'_\spanner,m'_\splitter) = h(\doc_1) \mop h(\vop{x_\splitter}) \mop (m_{V},m_\spanner,m_\splitter) \mop h(\vcl{x_\splitter}) \mop h(\doc_2) \Big\}\:.
  \end{align*}}
\begin{proof}
  Let $h\colon (\Sigma \cup \varop{V})^* \to M$ be the homomorphism defined
  by $h(\refWord) \eqdef (h_V(\refWord), h_\spanner(\refWord),
  h_\splitter(\refWord))$. We define $M^\acc$ as
  \macc

  Recall that $M_{V}^{\acc}= \{X \neq 0 \mid \forall v \in V, \text{it holds
    that} \vop{v} \in X \Leftrightarrow \vcl{v} \in X\}$. Furthermore, for a
  ref-word $\refWord$, it holds that $h(\refWord) \in M_\spanner^\acc$ (resp.,
  $h(\refWord) \in M_\splitter^\acc$) if and only if $\refWord \in
  \reflang^\spanner$ (resp., $\refWord \in \reflang^\splitter$). We have to show
  that, for every ref-word $\refWord$, it holds that $\refWord \in \canonlang$
  if and only if $h(\refWord) \in M^\acc$.

  \medskip
  
  (if): Let $\refWord$ be a ref-word and $\doc = \clr(\refWord)$. Assume that
  $h(\refWord) \in M^{\acc}$. By definition of $M_{V}$ and the fact that
  $h_{V}(r) \in M_{V}^\acc$, we can conclude that $\refWord$ is valid and
  satisfies the variable order condition. It remains to show that
  $\toTuple{\refWord} \in \canonsplitspanner(\doc),$ which implies that
  $\refWord \in \canonlang$. To this end, let $\doc' \in \docs$ and $s \in
  \splitter(\doc')$ such that $\doc'_s = \doc$. If no such $\doc'$ and $s$
  exist, $\toTuple{\refWord} \in \canonsplitspanner(\doc)$ and we are done.
  Otherwise, $\doc'$ can be decomposed as $\doc' = \doc_1 \cdot \doc \cdot
  \doc_2$. By definition of $M_\splitter^\acc$, we have that $m'_\splitter =
  h_\splitter(\doc_1) \mop h_\splitter(\vop{x_\splitter}) \mop
  h_\splitter(\refWord) \mop h_\splitter(\vcl{x_\splitter}) \mop
  h_\splitter(\doc_2) \in M_\splitter^{\acc}$. Let $\refWordPrime = \doc_1 \cdot
  \refWord \cdot \doc_2$. Thus, $\toTuple{\refWordPrime} =
  \shiftSpanBy{s}{\toTuple{\refWord}} \in \spanner(\doc')$ if and only if
  $h(\refWordPrime) \in M_\spanner^{\acc}$. Furthermore, due to $h_\spanner$
  ignoring $x_\splitter$, $h_\spanner(\refWordPrime) \in M_\spanner^{\acc}$ if
  and only if $m'_\spanner \in M_\spanner^{\acc}$. As $m'_\splitter \in
  M_\splitter^\acc$, we have by the definition of $M^\acc$ that
  $h_\spanner(\refWordPrime) \in M_\spanner^{\acc}$. This implies that
  $\toTuple{\refWordPrime} \in \spanner(\doc')$ and therefore
  $\toTuple{\refWord} \in \canonsplitspanner(\doc)$, concluding the if-part of
  the proof.

  \medskip
  
  (only if): Let $\refWord \in \canonlang$ and $\doc = \clr(\refWord)$. Thus,
  $\toTuple{\refWord} \in \canonsplitspanner(\doc)$ and $\refWord$ is valid and
  satisfies the variable order condition. We show that $m =
  (m_{V},m_\spanner,m_\splitter)=h(\refWord) \in M^{\acc}$. As $\refWord$ is
  valid and satisfies the variable order condition, we have that $m_{V} \in
  M_{V}^\acc$. It remains to show that for every $\doc_1,\doc_2 \in \docs$ it
  holds that $m'_\splitter \in M_\splitter^{\acc}$ implies that $m'_\spanner \in
  M_\spanner^{\acc}$, where $(m'_{V},m'_\spanner,m'_\splitter) = h(\doc_1) \mop
  h(\vop{x_\splitter}) \mop (m_{V},m_\spanner,m_\splitter) \mop
  h(\vcl{x_\splitter}) \mop h(\doc_2)$. To this end let $\doc_1,\doc_2 \in
  \docs$ be arbitrary documents and $m_\spanner'$ and $m_\splitter'$ be as above
  with $m'_\splitter \in M_\splitter^{\acc}$. We have to show that $m'_\spanner
  \in M_\spanner^{\acc}$. Let $\refWordPrime = \doc_1 \cdot \vop{x_\splitter}
  \cdot \doc \cdot \vcl{x_\splitter} \cdot \doc_2$ and recall that $\refWord\in
  \reflang^\splitter$ if and only if $h_\splitter(\refWord) \in
  M_\splitter^{\acc}$. Observe that $h_\splitter(\refWordPrime) = m'_\splitter
  \in M_\splitter^\acc$ and thus $\refWordPrime \in \reflang^\splitter$. Let
  $\doc' = \clr(\refWordPrime)$. Thus, $s =
  \spanFromTo{|\doc_1|}{|\doc_1\cdot\doc|} \in \splitter(\doc')$ and $\doc'_s =
  \doc$. Thus, $\shiftSpanBy{\toTuple{\refWord}}{s} \in \spanner(\doc)$ and
  $\refWordFrom{\doc}{\shiftSpanBy{\toTuple{\refWord}}{s}} \in
  \reflang^\spanner$. Observe that
  $\refWordFrom{\doc}{\shiftSpanBy{\toTuple{\refWord}}{s}} = \refWordPrime$ and
  therefore it follows that $m'_\spanner =
  h_\spanner(\refWordFrom{\doc}{\refWordPrime}) \in M_\spanner^{\acc}$,
  concluding the proof.
\end{proof}

\begin{corollary}\label{cor:canonregular}
  \canonsplitspanner is a regular document spanner.
\end{corollary}

\subsection{Complexity Upper Bound for Splittability in the General
  Case}\label{sec:splittability-general-upperbound}
The proof of the upper bound consists of two parts. We first show that testing
whether an element $m \in M$ belongs to $M^\acc$ is in \pspace
(Proposition~\ref{prop:minma}) and then give an \expspace algorithm for testing
splittability (Theorem~\ref{thm:splittabilityExpspace}).

\begin{proposition}\label{prop:minma}
  Let $m \in M$ be a monoid element. It can be tested in \pspace whether $m \in
  M^{\acc}$.
\end{proposition}
\begin{proof}
  Recall that
  \macc
  
  We give a \pspace algorithm which decides whether
  $(m_{V},m_\spanner,m_\splitter) \notin M^{\acc}$ by guessing a
  counterexample.\footnote{Recall that \pspace is closed under complement.} By
  definition of $M^{\acc}$, $(m_{V},m_\spanner,m_\splitter) \notin M^{\acc}$ if
  and only if
  \begin{enumerate}
  \item $m_{V} \notin M_{V}^{\acc}$; or
  \item there are $\doc_1,\doc_2 \in \docs$ with $h(\doc_1) \mop h(\vop{x_\splitter})
    \mop m \mop h(\vcl{x_\splitter}) \mop h(\doc_2) \in M_V \times (M_\spanner
    \setminus M_\spanner^\acc) \times M_\splitter^{\acc}$.
  \end{enumerate}
  Recall that $M_{V}^{\acc}= \{X \neq 0 \mid \forall v \in {V},
  \vop{v} \in X \Leftrightarrow \vcl{v} \in X\}$. Thus, the first condition can
  be checked in \ptime. Due to $h$ being a homomorphism, it must hold that
  $h(\sigma_1\cdots\sigma_n) = h(\sigma_1) \mop h(\sigma_n)$. Therefore, the
  second condition can be checked by guessing $\doc_1$ and $\doc_2$ symbol by
  symbol and computing $h(\doc_1)$ and $h(\doc_2)$ on the fly.
\end{proof}

We are now ready to given an upper bound for \splittability.

\begin{theorem}\label{thm:splittabilityExpspace}
  $\splittability[\cC]$ is in \expspace.
\end{theorem}
\begin{proof}
  Let $\spanner \in \cC$ and $\splitter \in \cC$ be a spanner and a splitter. By
  Theorem~\ref{thm:splittability}, $\spanner$ is splittable by $\splitter$ if
  and only if $\spanner$ is splittable by $\splitter$ by $\canonsplitspanner$.
  The high level idea of the proof is to compute a \vset-automaton $A$ for
  $\canonsplitspanner \circ \splitter$ and then test equivalence with \spanner.
  
  Recall that $|M|$ is exponential in the size of $A_\spanner$ and $A_\splitter$
  (cf. Lemma~\ref{lem:monoidConstruction}). To exploit the construction of
  Proposition~\ref{prop:pcircs-algebra-construction}, we turn $M$ into the
  \vset-automaton $A_M=(\Sigma, V, M, h(\varepsilon), M^\acc, \delta)$, where
  the transition function is defined by $\delta(m,\sigma)=m \mop h(\sigma)$. We
  use the monoid elements as states of the automaton. From the construction and
  definition of $M$ it is obvious that $\toSpanner{A}=\canonsplitspanner$ and
  that $A_M$ is linear in the size of $M$. By Proposition~\ref{prop:minma},
  $M^\acc$ can be constructed in \pspace. Now we apply
  Proposition~\ref{prop:pcircs-algebra-construction} to obtain an automaton $A$
  for $\canonsplitspanner \circ \splitter$, which is of polynomial size in $M$
  and thus exponential in the size of $\splitter$ and $\spanner$. Testing
  equivalence of $\spanner$ and $A$ can be done in space polynomial in
  $\spanner$ and $A$. As $A$ is of exponential size, this yields the \expspace
  bound claimed in the theorem statement.
\end{proof}

 \subsection{Complexity Upper Bound for Splittability under the Highlander
  Condition}\label{sec:splittabilityHighlander}
In this section we will show that the upper bound of splittability can be
improved to \pspace if the spanner and the splitter satisfy the highlander
condition. We begin by characterizing counter examples to splittability under
the highlander condition.

\begin{lemma}\label{lem:splittabilityCounterexample}
  Let \spanner and \splitter be a spanner and a splitter such that the
  highlander and cover conditions are satisfied. Then \spanner is splittable by
  \splitter if and only if there is no ref-word $\refWord=\doc_1 \cdot
  \vop{x_\splitter} \cdot \refWordPrime \cdot \vcl{x_\splitter} \cdot \doc_2 \in
  (\Sigma \cup \varop{\vars(\spanner)} \cup \varop{\vars(\splitter)})^*$ such
  that
  \begin{itemize}
  \item $\doc_1,\doc_2 \in \docs$;
  \item $\doc_1 \cdot \refWordPrime \cdot \doc_2 \in \reflang^\spanner$;
  \item $\doc_1 \cdot \vop{x_\splitter} \cdot \clr(\refWordPrime) \cdot
    \vcl{x_\splitter} \cdot \doc_2 \in \reflang^\splitter$; and
  \item $\refWordPrime \notin \reflang^{\canonsplitspanner}$.
  \end{itemize}
\end{lemma}
\begin{proof}
  Assume that \spanner is not splittable by \splitter. Due to
  Theorem~\ref{thm:splittability}, there must be a document $\doc$ and a tuple
  $\tup \in \spanner(\doc) \setminus (\canonsplitspanner \circ
  \splitter)(\doc)$. Due to $\tup \in \spanner(\doc)$, it holds that
  $\refWordFrom{\doc}{\tup} \in \reflang^\spanner$ and, due to the cover
  condition, there must be a split $\spanij \in \splitter(\doc)$ which covers
  $\tup$. Let $\doc_1,\doc_2 \in \Sigma^*$ and $\refWordPrime$ be a ref-word,
  such that $\doc_1 \cdot \refWordPrime \cdot \doc_2 =
  \refWordFrom{\doc}{\tup}$, $i = |\doc_1| + 1$, and $j = |\doc_1 \cdot
  \clr(\refWordPrime)| + 1$. Furthermore, let $\refWord=\doc_1 \cdot
  \vop{x_\splitter} \cdot \refWordPrime \cdot \vcl{x_\splitter} \cdot \doc_2$,
  thus $\clr(\refWord) = \doc$ and $\doc_{\spanij} = \clr(\refWordPrime)$. Due
  to $i = |\doc_1| + 1$ and $j = |\doc_1 \cdot \clr(\refWordPrime)| + 1$, it
  follows that $\doc_1 \cdot \vop{x_\splitter} \cdot \clr(\refWordPrime) \cdot
  \vcl{x_\splitter} \cdot \doc_2 \in \reflang^\splitter$. Therefore $\refWord$
  satisfies the first three conditions of the lemma statement. Assume, towards a
  contradiction, that $\refWordPrime \in \reflang^{\canonsplitspanner}$. This
  implies that $\tup \in (\canonsplitspanner \circ \splitter)(\doc)$, which is a
  contradiction to the assumption that $\tup \in \spanner(\doc) \setminus
  (\canonsplitspanner \circ \splitter)(\doc)$ showing that $\refWord$ also
  satisfies the last condition given in the lemma statement.

  On the other hand assume that there is a string $\refWord=\doc_1 \cdot
  \vop{x_\splitter} \cdot \refWordPrime \cdot \vcl{x_\splitter} \cdot \doc_2 \in
  (\Sigma \cup \varop{\vars(\spanner)} \cup \varop{\vars(\splitter)})^*$
  satisfying the conditions from the lemma statement. By $\doc_1 \cdot
  \refWordPrime \cdot \doc_2 \in \reflang^\spanner$, we have that $\tup =
  \toTuple{\doc_1 \cdot \refWordPrime \cdot \doc_2} \in \spanner(\clr(\refWord))$.
  By $\doc_1 \cdot \vop{x_\splitter} \cdot \clr(\refWordPrime) \cdot
  \vcl{x_\splitter} \cdot \doc_2 \in \reflang^\splitter$, we have that $\spanij
  \in \splitter(\clr(\refWord))$ covers $t$, where $i = |\doc_1|+1$ and
  $j=|\doc_1 \cdot \clr(\refWordPrime)| + 1$. As \spanner and \splitter satisfy
  the highlander condition, there can be no other span in
  $\splitter(\clr(\refWord))$ that covers $t$. Furthermore, as $\refWordPrime
  \notin \reflang^{\canonsplitspanner}$, we can conclude that $t \notin
  \canonsplitspanner \circ S(\clr(\refWord))$, contradicting that \spanner is
  splittable by \splitter using \canonsplitspanner. By
  Theorem~\ref{thm:splittability}, we can conclude that \spanner is not
  splittable by \splitter.
\end{proof}

\begin{theorem}\label{thm:splittabilityHighlander}
  Let $\spanner$ be a regular document spanner and \splitter be a regular
  document splitter, both given as \vset-automaton, such that the highlander
  condition is satisfied. Then, $\splittability[\vsa]$ is in \pspace.
\end{theorem}
\begin{proof}
  We first verify whether $\splitter$ covers $\spanner$. Note that the cover
  condition can be checked in \pspace (Proposition~\ref{prop:cover-pspace}) and
  is necessary for splittability (Observation~\ref{obs:splittabilityCover}).
  Thus, for the remainder of this proof, we can assume that the cover condition
  is satisfied.

  As \spanner and \splitter satisfy the highlander and cover condition, we can
  now use Lemma~\ref{lem:splittabilityCounterexample}. We provide a
  non-deterministic algorithm that runs in polynomial space for the complement
  problem, i.e., checking whether \spanner is not splittable by \splitter. We
  exploit Lemma~\ref{lem:splittabilityCounterexample}.

  The algorithm guesses a string $\refWord=\doc_1 \cdot \vop{x_\splitter} \cdot
  \refWordPrime \cdot \vcl{x_\splitter} \cdot \doc_2 \in (\Sigma \cup
  \varop{\vars(\spanner)} \cup \varop{\vars(\splitter)})^*$ letter by letter and
  computes $h_\spanner(\refWord)$, $h_\splitter(\refWord)$, and
  $h(\refWordPrime)$ on the fly. We note that $h_\spanner(\refWord)$ can be
  computed in polynomial space by starting with the monoid element $m_\spanner =
  h_\spanner(\varepsilon)$ and replacing $m_\spanner$ with $m_\spanner \cdot
  h_\spanner(\sigma)$ whenever a new letter $\sigma$ is guessed. The elements
  $h_\splitter(\refWord)$ and $h(\refWordPrime)$ can be computed analogously.

  Finally, by Lemma~\ref{lem:splittabilityCounterexample}, the facts that
  \spanner and \splitter satisfy the highlander and cover condition, and the
  definition of the monoids $M_\spanner$, $M_\splitter$, and $M$, we have that
  \spanner is not splittable by \splitter if $h_\spanner(\refWord) \in
  M_\spanner^\acc$, $h_\splitter(\refWord) \in M_\splitter^\acc$, and
  $h(\refWordPrime) \notin M^\acc$. We remind that $h_\spanner$ and
  $h_\splitter$ ignore ``foreign'' variables. By Proposition~\ref{prop:minma},
  the condition $h(\refWordPrime) \notin M^\acc$ can be checked in polynomial
  space. As the other two conditions can be easily checked in polynomial space,
  this concludes the proof.
\end{proof}

The following corollary is immediate by
Observation~\ref{obs:disjoint-highlander} and
Theorem~\ref{thm:splittabilityHighlander}.
\begin{corollary}\label{cor:splittabilityDisjoint}
  Let $\spanner$ be a regular document spanner and \splitter be a regular
  document splitter, both given as \vset-automaton, such that spanner is proper
  and splitter is disjoint. Then, deciding $\splittability[\vsa]$ is in \pspace.
  \qed
\end{corollary}

 \section{Complexity Lower Bounds}\label{sec:lower-bounds}
In this section, we will give lower bounds for \splitcorrectness, \splittability
and other related decision problems. To this end, recall that
$\splitcorrectness[\cC]$ and $\selfsplittability[\cC]$ are in \ptime if $\cC \in
\cCt$ and the highlander condition is satisfied. Here, we show that neither $\cC
\in \cCt$ nor the highlander condition on its own are sufficient to achieve
tractability.

We start by showing that \splitcorrectness, \splittability, and
\selfsplitcorrectness are \pspace-hard, even if the spanner is proper and both,
the spanner and the splitter are given as deterministic functional
\vset-automata. As we will see in the proof it is already \pspace-hard to decide
whether the cover condition is satisfied.

\begin{lemma}\label{lem:lowerbounds-det}
  $\selfsplittability[\dfvsa]$, $\splittability[\dfvsa]$, and $\cover[\dfvsa]$
  are \pspace-hard, even if \spanner is proper. 
\end{lemma}
\begin{proof}
  We give a reduction from the \pspace-complete problem of DFA concatenation
  universality \cite{Jiang93}. Given two DFAs $A_1,A_2$, DFA concatenation
  universality asks whether $\lang(\Sigma^*) = \lang(A_1) \cdot \lang(A_2)$.
  
  Let $A_1, A_2$ be regular languages, given as DFAs over the alphabet $\Sigma$.
  Furthermore, let $a \notin \Sigma$. Slightly abusing notation, we define the
  \dfvsa by a hybrid regex-formula, where the automata $A_i$ are plugged in. In
  particular, $A_\spanner = \Sigma^*\cdot y\{a\}$ and $A_\splitter = A_1 \cdot
  x\set{A_2\cdot a}.$ Let $\spanner = \toSpanner{A_\spanner}$ and $\splitter =
  \toSpanner{A_\splitter}$. Thus, $\spanner(\doc) = \emptyset = \splitter(\doc)$
  if $\doc \notin \lang(\docs \cdot a)$. Furthermore, if $\doc \in \lang(\docs
  \cdot a)$, $\spanner(\doc) = \spanFromTo{|\doc|}{|\doc|+1}$ and for all
  $\spanij \in \splitter(\doc)$ it holds that $i \leq |\doc|$ and $j =
  |\doc|+1$.
  
  We show that the following statements are equivalent:
  \begin{enumerate}
    \item \spanner is self-splittable by \splitter,
    \item \spanner is splittable by \splitter,
    \item $\lang(A_1) \cdot \lang(A_2) = \lang(\Sigma^*)$,
    \item \splitter covers \spanner.
  \end{enumerate}

  We observe that (1) implies (2). Thus, we only need to show that (2) implies (3),
  (3) implies (4), and (4) implies (1).

  \medskip

  \emph{(2) implies (3):} Assume that $\lang(A_1) \cdot \lang(A_2) \neq
  \lang(\Sigma^*)$. Thus there is a document $\doc \in \Sigma^*$ such that $\doc
  \notin \lang(A_1) \cdot \lang(A_2)$. Therefore, $\splitter(\doc \cdot a) =
  \emptyset$ but $\spanner(\doc \cdot a) = \{\spanFromTo{|\doc|+1}{|\doc|+2}\}
  \neq \emptyset$ and therefore \spanner can not be splittable by \splitter.

  \medskip

  \emph{(3) implies (4):} Assume that $\lang(A_1) \cdot \lang(A_2) =
  \lang(\Sigma^*)$. Let $\doc' \in (\Sigma \cup \{a\})^*$ and $\tup \in
  \spanner(\doc')$. Thus, $\doc' = \doc \cdot a$, for some document $\doc \in
  \docs$ and $\tup(y) = \spanFromTo{|\doc|+1}{|\doc|+2}$. Per assumption, there
  is a decomposition $\doc = \doc_1 \cdot \doc_2$, such that $\doc_i \in A_i$,
  for $i \in \{1,2\}$. Therefore, $s \eqdef \spanFromTo{|\doc_1|+1}{|\doc|+2}
  \in \splitter(\doc \cdot a) = \splitter(\doc')$ which implies that $s$ covers
  $\tup$.

  \medskip

  \emph{(4) implies (1):} We will show that $\spanner = \spanner \circ
  \splitter$. Let $\tup \in \spanner(\doc')$ be a tuple. Therefore, there is a
  document $\doc \in \docs$ such that $\doc' = \doc \cdot a$. As \splitter
  covers \spanner, there is a split $s \in \splitter(\doc')$ which covers
  $\tup$. Observe that per definition of \spanner, $\unshiftSpanBy{\tup}{s} \in
  \spanner(\doc'_s)$ and therefore $\tup \in (\spanner \circ \splitter)(\doc')$,
  implying that $\spanner \subseteq \spanner \circ \splitter$. For the other
  direction, let $\tup \in (\spanner \circ \splitter)(\doc')$. Therefore, there
  is a document $\doc \in \docs$ with $\doc' = \doc \cdot a$. Thus, there is a
  span $s \in \splitter(\doc')$ which covers $\tup$. It follows per definition
  of $\spanner$ that $\unshiftSpanBy{\tup}{s} \in \spanner(\doc'_s)$. Which
  implies that $\tup \in \spanner(\doc')$ and therefore $\spanner \circ
  \splitter \subseteq \spanner$.
\end{proof}

It follows directly that $\splitcorrectness[\dfvsa]$ is \pspace-hard.
\begin{corollary}\label{cor:lowerbound-splitcorrectness-det}
  $\splitcorrectness[\dfvsa]$ is \pspace-hard, even if \spanner is proper.\qed
\end{corollary}

Smit~\cite[Proposition 3.3.7]{Smit20} shows that $\splitcorrectness[\dfvsa]$ and
$\selfsplittability[\dfvsa]$ remain \pspace-hard if $\spanner$ is a Boolean
spanner (and therefore not proper) and $\splitter$ is disjoint. It is
straightforward to extend the proof of Smit along the lines of
Lemma~\ref{lem:lowerbounds-det} to show that $\splittability[\dfvsa]$ and
$\cover[\dfvsa]$ are also \pspace-hard for disjoint splitter. We refer to
Appendix~\ref{apx:smitProof} for the proof.

\begin{restatable}{lemma}{smitProof}\label{lem:lowerbounds-det-disjoint}
  $\selfsplittability[\dfvsa]$, $\splittability[\dfvsa]$, and $\cover[\dfvsa]$
  are \pspace-hard, even if \splitter is disjoint.
\end{restatable}
\begin{corollary}\label{cor:lowerbound-splitcorrectness-det-disjoint}
  $\splitcorrectness[\dfvsa]$ is \pspace-hard, even if \splitter is disjoint.\qed
\end{corollary}

Recall that $\selfsplittability[\usvsa]$ is in \ptime if the spanner is proper
and the splitter is disjoint (cf. Lemma~\ref{lem:splitcorrectness-upper-bound}).
We will show now that tractability is also lost if the spanner and the splitter
are not required to be unambiguous and sequential. That is, we show that
\selfsplittability and \splittability remain \pspace-hard even if the highlander
condition is satisfied\footnote{Recall that the highlander condition is
  satisfied if \spanner is proper and \splitter is disjoint (cf.
  Observation~\ref{obs:disjoint-highlander}).} and the spanner and splitter are
given as functional regex-formulas or functional \vset-automata.

\begin{lemma}\label{lem:lowerbounds-highlander}
  $\selfsplittability[\cC]$ and $\splittability[\cC]$, for $\cC \in
  \{\frgx,\fvsa\}$, are \pspace-hard, even if the splitter \splitter is
  disjoint and the spanner \spanner is proper.
\end{lemma}
\begin{proof}
  The reductions are from the containment problem for regular expressions and
  NFAs which are both known to be \pspace-complete. 

  Let $\lang_1$ and $\lang_2$ be regular languages and let $\spanner =
  y\set{\lang_1}$ and $\splitter = x\set{\lang_2}$. We show the following
  statements are equivalent:

  \begin{enumerate}
  \item \spanner is self-splittable by \splitter,
  \item \spanner is splittable by \splitter,
  \item $\lang_1 \subseteq \lang_2$.
  \end{enumerate}
  
  The lemma statement follows directly from the fact that containment of regular
  languages is \pspace-complete for NFAs and regular expressions.
  It remains to show the equivalence of~(1),~(2), and~(3). We observe that~(1)
  implies~(2) per definition.

  \medskip

  \emph{(2) implies (3):} Assume that \spanner is splittable by \splitter. Let
  $\doc \in \lang_1$ be a document. By definition of \spanner it follows that
  $\spanFromTo{1}{|\doc|+1} \in \spanner(\doc)$. Since \spanner is splittable by
  \splitter and $\spanFromTo{1}{|\doc|+1}$ is only covered by itself, it follows
  that $\spanFromTo{1}{|\doc|+1} \in \splitter(\doc)$ and
  $\spanFromTo{1}{|\doc|+1} \in \splitspanner(\doc_{\spanFromTo{1}{|\doc|+1}}) =
  \splitspanner(\doc)$ for some spanner \splitspanner. Therefore, by definition
  of \splitter, we have that $\doc \in \lang_2$.

  \medskip

  \emph{(3) implies (1):} Let $\lang_1 \subseteq \lang_2$. Observe, that \spanner
  only selects the span $\spanFromTo{1}{|\doc|+1}$. Therefore, \spanner is
  self-splittable by \splitter:
  \[
    \begin{array}[b]{rcl}
      \spanFromTo{1}{|\doc|+1} \in \spanner(\doc) & \Leftrightarrow & \doc \in \lang_1 \\
      &\Leftrightarrow & \doc \in \lang_1 \;\text{ and }\; \doc \in \lang_2 \\
      &\Leftrightarrow & \spanFromTo{1}{|\doc|+1} \in \spanner(\doc) \;\text{ and }\;
                        \spanFromTo{1}{|\doc|+1} \in \splitter(\doc) \\
      &\Leftrightarrow & \spanFromTo{1}{|\doc|+1} \in (\spanner \circ
                         \splitter)(\doc)
    \end{array}\qedhere
  \]
\end{proof}

Again, it follows directly that $\splitcorrectness[\dfvsa]$ is \pspace-hard.
\begin{corollary}\label{cor:lowerbound-splitcorrectness-highlander}
  $\splitcorrectness[\frgx]$ and $\splitcorrectness[\fvsa]$ are \pspace-hard,
  even if the splitter is disjoint and the spanner is proper. \qed
\end{corollary}

 \section{Connection To Language Primality}\label{sec:primality}
In the previous sections we discussed split-correctness and splittability, where
either $\spanner$, $\splitspanner$, and $\splitter$ are given, and the question
is whether $\spanner$ is splittable by $\splitter$ via $\splitspanner$. In the
case of splittability, only $\spanner$ and $\splitter$ are given and the
question is whether there is a spanner $\splitspanner$ such that $\spanner$ is
splittable by \splitter via \splitspanner. One natural question is the case
where only $\spanner$ is given and it is asked whether $\splitter$ and
$\splitspanner$ exist, such that \spanner is splittable by \splitter via
\splitspanner. In general, the answer to this question is yes, as every spanner
is self-splittable by the splitter that only selects the whole document, i.e.
$\splitter = x\{\Sigma^*\}$. We therefore parameterize the decision problem with
a class $\mathcal{S}$ of splitters.

\algproblem{$\mathcal{S}\text{-}\splitexistence[\cC]$}{Spanner $\spanner \in
  \cC$.}{Is there a splitter $\splitter \in \mathcal{S}$ such that $\spanner$ is
  splittable by $\splitter$? }

As we show in Observation~\ref{obs:prepostprimality}, \splitexistence is
strongly connected to a classical problem from Formal Language Theory, which is
called Language Primality. To this end, we define \emph{middle extractors},
which capture $N$-gram extractors or splitters extracting pairs of consecutive
sentences. A splitter \splitter is a \emph{middle extractor}, if $\splitter =
\lang_{1} \cdot x\{\lang_{2}\} \cdot \lang_{3}$, where $\lang_{i} \neq
\{\varepsilon\}$, for $1 \leq i \leq 3$ are regular languages. We denote the
class of middle extractors by $\splitterprepost$. Observe that the splitters
used in the proofs of Lemma~\ref{lem:lowerbounds-det} and
Lemma~\ref{lem:lowerbounds-highlander} are middle extractors and therefore, the
problems are \pspace-hard for middle extractors.

The \prim problem asks, given a language $\lang$, whether $\lang$ is prime,
i.e., whether it cannot be decomposed into two languages $\lang_1$ and $\lang_2$
such that $\lang = \lang_1 \cdot \lang_2$ and $\lang_1 \neq \{\varepsilon\} \neq
\lang_2$. The complexity of \prim has been considered an open problem since
the late 90's (cf. Salomaa~\cite[Problem~2.1]{Salomaa08}). Martens, Niewerth and
Schwentick~\cite{MartensNS10} showed that \prim is \pspace-complete, even if the
language is given as a deterministic finite state automaton. However, to the
best of our knowledge, the complexity of \prim for other representations of the
input remains open. Here, we define the complement of \prim. Furthermore, we add
an additional parameter $k$ specifying into how many languages we want to
decompose the language $\lang$.

\algproblem{\decomp k}{A regular language $\lang$.}{Is
  there a decomposition of $\lang$ into $\lang_1,\ldots,\lang_k$ such that
  $\lang = \lang_1 \cdots \lang_k$, and $\lang_i \neq \{\varepsilon\}$ for all
  $1 \leq i \leq k$?}

Clearly \decomp{2} is the complement of \prim. There is a connection between
\splitexistence and \decomp{3} that is most easily seen in the case of Boolean
spanners:

\begin{observation}\label{obs:prepostprimality}
  Let \lang be a Boolean spanner (i.e., a regular language). Then $\lang$ is
  \decomp{3} if and only if $\lang \in \splitexistenceprepost$.
\end{observation}

Of course, we are not interested in studying Boolean spanners, but the
observation above gives little hope to settle the complexity of
\splitexistenceprepost without settling the complexity of \decomp{3}. We note
that the complexity of \decomp{k} is still open even for deterministic automata
in the case $k>2$.

 \section{The Framework in the Context of the Relational
  Algebra}\label{sec:reasoning}
In a complex pipeline that involves multiple spanners and splitters, it may be
beneficial to reason about the manipulation or replacement of operators for the
sake of query planning (in a similar way as we reason about query plans in a
database system). In this section, we consider questions of this sort. As a
basis for optimizing query plans, we show that the composition of spanners and
splitters is associative (Section~\ref{sec:associative}) and that splittability
as well as self-splittability is transitive (Section~\ref{sec:transitive}).
Furthermore, we give a sufficient condition which for distributivity of spanner
composition over join (Section~\ref{sec:distributejoincomp}). Afterwards we
study the problem of deciding on the splittability in the presence of
\e{black-box} spanners that are known to follow \e{split constraints}
(Section~\ref{sec:split-constrained}). Furthermore, we study the complexity of
deciding whether one splitter subsumes another, that is whether a splitter can
always be executed after another (Section~\ref{sec:subsumption}). We conclude
this section by studying splittability under precondition on the input documents
(Section~\ref{sec:schemaConstraints}). Note that all, but the complexity
results, are independent of the representation of the spanner. That is, they
also hold for non-regular spanners, like core spanners~\cite{FreydenbergerH18}
and context-free spanners~\cite{Peterfreund20}. 

\subsection{Associativity of Composition}\label{sec:associative}
\begin{theorem}\label{thm:associative}
  Given a spanner \spanner and two splitters $\splitter_1$ and $\splitter_2$,
  then it holds that $\spanner \circ (\splitter_1 \circ \splitter_2) = (\spanner
  \circ \splitter_1) \circ \splitter_2$.
\end{theorem}
\begin{proof}
  We use the algebraic characterization from Lemma~\ref{lem:pcircs-algebra} and
  denote the variables of the splitters $\splitter_1$ and $\splitter_2$ by $x_1$
  and $x_2$, respectively.

  \begin{align*}
    (\spanner \circ \splitter_1) \circ \splitter_2 \;
      &\eqnumber 1\; \pi_{\vars(\spanner)} \Big(\big(\Sigma^* \cdot x_2\set{\spanner \circ \splitter_1} \cdot \Sigma^*\big) \join \splitter_2 \Big) \\
      &\eqnumber 2\; \pi_{\vars(\spanner)} \Big(\big(\Sigma^* \cdot x_2\bigset{
         \pi_{\vars(\spanner)} ((\Sigma^* \cdot x_1\set{\spanner} \cdot \Sigma^*) \join \splitter_1)       
      } \cdot \Sigma^*\big) \join \splitter_2 \Big)\\
      &\eqnumber 3\; \pi_{\vars(\spanner)} \Big(\big(\Sigma^* \cdot x_2\bigset{
         (\Sigma^* \cdot x_1\set{\spanner} \cdot \Sigma^*) \join \splitter_1       
        } \cdot \Sigma^*\big) \join \splitter_2 \Big)\\
      &\eqnumber 4\; \pi_{\vars(\spanner)} \bigg(\Big( \big(\Sigma^* \cdot x_2\bigset{(\Sigma^* \cdot x_1\set{\spanner} \cdot \Sigma^*)} \cdot \Sigma^*\big)      \join (\Sigma^* \cdot x_2\set{\splitter_1}
    \cdot \Sigma^*)  \Big) \join \splitter_2 \bigg)\\
      &\eqnumber 5\; \pi_{\vars(\spanner)} \Big(\big(\big(\Sigma^* \cdot x_1 \set{\spanner} \cdot \Sigma^*\big) \join \big(\Sigma^* \cdot x_2\set{\splitter_1} \cdot \Sigma^*\big)\big) \join \splitter_2 \Big) \\
      &\eqnumber 6\; \pi_{\vars(\spanner)} \Big(\big(\Sigma^* \cdot x_1 \set{\spanner} \cdot \Sigma^*\big) \join \pi_{x_1}\big(\big(\Sigma^* \cdot x_2\set{\splitter_1} \cdot \Sigma^*\big) \join \splitter_2\big)\Big) \\
      &\eqnumber 7\; \pi_{\vars(\spanner)} \Big(\big(\Sigma^* \cdot x_1 \set{\spanner} \cdot \Sigma^*\big) \join (\splitter_1 \circ \splitter_2) \Big)\\
      &\eqnumber 8\; \spanner \circ (\splitter_1 \circ \splitter_2)
  \end{align*}

  The equalities~(1),~(2),~(7), and~(8) hold by the algebraic characterization
  of Lemma~\ref{lem:pcircs-algebra}. The equalities~(3) and~(6) hold by the
  definition of projection and join in the relational algebra, i.e., it is
  enough to project only once and the intermediate projections do not have an
  effect, as the variables removed by the projection are not part of the natural
  join. The Equality~(4) follows from Equality~(9) below that, as we will show,
  holds true for all spanners $\spanner_1$ and $\spanner_2$ by using $\spanner_1
  \eqdef (\Sigma^* \cdot x_1 \set{\spanner} \cdot \Sigma^*)$, $\spanner_2 \eqdef
  \splitter_1$, and $x \eqdef x_2$.
  \[
    \Sigma^* \cdot x\set{\spanner_1 \join \spanner_2} \cdot \Sigma^* \;\;\eqnumber 9\;\; \Sigma^* \cdot x \set{\spanner_1} \cdot \Sigma^* \join \Sigma^* \cdot x \set{\spanner_2} \cdot \Sigma^*
  \]
  The Equality~(5) follows from the observation that in the lefthand side
  of the join, the only restriction of $x_2$ is that the span of $x_2$ has to
  cover the span of $x_1$. However, this restriction is already imposed by the
  righthand side of the join, where $x_2$ has to cover the part of the document
  matched by $\splitter_1$ and therefore the span of $x_1$. Therefore, removing
  $x_2$ on the lefthand side of the join does not alter the result.

  \medskip
  
  It remains to show Equality~(9).
  Let $\doc$ be a document and $\tup$ be a tuple such that
  $\tup \in (\Sigma^* \cdot x\set{\spanner_1 \join \spanner_2} \cdot
  \Sigma^*)(\doc)$. Let $s=\tup(x)$ be the span assigned to $x$ and
  $\tup' = \unshiftSpanBy{\tup}{s}$. By definition of
  concatenation and variable enclosing, it holds that
  \[
    \tup' \;\in\; x\set{\spanner_1 \join \spanner_2}(\doc_s)
    \quad\text{and}\quad \projectTup{\tup'}{\vars(\spanner_1 \join \spanner_2)}
    \;\in\; (\spanner_1 \join \spanner_2)(\doc_s)
  \]
  and therefore it holds that $\projectTup{\tup'}{\vars(\spanner_i)} \in
  \spanner_i(\doc_s)$ and $\projectTup{\tup'}{\vars(\spanner_i) \cup \{x\}} \in
  x\set{\spanner_i(\doc_s)}$. We can conclude that
  $\projectTup{\tup}{\vars(\spanner_i) \cup \{x\}} \in (\Sigma^*\cdot
  x\set{\spanner_i(\doc)}\cdot \Sigma^*)(\doc)$, and finally
  \[
    \tup \;\;\in\;\; \big((\Sigma^* \cdot x \set{\spanner_1} \cdot \Sigma^*)
    \join (\Sigma^* \cdot x \set{\spanner_2} \cdot \Sigma^*)\big)(\doc)\;.
  \]

  The other direction can be shown symetrically. Let $\doc$ be a document, $\tup
  \in \big((\Sigma^* \cdot x \set{\spanner_1} \cdot \Sigma^*) \join (\Sigma^*
  \cdot x \set{\spanner_2} \cdot \Sigma^*)\big)(\doc)$ be a tuple and $s =
  \tup(x)$. Then $\projectTup{\tup}{\vars(\spanner_i) \cup \{x\}} \in (\Sigma^*
  \cdot x\set{\spanner_i} \cdot \Sigma^*)(\doc)$ and therefore
  $\projectTup{(\unshiftSpanBy{\tup}{s}}{\vars(\spanner_i)\cup\{x\}} \in x
  \set{\spanner_i}$. We can conclude that $\unshiftSpanBy{\tup}{s} \in x
  \set{\spanner_1 \join \spanner_2}$ and therefore $\tup \in \Sigma^* \cdot
  x\set{\spanner_1 \join \spanner_2} \cdot \Sigma^*$, which concludes the proof
  of~(9) and the theorem.
\end{proof}

\subsection{Transitivity}\label{sec:transitive}
The fact that spanner decompotion is associative allows us to show that
splittability and self-splittability are \emph{transitive}.

\begin{theorem}\label{thm:transitive}
  Let $\spanner$ be a document spanner and $\splitter_1$ and $\splitter_2$ be
  document splitters such that $\spanner$ is splittable by $\splitter_1$ and
  $\splitter_1$ is splittable by $\splitter_2$, then $\spanner$ is splittable by
  $\splitter_2$. If furthermore $\spanner$ is self-splittable by $\splitter_1$
  and $\splitter_1$ is self-splittable by $\splitter_2$ then $\spanner$ is
  self-splittable by $\splitter_2$.
\end{theorem}
\begin{proof}
  Assume that $\spanner$ is splittable by $\splitter_1$ and $\splitter_1$ is
  splittable by $\splitter_2$, then there is a spanner $\spanner'$
  such that $\spanner = \spanner' \circ \splitter_1$. Furthermore,
  there is a splitter $\splitter'$ such that $\splitter_1 = \splitter' \circ
  \splitter_2$.
  As the composition of document spanners is associative, we can conclude that
  \[
    \spanner \;\;=\;\; \spanner' \circ \big(\splitter' \circ
    \splitter_2\big) \;\;=\;\; \big(\spanner' \circ \splitter'\big) \circ \splitter_2\;.
  \]
  Therefore $\spanner$ is splittable by $\splitter_2$ via $\spanner' \circ
  \splitter'$.

  \medskip

  Let now $\spanner$ be self-splittable by $\splitter_1$ and $\splitter_1$ be
  self-splittable by $\splitter_2$. Then we have $\spanner'=\spanner$ and
  $\splitter'=\splitter$ in the equation above and using $\spanner = \spanner
  \circ \splitter_1 = \spanner' \circ \splitter'$ we can conclude that $\spanner
  = \spanner \circ \splitter_2$, which shows that \spanner is self-splittable by
  $\splitter_2$.
\end{proof}

\subsection{Distributivity of Composition and
  Join}\label{sec:distributejoincomp}
Another important question is whether applying a splitter commutes with other
operations of the algebra, especially the join operation.
We now give a sufficient precondition such that 
\[ (\spanner_1 \join \spanner_2) \circ \splitter \;\;=\;\; (\spanner_1 \circ
  \splitter) \join (\spanner_2 \circ \splitter)\;.\] The problem is that the two
spans on the righthand side of the equation could be different. If they are, the
equation needs not to be true, though it is still possible in some corner cases.
An obvious idea is to require that $\spanner_1 \circ \splitter$ and $\spanner_2
\circ \splitter$ satisfy the highlander condition. However, as we show in
Example~\ref{ex:noDistributivity}, this might not be enough, as it is possible
that there are two overlapping spans covering tuples from $\spanner_1$ and
$\spanner_2$, respectively, such that $x$ is in the intersection of both spans.
Even requiring that the spanners are proper and the splitter is disjoint might
not be enough if $x$ is assigned the empty span. This explains the rather
complicated precondition of the following theorem.

\begin{example}\label{ex:noDistributivity}
  Let $\spanner_1 \eqdef \Sigma^* \cdot x_1\{ a \} \cdot x_2\{ b \} \cdot \Sigma^*$,
  $\spanner_2 \eqdef \Sigma^* \cdot x_2\{ b \} \cdot x_3\{ a \} \cdot \Sigma^*$, and  
  $\splitter \eqdef \Sigma^* \cdot x\{\Sigma \cdot \Sigma\} \cdot\Sigma^*$.
  We observe that $\spanner_1$ (resp., $\spanner_2$) and $\splitter$ satisfy the
  highlander condition. 

  Let $\spanner \eqdef \spanner_1 \join \spanner_2$ be the join of both spanners
  and let $\doc = aba$. It follows that $\splitter(\doc) =
  \set{\spanFromTo{1}{3}, \spanFromTo{2}{4}}$ and $\spanner(\doc) = \{\tup\},$
  where $\tup(x_1) = \spanFromTo{1}{2}$, $\tup(x_2) = \spanFromTo{2}{3}$, and
  $\tup(x_3) = \spanFromTo{3}{4}$. As there is not split $s \in \splitter(\doc)$
  that covers $\tup \in \spanner(\doc)$ it follows directly from
  Observation~\ref{obs:splittabilityCover} that $\spanner$ is not splittable by
  \splitter and therefore $\spanner \circ \splitter \neq \spanner$. However,
  both spanners, $\spanner_1$ and $\spanner_2$, are self-splittable by \splitter
  which implies that $(\spanner_1 \circ \splitter) \join
  (\spanner_2 \circ \splitter) = \spanner_1 \join \spanner_2 = \spanner$. It
  follows directly that
  \[\pushQED{\qed}
    (\spanner_1 \join \spanner_2) \circ \splitter \;\;=\;\; \spanner \circ \splitter
    \;\;\neq\;\; \spanner \;\;=\;\; \spanner_1 \join \spanner_2 \;\;=\;\; (\spanner_1 \circ \splitter)
    \join (\spanner_2 \circ \splitter).\qedhere\popQED
  \]
\end{example}

\begin{theorem}\label{thm:distributivity}
  Let $\splitter$ be a disjoint document splitter and $\spanner_1$ and
  $\spanner_2$ be a document spanners such that
  $X \eqdef \vars(\spanner_1) \cap \vars(\spanner_2) \neq \emptyset$ and the
  spanner $\pi_X(\spanner_1) \join \pi_X(\spanner_2)$ is proper. Then
  \[
    (\spanner_1 \join \spanner_2) \circ \splitter \;\;=\;\; (\spanner_1 \circ
    \splitter) \join (\spanner_2 \circ \splitter)\;.
  \]
\end{theorem}
\begin{proof}
  Let $\doc$ be a document and $\tup$ be a tuple such that $\tup \in
  \big((\spanner_1 \join \spanner_2) \circ \splitter\big)(\doc)$. Then there is
  a decomposition $\doc=\doc_1 \cdot \doc_2 \cdot \doc_3$ such that
  $s=\spanFromTo{|\doc_1|+1}{|\doc_1 \cdot \doc_2|+1} \in \splitter(\doc)$, and
  $\unshiftSpanBy{\tup}{|\doc_1|} \in (\spanner_1 \join \spanner_2)(\doc_2)$. We
  can conclude that $\projectTup{\tup'}{\vars(\spanner_i)} \in \spanner_i(\doc_2)$,
  therefore $\projectTup{\tup}{\vars(\spanner_i)} \in (\spanner_i \circ
  \splitter)(\doc)$, and finally $\tup \in \big((\spanner_1 \circ \splitter)
  \join (\spanner_2 \circ \splitter)\big)(\doc)$.

  For the other direction let $\doc$ be a document and $\tup$ be a tuple such
  that $\tup \in \big((\spanner_1 \circ \splitter) \join (\spanner_2 \circ
  \splitter)\big)(\doc)$. For $1 \leq i \leq 2$, it must hold that
  $\projectTup{\tup}{\vars(\spanner_i)} \in (\spanner_i \circ \splitter)(\doc)$.
  Thus there are spans $s_1$ and $s_2$, such that
  $\unshiftSpanBy{\projectTup{\tup}{\vars(\spanner_i)}}{s_i} \in
  \spanner_i(\doc_{s_i})$. Due to $\pi_X(\spanner_1) \join \pi_X(\spanner_2)$
  being proper, the minimal span $s$ that covers $\projectTup{\tup}{X}$ is not
  empty. As furthermore $s$ is covered by both $s_1$ and $s_2$ and $\splitter$
  is disjoint, we can conclude that $s_1=s_2$. Therefore, we have that
  $\unshiftSpanBy{\tup}{s_1} \in (\spanner_1 \join \spanner_2)(\doc_{s_1})$ and
  finally $\tup \in \big((\spanner_1 \join \spanner_2) \circ
  \splitter\big)(\doc)$, concluding the proof.
\end{proof}

Note that in the previous theorem it is sufficient if either the spanner
$\pi_X(\spanner_1)$ or the spanner $\pi_X(\spanner_2)$ is proper.

\subsection{Split-Constrained Black Boxes}\label{sec:split-constrained}

We begin with motivating examples.

\begin{example}\label{example:blackbox1}
  In this example and the next, we'll denote by $\spanner(x,y)$ that spanner
  $\spanner$ uses the variables $x$ and $y$. Consider the spanner $\spanner$
  that seeks to extract adjectives for Galaxy phones from reports. We define
  this spanner by joining three spanners:
  
  The spanner $\spanner_1(x,y)$ is given by the regex formula
  \[
    \Sigma^*\cdot x\{\textsf{Galaxy [A-Z]$\setminus$d}^*\}\cdot\Sigma^*\cdot
    y\{\Sigma^*\}\cdot\Sigma^*
  \]
  \noindent that extracts mentions of Galaxy brands (e.g., Galaxy A6 and Galaxy
  S8) followed by substrings $y$ that occur right before a period.

  The spanner $\spanner_2(x,x')$ is a coreference resolver (e.g., the \e{sieve}
  algorithm~\cite{DBLP:conf/emnlp/RaghunathanLRCSJM10}) that finds spans $x'$
  that coreference spans $x$.
  The spanner $\spanner_3(x',y)$ finds pairs of noun phrases $x'$ and attached
  adjectives $y$ (e.g., based on a Recursive Neural
  Network~\cite{DBLP:conf/icml/SocherLNM11}).
  
  For example, consider the review \e{``I am happy with my Galaxy A6. It is
    stable.''} Here, in one particular match, $x$ will match (the span of)
  \e{Galaxy A6}, $x'$ will match \e{it} (which is an anaphora for \e{Galaxy
    A6}), and $y$ will match \e{stable}. (Other matches are possible too.)

  How should a system find an efficient query plan to this join on a long
  report? Naively materializing each relation might be too costly:
  $\spanner_1(x,y)$ may produce too many matches, and $\spanner_2(x,x')$ and
  $\spanner_3(x',y)$ may be computationally costly. Nevertheless, we may have
  the information that $\spanner_2$ is splittable by paragraphs and that
  $\spanner_3$ is splittable by sentences (hence, by paragraphs). This
  information suffices to determine that the entire join $\spanner_1(x,y)\join
  \spanner_2(x,x') \join \spanner_3(x',y)$ is splittable, hence parallelizable,
  by paragraphs. \qed
\end{example}

\begin{example}\label{example:blackbox2}
  Now consider the spanner that joins two spanners: $\spanner(x)$ extracts
  spans $x$ followed by the phrase \e{``is kind''} (e.g., \e{``Barack Obama is
    kind''}). The spanner $\spanner'(x)$ extracts all spans $x$ that match
  person names. Clearly, the spanner $\spanner(x)$ does not split by a natural
  splitter, since it includes, for instance, the entire prefix of the document
  before \e{``is kind''}. However, by knowing that $\spanner'(x)$ splits by
  sentences, we know that the join $\spanner(x)\join P'(x)$ splits by sentences.
  Moreover, by knowing that $\spanner'(x)$ splits by 3-grams, we can infer that
  $\spanner(x)\join \spanner'(x)$ splits by 5-grams. Here, again, the holistic
  analysis of the join infers splittability in cases where intermediate spanners
  are not splittable. \qed
\end{example}
  
We now formalize the splittability question that the examples give rise to. A
\e{spanner signature} $\scs$ is a collection
$\set{\bbspanner_1,\dots,\bbspanner_k}$ of \e{spanner symbols}, where each
$\bbspanner_i$ is associated with a set $\dom(\bbspanner_i)$ of span variables.
Furthermore, let $X_i \eqdef \dom(\bbspanner_i) \cap \big( \bigcup_{i < j \leq
  k} \dom(\bbspanner_j))$. We assume that $X_i \neq \emptyset$, for all $1 \leq
i \leq k$. An \e{instance} $I$ of $\scs$ associates with each spanner symbol
$\bbspanner_i$ an actual spanner $\spanner_i$ such that
$\dom(\spanner_i)=\dom(\bbspanner_i)$ and $\pi_{X_i}(\spanner_i)$ is proper. In
Example~\ref{example:blackbox1}, $\bbspanner_1$ would correspond to the
regex-formula $\spanner_1$, with $\dom(\bbspanner_1) = \{x,y\}$. Furthermore,
$\bbspanner_2$ and $\bbspanner_3$ would correspond to the name of a coreference
resolver $\spanner_2$ and an adjective extractor $\spanner_3$, respectively,
with $\dom(\bbspanner_1)=\set{x,x'}$ and $\dom(\bbspanner_2)=\set{x',y}$.

Let $\scs$ be a spanner signature and $I$ an instance of $\scs$. We denote by
$I_{\join}$ the spanner that is given by
\[
  I_{\join} \;\;\eqdef\;\; \spanner_1\join\dots\join \spanner_k\;.
\]
\noindent We note that this is well-defined due to the associativity and
commutativity of the $\join$-operator.

A \e{split constraint} over a spanner signature $\scs$ is an expression of the
form ``$\bbspanner_i$ is self-splittable by the splitter $\splitter$,'' which we
denote by $\bbspanner_i\sqsubseteq \splitter$. An instance $I$ of $\scs$
\e{satisfies} a set $C$ of split constraints, denoted $I\models C$, if for every
constraint $\bbspanner_i\sqsubseteq \splitter$ in $C$ it is the case that $P_i$
is self-splittable by $\splitter$. The problem of \e{split-correctness with
  black boxes} is the following:

\algproblem{Black Box Splittability}{A spanner signature
  $\scs$, a set $C$ of split constraints, and a splitter $\splitter$.}{Is
  $I_{\join}$ self-splittable by $\splitter$ whenever $I$ is an instance of
  $\scs$ such that $I\models C$?} 

A natural question to ask is the following. Assume that all spanners are
self-splittable by the same splitter $\splitter$, that is $\bbspanner
\sqsubseteq \splitter$, for every $\bbspanner \in \scs$. Does this imply that
$I_{\join}$ is self-splittable by $\splitter$? In general, the answer to this
question is no, as shown by the spanners and splitter defined in
Example~\ref{ex:noDistributivity}. The next result shows that in the presence of
disjoint splitters the join operator preserves self-splittability.

\begin{theorem}\label{theorem:blackbox}
  Let $\splitter$ be a disjoint splitter, let $\scs$ be a spanner signature, and
  let $C$ be a set of split constraints, such that $\bbspanner_i \sqsubseteq
  \splitter \in C$, for all $1 \leq i \leq k$. Then $I_{\join}$ is self-splittable
  by $\splitter$ if $I\models C$.
\end{theorem}
\begin{proof}
  Let $I$ be an instance of $\scs$, such that $I \models \scs$ and let $P_i$ be
  the spanner interpreting $\bbspanner_i$. We have to show, that $I_{\join} =
  I_{\join} \circ \splitter$.

  Recall that per definition of $\scs$, $X_i = \dom(\bbspanner_i) \cap \big(
  \bigcup_{i < j \leq k} \dom(\bbspanner_j)\big)$, and $X_i \neq \emptyset$, for all
  $1 \leq i \leq k$. Furthermore, per definition of $I$, $\pi_{X_i}(\spanner_i)$
  is proper and $\spanner_i$ is self-splittable by \splitter, for all $1 \leq i
  \leq k$. Thus, using associativity of $\join$ and
  Theorem~\ref{thm:distributivity}, it follows that
  \begin{align*}
    I_{\join} \circ \splitter \;\;
    &=\;\; \big( \spanner_1 \join \dots \join \spanner_k \big) \circ \splitter \\
    &=\;\; \Big( \spanner_1 \join \big(\spanner_2 \join (\dots \join \spanner_k) \big) \Big) \circ \splitter \\
    &=\;\; (\spanner_1 \circ \splitter) \join
      \Big(\big(\spanner_2 \join (\spanner_3 \join( \dots \join \spanner_k)) \big) \circ \splitter\Big) \\ &\mathrel{\makebox[\widthof{=}]{\raisebox{0pt}[0pt][0pt]{\vdots}}} \\
    &=\;\; (\spanner_1 \circ \splitter) \join \dots \join (\spanner_k \circ \splitter) \\
    &=\;\; \spanner_1 \join \dots \join \spanner_k \\
    &=\;\; I_{\join}\; .
  \end{align*}
  This concludes the proof.
\end{proof}
Observe that the requirement that $\pi_{X_i}(\spanner_i)$ is proper is always
satisfied if $\spanner_i$ does not assign the empty span to variables and it
holds, for every document $\doc \in \docs$ and every tuple $\tup \in
\spanner_i(\doc)$, that $X_i \subseteq \dom(\tup)$.

\subsection{Subsumption}\label{sec:subsumption}
Another form of optimization is \emph{subsumption}. A splitter $\splitter$
\emph{subsumes} a splitter $\splitter'$ if $\splitter = \splitter' \circ
\splitter$.

\begin{theorem}
  Let $A_\splitter$, $A_{\splitter'}$, all coming from the class $\cC$.
  Then deciding if $\toSpanner{A_\splitter}$ subsumes $\toSpanner{A_{\splitter'}}$
  is \pspace-hard if $\cC = \frgx$ and in \pspace if $\cC = \vsa$.
\end{theorem}
\begin{proof}
  The upper bound follows from
  Proposition~\ref{prop:pcircs-algebra-construction} and
  Corollary~\ref{cor:spannerContainment}. The lower bound follows by reduction
  from regular expression universality. Let $E$ be a regular expression and let
  $\splitter' \eqdef x\{E\}$. We show that $\splitter = x\set{\Sigma^*}$
  subsumes $\splitter'$ if and only if $\lang(E) = \Sigma^*$. To this end,
  assume that $\lang(E) = \Sigma^*$. Thus, $\splitter = \splitter'$ and due to
  idempotency of $\splitter$, $\splitter = \splitter' \circ \splitter$. For the
  other direction, assume that $\lang(E) \neq \Sigma^*$. Thus, there is a
  document $\doc \in \docs$ with $\doc \notin \lang(E)$. Therefore
  $\splitter'(\doc) = \emptyset$, but $\splitter(\doc) =
  \spanFromTo{1}{|\doc|+1}$. Thus, $\splitter(\doc) \neq (\splitter' \circ
  \splitter)(\doc)$ concluding that $\splitter$ is does not subsume
  $\splitter'$.
\end{proof}

\begin{corollary}\label{cor:subsume}
  Let $A_\splitter$, $A_{\splitter'}$ all coming from the class $\cC$. Then
  deciding whether $\toSpanner{A_\splitter}$ subsumes
  $\toSpanner{A_{\splitter'}}$ is \pspace-complete if $\cC \in \cCg \setminus \cCt$.
\end{corollary}

\begin{theorem}
  Deciding whether $\splitter$ subsumes $\splitter'$ is in \ptime if \splitter
  and $\splitter'$ are given as \usvsa, $\splitter$ is disjoint, and
  $\splitter'$ is proper.
\end{theorem}
\begin{proof}
  The upper bound is immediate from
  Proposition~\ref{prop:pcircs-algebra-construction} and
  Theorem~\ref{thm:spannerContainment-det}.
\end{proof}

\begin{corollary}
  Let $\splitter$, $\splitter'$, all coming from the class $\cC$, let
  $\splitter$ be disjoint and $\splitter'$ be proper. Then deciding whether
  $\splitter$ subsumes $\splitter'$ is in \ptime if $\cC \in \cCt$.
\end{corollary}

\subsection{Schema Constraints}\label{sec:schemaConstraints}
Sometimes a spanner is not splittable by a given splitter, because of a reason
that seems marginal. For instance, the spanner may first check that the document
conforms to some standard format, such as Unicode, UTF-8, CSV, HTML, etc. This
is no issue, if the document collection is verified to conform to the standard
prior to splitting. In this section, we will introduce \emph{schema
  constraints}, which extend the general framework in order to embark this.

A \emph{schema constraint} $\lang$ is a---not necessary regular---language.
We say that two spanners $\spanner, \spanner'$ are equivalent under a
schema constraint $\lang$ if and only if for all documents $\doc \in \lang$ it
holds that $\spanner(\doc) = \spanner'(\doc)$. We denote this by
$\sconstL{\spanner}{\spanner'}$. We say that \spanner is splittable by \splitter
via \splitspanner under the schema constraint $\lang$ if and only if
$\sconstL{\spanner}{\splitspanner \circ \splitter}$. A schema constraint $\lang$
is regular, if $\lang$ is regular.

\begin{lemma}\label{lem:schemaConstraint}
  Let \spanner, \splitspanner be spanners, \splitter be a splitter and $\lang$ be a
  schema constraint. Then $\sconstL{\spanner}{\splitspanner \circ \splitter}$ if
  and only if $\spanner \join \lang = \big(\splitspanner \circ
  (\splitter\join\lang)\big)$.
\end{lemma}
\begin{proof}
  Per definition of $\sconstL{\;}{\;}$, it holds that
  $\sconstL{\spanner}{\splitspanner \circ \splitter}$ if and only if $\spanner
  \join \lang = (\splitspanner \circ \splitter) \join \lang$. Therefore, we have
  to show that $(\splitspanner \circ \splitter) \join \lang = \splitspanner
  \circ (\splitter \join \lang)$.
  \begin{align*}
    (\splitspanner \circ \splitter) \join \lang \;\;
    &\eqnumber 1\;\; \pi_{\vars(\splitspanner)} \big((\Sigma^* \cdot x\set{\splitspanner} \cdot \Sigma^*) \join \splitter \big) \join \lang \\
    &\eqnumber 2\;\; \pi_{\vars(\splitspanner)} \Big(\big((\Sigma^* \cdot x\set{\splitspanner} \cdot \Sigma^*) \join \splitter\big) \join \lang \Big)\\
    &\eqnumber 3\;\; \pi_{\vars(\splitspanner)} \big((\Sigma^* \cdot x\set{\splitspanner} \cdot \Sigma^*) \join (\splitter \join \lang) \big)\\
    &\eqnumber 4\;\; \splitspanner \circ (\splitter \join \lang)
  \end{align*}
  The equalities~(1) and~(4) are by the algebraic characterization of
  Lemma~\ref{lem:pcircs-algebra}. The Equality~(2) is by the fact that $\lang$
  does not use any variables and we are therefore allowed to change the order of
  projection and join. Finally, the Equality~(3) holds because of the
  associativity of joins.
\end{proof}

It follows directly from Lemma~\ref{lem:schemaConstraint} that schema
constraints do not extend the expressivity of the general framework. Using
a slightly modified product construction, we can extend the complexity results
for \splitcorrectness, \selfsplittability, and \splittability (cf.
Theorems~\ref{thm:splitcorrectness},~\ref{thm:mainSplittability}) to also hold in
the presence of regular schema constraints.

\begin{lemma}\label{lem:schemaConstraintConstruction}
  Given \vset-automata $A_\spanner$ and $A_\lang$ representing a spanner
  $\spanner$ and a regular schema constraint $\lang$, respectively, a \vset-automaton $A$ can be constructed in polynomial time, such that
  \begin{enumerate}
  \item $\toSpanner{A} = \toSpanner{A_\spanner} \join \lang$;
  \item $A \in \svsa$ if $A_\spanner \in \svsa$; and
  \item $A \in \uvsa$ if $A_\spanner,A_\lang \in \uvsa$.
  \end{enumerate}
\end{lemma}
\begin{proof}
  Let $A_\spanner = (\Sigma, V, Q_\spanner, q_{0,\spanner}, Q_{F,\spanner},
  \delta_\spanner) \in \cC$ and $A_\lang = (\Sigma, \emptyset, Q_\lang,
  q_{0,\lang}, Q_{F,\lang}, \delta_\lang) \in \cC$ be as given. We define the
  automaton $A \eqdef (\Sigma, V, Q, q_{0}, Q_{F}, \delta)$, where $Q \eqdef
  Q_\spanner \times Q_\lang$, $q_0 \eqdef (q_\spanner,q_\lang)$, $Q_F \eqdef
  Q_{F,\spanner} \times Q_{F,\lang}$, and
  \[
    \begin{array}{r@{\quad}l@{\;}l@{\;\;}}
      \delta \quad\eqdef
      & \big\{\big((q_\spanner,q_\lang),\sigma,(q'_\spanner,q'_\lang)\big)
      & \mid \sigma \in \Sigma \cup \{\varepsilon\},
        (q_\spanner,\sigma,q'_\spanner) \in \delta_\spanner,
        (q_\lang,\sigma,q'_\lang) \in \delta_\lang \big\} \;\;\cup\\[.2ex]
      & \big\{\big((q_\spanner,q_\lang),v,(q'_\spanner,q_\lang)\big)
      & \mid v \in \varop{V}, (q_\spanner,v,q'_\spanner) \in \delta_\spanner,
        q_\lang \in Q_\lang \big\}\;.
    \end{array}
  \]
  The only difference to the usual product construction is, that transitions
  related to variable operations are only processed by $A_\spanner$ and ignored
  by $A_\lang$. It is easy to see that $A \in \vsa$ can be constructed in polynomial time.
  Furthermore, $\reflang(A) = \reflang(A_\spanner) \cap \{\refWord \mid
  \clr(\refWord) \in \lang\}$. Therefore it must hold that \[\toSpanner{A} \;\;=\;\;
  \toSpanner{\reflang(A)} \;\;=\;\; \reflang(A_\spanner) \cap \{\refWord \mid
  \clr(\refWord) \in \lang\} \;\;=\;\; \toSpanner{A_\spanner} \join \lang\;,\] concluding
  the proof of statements~(1) and~(2).
  
  It only remains to show that $A \in \uvsa$ if $A_\spanner,A_\lang \in \uvsa$.
  To this end, assume that $A$ is not unambiguous. As observed before,
  $\reflang(A) = \reflang(A_\spanner) \cap \{\refWord \mid \clr(\refWord) \in
  \lang\}$ and therefore $\reflang(A) \subseteq \reflang(A_\spanner)$. Thus, due
  to $A_\spanner \in \uvsa$, $A$ must satisfy the variable order condition.
  Assume there are two distinct runs of $A$ that violate unambiguity
  condition~\ref{cond:unambig}. Due to $A_\spanner \in \uvsa$, both runs must
  coincide in the $A_\spanner$ component of $A$. However, by the same argument,
  both runs must also coincide in the $A_\lang$ component of $A$, leading to the
  desired contradiction. This concludes the proof.
\end{proof}

Due to Lemmas~\ref{lem:schemaConstraint}
and~\ref{lem:schemaConstraintConstruction}, the complexity results for
\splitcorrectness, \selfsplittability, and \splittability (cf.
Theorems~\ref{thm:splitcorrectness},~\ref{thm:mainSplittability}) also hold in
the presence of schema constraints. Note that this also includes the \ptime
fragment if the schema constraint $\lang \in \cC$ is represented by a class of
document spanners $\cC \in \cCt$.

Schema constraints also give rise to other problems that can be studied. For
instance, it may be the case that we already have a spanner and splitter
available that we do not want to change, but we want to know whether there
exists a schema constraint $\lang$ such that the spanner is splittable by the
splitter under the schema constraint. In general, the answer to this is always
positive, splittability holds for any combination of a spanner and a splitter
under the schema constraint $\lang = \emptyset$. Therefore, we say that \emph{a
  schema constraint $\lang$ covers \spanner} if and only if the splitter
$\splitter_\lang \eqdef x\{\lang\}$ covers $\spanner$.

Next we observe that, for each spanner \spanner, there is a minimal schema
constraint $\lang_\spanner \eqdef \{\doc \mid \spanner(\doc) \neq \emptyset\}$
such that split-correctness holds under $\lang_\spanner$ if it holds for any
schema constraint which covers the spanner. We first observe that
$\lang_\spanner$ is indeed contained in any schema condition which covers
$\spanner$.
\begin{observation}\label{obs:minimal-schema-constraint}
  Let $\spanner$ be a spanner and let $\lang$ be a schema constraint which
covers $\spanner$. Then, $\splitter_\lang = x\{\lang\}$ covers $\spanner$ and
  therefore, $\lang_\spanner \subseteq \lang$.
\end{observation}

The following observation follows directly from
Observation~\ref{obs:minimal-schema-constraint} and
Lemma~\ref{lem:schemaConstraint}.

\begin{observation}\label{obs:spittcorrectness-under-minimal-schema-constraint}
  Let $\spanner$ and $\splitspanner$ be spanners, $\splitter$ be a splitter, and
  $\lang$ be a schema constraint which covers $\spanner$. Then
  $\sconstL{\spanner}{\splitspanner \circ \splitter}$ implies that
$\sconst{\spanner}{\splitspanner \circ \splitter}{\lang_\spanner}$.
\end{observation}

As we show next, given a sequential \vset-automaton $A_\spanner \svsa$, a
\vset-automaton $A$ that represents the minimal schema constraint can be
constructed in polynomial time.

\begin{lemma}\label{lem:minimalSchemaConstraintConstruction}
  Let $A_\spanner \in \svsa$ be a sequential \vset-automaton. Then an automaton
  $A\in\svsa$ with $\toSpanner{A} = \pi_\emptyset\toSpanner{A_\spanner}$ can be
  constructed in polynomial time.
\end{lemma}
\begin{proof}
  Let $A_\spanner = (\Sigma, V, Q_\spanner, q_{0,\spanner}, Q_{F,\spanner},
  \delta_\spanner)$. We define $A \eqdef (\Sigma, \emptyset, Q_\spanner, 
  q_{0,\spanner}, Q_{F,\spanner}, \delta)$, where
  \[
    \begin{array}{r@{\quad}l@{\;}l@{\;\;}}
      \delta \quad\eqdef
      & \big\{(p,\sigma,q)
      & \mid \sigma \in \Sigma \cup \{\varepsilon\}, 
        (p,\sigma,q) \in \delta_\spanner \big\} \;\;\cup\\[.2ex]
      & \big\{(p,\varepsilon,q)
      & \mid (p,v,q) \in \delta_\spanner, v \in \varop{V} \big\}.
    \end{array}
  \]
  Observe that $A \in \svsa$ can be constructed in polynomial time. Furthermore,
  due to the assumption that $A_\spanner$ is sequential, it follows that there
  is a tuple $\tup \in \toSpanner{A_\spanner}(\doc)$ if and only if $\emptytup
  \in \toSpanner{A}$. Therefore, it must hold that $\toSpanner{A} =
  \pi_\emptyset\toSpanner{A_\spanner}$.
\end{proof}   

Due to Observation~\ref{obs:minimal-schema-constraint} and
Observation~\ref{obs:spittcorrectness-under-minimal-schema-constraint}, we can
decide whether there exists a schema constraint $\lang$ which covers $\spanner$
such that $\sconstL{\spanner}{\splitspanner \circ \splitter}$ by checking
whether $\sconst{\spanner}{\splitspanner \circ
  \splitter}{\reflang(\pi_\emptyset\spanner)}$. Furthermore, it follows directly
from Lemma~\ref{lem:minimalSchemaConstraintConstruction} that
$\reflang(\pi_\emptyset\spanner)$ can indeed be constructed in polynomial time.
However, given an unambiguous (resp., deterministic) and sequential
\vset-automaton, one can not guarantee that the automaton $A$, as constructed
in Lemma~\ref{lem:minimalSchemaConstraintConstruction}, is unambiguous. Thus,
all but the \ptime complexity result for \splitcorrectness, \selfsplittability,
and \splittability (cf.
Theorems~\ref{thm:splitcorrectness},~\ref{thm:mainSplittability}) also hold if
one asks whether there exists a schema constraints $\lang$ which covers
$\spanner$, such that $\sconstL{\spanner}{\splitspanner \circ \splitter}$.

 \section{Concluding Remarks}\label{sec:conclusions}
We embarked on an exploration of the task of automating the
distribution of information-extraction programs across
splitters. Adopting the formalism of document spanners and the concept
of parallel-correctness, our framework focuses on two computational
problems, \splitcorrectness and \splittability, as well as their
special case of \selfsplittability. We presented an analysis of these
problems and studied their complexity within the class of regular
spanners. We have also discussed several natural extensions of the
framework, considering the reasoning about splittability, schema
constraints, and black-box spanners with split constraints. Our
principal objective is to open up new directions for research within
the framework, and indeed, several open problems are left for future
investigation. We discuss these problems in the remainder of this section.

One open problem is the exact complexity of \splittability, as we do not have
matching upper- and lower-bounds in the general case. The complexity is also
open if the input is restricted to unambiguous sequential \vset-automata and the
highlander condition holds.

We know more about \splitcorrectness and \selfsplittability, but there are some
basic open problems there as well. For instance, when considering more expressive
languages for spanners (e.g., the class of \e{core}
spanners~\cite{FaginKRV15-jacm,FreydenbergerH18} that allow for string
equalities or context-free spanners~\cite{Peterfreund20}), all problems reopen.

A variant of \splittability that we barely touched upon is that of deciding,
given a spanner \spanner, whether it can be decomposed in a nontrivial way. We
showed (Observation~\ref{obs:prepostprimality}) that this variant closely
relates to the \prim problem---can a given regular language
be decomposed as the concatenation of non-trivial regular languages?
Interestingly, Martens et al.~\cite{MartensNS10} showed that \prim
is also related to the work of Abiteboul et al.~\cite{AbiteboulGM-jcss11} on
typing in distributed XML, which is quite reminiscent, yet different from, our
work.

For the extensions of reasoning about splitters, and deciding on splittability
with black-box spanners, we barely scratched the surface. Specifically, we
believe that reasoning about split constraints over black-box extractors can
have a profound implication on the usability of IE systems to developers at
varying degrees of expertise, while embracing the advances of the Machine
Leaning and Natural Language Processing communities on learning complex
functions such as artificial neural networks.

\bibliographystyle{abbrv}

\providecommand\noopsort[1]{}

\newpage
\appendix
\section{Proof of
  Lemma~\ref{lem:monoidConstruction}}\label{sec:generalMonoid}

\monoidConstructionLemma*

We start by giving some intuition about the proof idea. To this end, let $A =
(\Sigma, V, Q, q_0, Q_F, \delta)\in \vsa$. We define the monoid $M_V$ that can
test whether a ref-word, using variables from $V$, satisfies the variable order
condition:
\begin{align*}
  M_V \;\eqdef{}&\; \Big(2^{\varop{V}} \cup \{0\},\mop_V,\emptyset\Big) \\
  X \mop_V Y \;\eqdef{}&
  \begin{cases}
    X \cup Y & \text{if } X \cap Y = \emptyset \text{ and } x \prec y \text{ for
      all } x \in X, y \in Y \\
    0 & \text{otherwise}
  \end{cases}
\end{align*}
Building up on $M_V$, we define $M_A^\prec$ as
\[
  M_A^\prec \;\eqdef\; \Big(M_V \cup (M_V \times M_A \times M_V),\;
  \mop_A^\prec\;, \emptyset\Big)\;.
\]
The intuitive idea behind our construction is that we use the monoid $M_V$
to process substring consisting entirely of variable operations. The monoid
$M_V$ conveniently already checks that the variable operation occur in the
correct order and we can derive the whole set of processed variable operations
from the monoid element obtained after processing a substring of variable
operations. In fact, if the operations contain no duplicates and are in the
correct order, the monoid element is the desired set. Otherwise it is 0 to
denote that the processed ref-word is invalid.

Monoid elements $m$ from $M_A^\prec$ that are from $M_V$ correspond to
substrings containing only variable operations. Monoid elements of the form
$m=(m_{v_1},m_a,m_{v_2})$ correspond to a substring containing variable
operations and symbols. Here $m_{v_1}$ and $m_{v_2}$ correspond to the variable
operations before the first and after the last symbol from $\Sigma$,
respectively, while $m_a$ corresponds to possible runs of the automaton for the
substring $\refWordPrime$ from the first to the last $\Sigma$-symbol. However,
we cannot simply compute $h_A(\refWordPrime)$, as we also have to consider runs
of the automaton that process the variable operations that occur inside
$\refWordPrime$ in a different order.

At some point we need to connect monoid elements from $M_V$ with monoid elements
from $M_A$. We therefore define a function $f \colon M_V \to M_A$ that, given
some $m_v \in M_V$, computes all possible runs in $A$ that use exactly the
variable operations encoded by $m_v$.

We give the formal proof now.
\begin{proof}
  Let $A = (\Sigma, V, Q, q_0, Q_F, \delta)\in \vsa$. We define $M_A^\prec$ as
  \[
    M_A^\prec \;\eqdef\; \Big(M_V \cup (M_V \times M_A \times M_V),\;
    \mop_A^\prec,\; \emptyset\Big)\;.
  \]
  It is obvious that $M_A^\prec$ can be constructed with polynomial space in
  $|A|$, as $M_A$ and $M_V$ can be constructed with polynomial space in $|A|$.
  Therefore, $M_A^\prec$ is of exponential size in $|A|$. First, we define for
  any subset $\Gamma$ of $\varop{V}$ the language $R^{\Gamma} \subseteq
  \Gamma^{|\Gamma|}$ as the language containing all strings $v_1 \cdots
  v_{|\Gamma|}$ of variable operations such that each variable operation in $V$
  occurs exactly once and $i < j$ implies that for no variable $x$ it holds that
  $v_i = \vcl{x}$ and $v_j = \vop{x}$. With other words, $\reflang^\Gamma$
  contains all strings of variable operations over $\Gamma$ that can be
  completed to a valid ref-word by adding a prefix and a suffix. Both, the
  prefix and/or the suffix can be empty. We remind that $m_v \in M_V$ is a set
  of variable operations, except for the case $m_v=0$. 

  Now we are ready to define the function $f \colon M_V \to M_A$.
  \[
    f(m_v) \;\eqdef\; \begin{cases}
      \emptyset & \text{if } m_v=0, \text{and}  \\
      \big\{(q_1,q_2) \mid \text{there is a string } \refWord \in
    \reflang^{m_v} \text{, such that } q_2 \in \delta^*(q_1,\refWord) \big\} &
    \text{otherwise.}
    \end{cases}
  \]
  We remind that $m_v$ is a set of variable operations, except for the case
  $m_v=0$. Finally, we can define the multiplication operation of $M_A^\prec$. There
  are four different cases depending on whether the operands are from $M_V$
  or from $M_V \times M_A \times M_V$.
  \begin{align*}
    m_{v_1} \mop_A^\prec m_{v_2} \;\eqdef\;{}& m_{v_1} \mop_V m_{v_2} \\
    m_{v_2} \mop_A^\prec \big(m_{v_2},m_a,m_{v_3}\big) \;\eqdef\;{} & \big(m_{v_1} \mop_V m_{v_2},\; m_a,\; m_{v_3}\big) \\
    \big(m_{v_1},m_a,m_{v_2}\big) \mop_A^\prec m_{v_3} \;\eqdef\;{} & \big(m_{v_1},\; m_a,\; m_{v_2} \mop_V m_{v_3}\big) \\
    \big(m_{v_1},m_{a_1},m_{v_2}\big) \mop_A^\prec \big(m_{v_3},m_{a_2},m_{v_4}\big) \;\eqdef\;{} & \big(m_{v_1},\; m_{a_1} \mop_A f(m_{v_2} \mop_V m_{v_3}) \mop_A m_{a_2},\;m_{v_4}\big)
  \end{align*}
  
  We remind that $\mop_V$ denotes the multiplication of $M_V$ and
  $\mop_A$ denotes the multiplication of $M_A$. It remains to show that $M_A^\prec$
  accepts $\reflang^\toSpanner{A}$. We use the homomorphism, induced
  by
  \[
    h_A^\prec(a) \;\eqdef\;
    \begin{cases}
      h_V(a) & \text{if } a \in \varop{V} \\
      \big(\emptyset,h_A(a),\emptyset\big) & \text{if } a \in \Sigma\;, \\
    \end{cases}
  \]
  that maps variable operations to the corresponding elements of $m_V$ and
  symbols to the corresponding elements from $m_A$. We define $M_A^{\prec\acc}$ as
  \begin{align*}
    M_A^{\prec\acc} \;\eqdef\;{} &\big\{ m \in M_V \;\mid\; f(m) \in M_A^\acc\big\}\; \cup \\
                       & \big\{ (m_{v_1}, m_a, m_{v_2}) \in M_V \times M_A \times M_V \;\mid\;
                         f(m_{v_1}) \mop_A m_a \mop_A f(m_{v_2}) \in M_A^\acc \big\}\;.
  \end{align*}
  The top row corresponds to the case that the document is empty, i.e., the
  ref-word consists only of variable operations, while the bottom row
  corresponds to non-empty documents. To determine whether a ref-word should be
  accepted, we have to incorporate the variable operations before the first and
  after the last symbol from $\Sigma$. Then, we can use $M_A^\acc$ to
  check whether we should accept.
  
  It remains to show that $\{ \refWord \mid h_A^\prec(\refWord) \in
  M_A^{\prec\acc}\} = \reflang^{\toSpanner{A}}$. Let $\refWordPrime \in
  \reflang^{\toSpanner{A}}$, $\tup \eqdef \toTuple{\refWordPrime}$, and $\doc
  \eqdef \clr(\refWordPrime)$. Thus, it must hold that $\tup \in
  \toSpanner{A}(\doc)$ and there is a valid ref-word $\refWord \in \reflang(A)$
  which is accepted by $A$, such that $\toTuple{\refWord} = \tup$ and
  $\clr(\refWord) = \doc$. Per definition of $\reflang^{\toSpanner{A}}$ it
  follows that $\refWordFrom{\doc}{\tup} = \refWordPrime$. We have to show that
  $h_A^\prec(\refWordPrime) = h_A^\prec(\refWordFrom{\doc}{\tup}) \in
  M_A^{\prec\acc}$. We decompose $\refWord$ as
  \[
    \refWord \;=\; V_0 \cdot \doc_1 \cdot V_1 \cdot \doc_2 \cdot V_2 \cdots V_{k-1}
    \cdot \doc_k \cdot V_k\;
  \]
  and $\refWordFrom{\doc}{\tup}$ as
  \[
    \refWordFrom{\doc}{\tup} \;=\; V_0' \cdot \doc_1' \cdot V_1' \cdot \doc_2' \cdot V_2' \cdots V_{\ell-1}'
    \cdot \doc_k' \cdot V_\ell'\;
  \]
  where $V_i,V_i' \in \varop{V}^*$ and $\doc_j,\doc_j' \in \Sigma^*$. As
  both ref-words encode the same tuple for the same document, we have that
  $k=\ell$, $\doc_i=\doc_i'$, and $V_j'$ is a permutation of the symbols in
  $V_j$ for $0 \leq i \leq k$ and $1 \leq j \leq k$. By definition of $M_A$ and
  $M_A^\prec$, we get that
  \begin{align*}
    h_A(\refWord)& \;=\; h_A(V_0) \;\mop_A\; h_A(\doc_1) \;\mop_A\; \cdots \;\mop_A\; h_A(\doc_k) \;\mop_A\; h_A(V_k) \;\in\; M_A^\acc\\
    h_A^\prec(\refWordFrom{\doc}{\tup})& \;=\; h_A^\prec(V_0') \;\mop_A^\prec\; h_A^\prec(\doc_1) \;\mop_A^\prec\; \cdots \;\mop_A^\prec\; h_A^\prec(\doc_k) \;\mop_A^\prec\; h_A^\prec(V_k') \\
                 &\hspace{-1.7cm}\overset{\mathclap{\small{(1)}}}{=}\; h_V(V_0') \;\mop_A^\prec\; \big(\emptyset,h_A(\doc_1),\emptyset\big) \;\mop_A^\prec\; \cdots \;\mop_A^\prec\; \big(\emptyset, h_A(\doc_k), \emptyset\big) \;\mop_A^\prec\; h_V(V_k') \\
                 &\hspace{-1.7cm}\overset{\mathclap{\small{(2)}}}{=}\;\Big(h_V(V_0'),\; h_A(\doc_1) \;\mop_A\; f(h_V(V_1')) \;\mop_A\; h_A(\doc_2) \;\mop_A\; \cdots \;\mop_A\; f(h_V(V_{k-1}')) \;\mop_A\; h_A(\doc_k),\; h_V(V_k')\Big)
  \end{align*}
  The equality~(1) holds by the definition of $\mop_A^\prec$, which for substrings
  consisting only of variable operations just uses $\mop_V$ and for
  substrings containing only $\Sigma$-symbols uses basically $m_A$. We note that
  $f(\emptyset)=h_A(\varepsilon)$, as $R^\emptyset=\{\varepsilon\}$. The
  equality~(2) can be derived by iteratively applying the definition of
  $\mop_A^\prec$ as often as possible.
  
  By definition of $M_A^{\prec\acc}$, we get that $h_A^\prec(\refWordFrom{\doc}{\tup}) \in
  M_A^{\prec\acc}$ if and only if $m_a$ defined as
  \[
    m_a \;\eqdef\; f(h_V(V_0')) \;\mop_A\; h_A(\doc_1) \;\mop_A\; f(h_V(V_1')) \;\mop_A\;
    h_A(\doc_2) \;\mop_A\; \cdots \;\mop_A\; h_A(\doc_k) \;\mop_A\; f(h_V(V_k'))
  \] is in $M_A^\acc$. As $V_i'$ respects the variable ordering, $h_V(V_i') \neq
  0$ is the set containing all variable operations from $V_i'$. By definition of
  $f$ and the fact that $V_i'$ contains exactly the same variable operations as
  $V_i$, we can conclude that $h_A(V_i) \subseteq f(h_V(V_i'))$ for $0 \leq i
  \leq k$.\footnote{We remind that elements of $M_A$ are sets of pairs of
    states, which we can compare using $\subseteq$.} As the multiplication
  $\mop_A$ is monotone\footnote{That is $m_1 \subseteq m_1'$ and $m_2 \subseteq
    m_2'$ imply $m_1 \mop_A m_2 \subseteq m_1' \mop_A m_2'$ for all
    $m_1,m_2,m_1',m_2' \in M_A$.} and $h_A(V_i) \subseteq f(h_V(V_i'))$, we get
  that $h_A(\refWord) \subseteq m_a$. As, furthermore, $A$ accepts $\refWord$,
  it holds that $h_A(\refWord) \in M_A^\acc$ and due to $M_A^\acc$ being upwards
  closed\footnote{That is $m \subseteq m'$ and $m \in M_A^\acc$ implies that $m'
    \in M_A^\acc$.} we can conclude that $m_a \in M_A^\acc$ and therefore
  $h_A^\prec(\refWordFrom{\doc}{\tup}) \in M_A^{\prec\acc}$. This concludes one
  direction of the proof.

  Let now $\refWord$ be some ref-word, such that $h_A^\prec(\refWord) \in
  M_A^{\prec\acc}$. We have to show that there exists a valid ref-word
  $\refWordPrime \in \reflang(A)$ such that $\clr(\refWord)=\clr(\refWordPrime)$
  and $\toTuple{\refWord}=\toTuple{\refWordPrime}$.

  We decompose $\refWord$ as
  \[
    \refWord \;=\; V_0 \cdot \doc_1 \cdot V_1 \cdot \doc_2 \cdot V_2 \cdots V_{k-1}
    \cdot \doc_k \cdot V_k\;.
  \]
  Observe that $k = 0$, if $h_A^\prec(\refWord) \in M_V$, and $k > 0$ otherwise. By the
  definition of $M_A^{\prec\acc}$ we know that
  \[
    m_\refWord \;\;\eqdef\;\; f(h_V(V_0)) \;\mop_A\; h_A(\doc_1) \;\mop_A\; f(h_V(V_1))
    \;\mop_A\; \cdots \;\mop_A\; h_A(\doc_k) \;\mop_A\; f(h_V(V_k)) \;\;\in\;\;
    M_A^\acc\;.
  \]
  Let now $q_0^V,q_1^\doc,q_1^V,q_2^\doc,\dots,q_k^\doc,q_k^V,q_{k+1}^\doc$ be
  states such that $q_0^V$ is the initial state and $q_{k+1}^\doc$ is some final
  state of $A$ and for $0 \leq i \leq k$ and $1 \leq j \leq k$ it holds that
  \begin{itemize}
  \item $(q_i^V,q_{i+1}^\doc) \in f(h_V(V_i))$; and
  \item $(q_j^d,q_j^V) \in h_A(\doc_j)$.
  \end{itemize}
  Due to $m_\refWord \in M_A^\acc$, and the definition of $m_\refWord$, there
  have to be states $(q_0^V,q_{k+1}^d) \in m_\refWord$, such that $q_0^V$ is
  initial and $q_{k+1}^\doc$ is final. The other states have to exist as
  $\mop_A$ is defined as the composition of relations.

  By the definition of $f$ and the fact that $(q_i^V,q_{i+1}^\doc) \in
  f(h_V(V_i))$, for every $0 \leq i \leq k$, there must be a strings $V_0'\ldots
  V_k' \in \varop{V}^*$ of variable operations, such that $V_i' \in
  \reflang^{h_V(V_i)}$ and $q_{i+1}^\doc \in
  \delta^*(q_i^V,V_i')$, for every $0 \leq i \leq k$. We define
  $\refWordPrime$ as
  \[
    \refWordPrime \; \eqdef \; V_0' \cdot \doc_1 \cdot V_1' \cdot \cdots \cdot
    V_{k-1}' \cdot \doc_k \cdot V_k'\;.
  \]
  
  By the construction $\refWordPrime$ is a valid ref-word such that
  $\clr(\refWord)=\clr(\refWordPrime)$ and $\toTuple{\refWord}=\toTuple{\refWordPrime}$.
  Furthermore, we have that $\delta^*(q_0,\refWordPrime) \cap Q_F \neq
  \emptyset$ and therefore $\refWordPrime$ is accepted by $A$, concluding
  the proof.
\end{proof}

 \section{Proof of Lemma~\ref{lem:lowerbounds-det-disjoint}}\label{apx:smitProof}

\smitProof*

\begin{proof}
  In order to proof this result, we use a reduction by Smit~\cite[Proposition
  3.3.7]{Smit20}, who showed that $\splitcorrectness[\dfvsa]$ is \pspace-hard,
  even if $\splitter$ is disjoint.
  
  We give a reduction from the \pspace~complete problem of DFA union
  universality~\cite{kozen77}. Given deterministic finite automata
  $A_1,\ldots,A_n$ over the alphabet $\Sigma$, the union universality problem
  asks whether
  \[  
    \lang(\Sigma^*) \;\subseteq\; \bigcup_{1 \leq i \leq n} \lang(A_i)\;. \qquad (\dag)
  \] 

  Let $A_1,\ldots,A_n$ be DFAs over the alphabet $\Sigma$ and let $a \notin
  \Sigma$ be a new alphabet symbol. Slightly abusing notation, we define the
  \dfvsa by a hybrid regex-formula, where the automata $A_i$ are plugged in. In
  particular, $A_\spanner = a \lor a^{n+1}\Sigma^*$ and $A_\splitter = x\{a\}
  \lor a\cdot x\{a\} \cdot a^{n-1}\cdot A_1 \lor a^2\cdot x\{a\}\cdot
  a^{n-2}\cdot A_2 \lor \cdots \lor a^{n}\cdot x\{a\}\cdot A_n$. Furthermore,
  let $\spanner = \toSpanner{A_\spanner}$ and $\splitter =
  \toSpanner{A_\splitter}$. We show that the following statements are
  equivalent:
  \begin{enumerate}
    \item \spanner is self-splittable by \splitter,
    \item \spanner is splittable by \splitter,
\item $\dag$ holds,
    \item \splitter covers \spanner.
  \end{enumerate}

  We observe that (1) implies (2). Thus, we only need to show that (2) implies (3),
  (3) implies (4), and (4) implies (1).

  \medskip

  \emph{(2) implies (3):} Assume that $\dag$ does not hold. Therefore, there is
  a document $\doc \in \docs$ with $\doc \notin \lang(A_i)$, for every $1 \leq i
  \leq n$. Thus, $\splitter(a^{n+1} \cdot \doc) = \emptyset$, but $\emptytup \in
  \spanner(a^{n+1}\cdot \doc)$, which leads to the desired contradiction that
  \spanner can not be splittable by \splitter.

  \medskip

  \emph{(3) implies (4):} Assume that $\dag$ holds. Let $\doc' \in (\Sigma \cup
  \{a\})^*$ be a document and $\tup \in \spanner(\doc')$ be a tuple. Recall that
  $A_\spanner = a \lor a^{n+1}\Sigma^*$ and $A_\splitter = x\{a\} \lor a\cdot
  x\{a\} \cdot a^{n-1}\cdot A_1 \lor a^2\cdot x\{a\}\cdot a^{n-2}\cdot A_2 \lor
  \cdots \lor a^{n}\cdot x\{a\}\cdot A_n$. As $\spanner$ does not use any
  variables, we have that $\tup = \emptytup$. We make a case distinction on
  $\doc'$:
  \begin{itemize}
  \item $\doc' = a$,
  \item $\doc' \in \lang(a^{n+1}\Sigma^*)$,
  \item $\doc' \notin \{a\} \cup \lang(a^{n+1}\Sigma^*)$.
  \end{itemize}
  If $\doc' = a$, we have that $\splitter(\doc') = \{\spanFromTo{1}{2}\}$ and
  therefore the cover condition is satisfied. On the other hand, if $\doc' \in
  \lang(a^{n+1}\Sigma^*)$ there is a document $\doc \in \docs$ such that $\doc'
  = a^{n+1}\doc$. Thus, there is an index $1 \leq i \leq n$, such that $\doc \in
  \lang(A_i)$ and therefore $\spanFromTo{i+1}{i+2} \in \splitter(\doc')$,
  covering $\emptytup$. In the last case, $\spanner(\doc') = \emptyset$ which
  contradicts the assumption that $\tup \in \spanner(\doc')$.

  \medskip

  \emph{(4) implies (1):} We will show that $\spanner = \spanner \circ
  \splitter$. Let $\doc \in (\Sigma \cup \{a\})^*$ be a document and let $\tup
  \in \spanner(\doc)$ be a \doc-tuple. Again, as $\spanner$ does not use any
  variables, we have that $\tup = \emptytup$. As \splitter covers \spanner,
  there is a split $s \in \splitter(\doc)$ which covers $\tup$. Using $\emptytup
  =\shiftSpanBy{\emptytup}{s}$ and $\emptytup \in (\spanner \circ
  \splitter)(\doc)$, we can conclude that $\spanner \subseteq \spanner \circ
  \splitter$. For the other direction, let $\doc \in \docs$ be a document and
  $\tup \in (\spanner \circ \splitter)(\doc)$. As \spanner does not use any
  variables, we have that $\tup=\emptytup$ and by definition of \spanner we have
  that $\emptytup \in \spanner(\doc)$ showing $\spanner \circ \splitter
  \subseteq \spanner$.
\end{proof}

\end{document}